\documentclass[twocolumn]{aastex63}
\usepackage{amsmath}
\usepackage{makecell}

\newcommand{\siiv}{Si\,{\sc iv}}

\newcommand{\civ}{C\,{\sc iv}}
\newcommand{\ciii}{C\,{\sc iii]}}
\newcommand{\mgii}{Mg\,{\sc ii}}

\newcommand{\feii}{Fe\,{\sc ii}}
\newcommand{\oiii}{[O\,{\sc iii}]}

\submitjournal{AJ}

\shorttitle{Lensed Quasars and Quasar Pairs at $z\sim5$}
\shortauthors{Yue et al.}
\graphicspath{{./}{figures/}}

\begin{document}

\title{A Survey for High-redshift Gravitationally Lensed Quasars and Close Quasars Pairs.\\
I. the Discoveries of an Intermediately-lensed Quasar and a Kpc-scale Quasar Pair at $z\sim5$}

\correspondingauthor{Minghao Yue}
\email{yuemh@arizona.edu}

\author[0000-0002-5367-8021]{Minghao Yue}
\affiliation{Steward Observatory, University of Arizona, 933 North Cherry Avenue, Tucson, AZ 85721, USA}
\affiliation{MIT Kavli Institute for Astrophysics and Space Research, 77 Massachusetts Ave., Cambridge, MA 02139, USA}

\author[0000-0003-3310-0131]{Xiaohui Fan}
\affiliation{Steward Observatory, University of Arizona, 933 North Cherry Avenue, Tucson, AZ 85721, USA}

\author[0000-0001-5287-4242]{Jinyi Yang}
\altaffiliation{Strittmatter Fellow}
\affil{Steward Observatory, University of Arizona, 933 North Cherry Avenue, Tucson, AZ 85721, USA}

\author[0000-0002-7633-431X]{Feige Wang}
\altaffiliation{NHFP Hubble Fellow}
\affiliation{Steward Observatory, University of Arizona, 933 North Cherry Avenue, Tucson, AZ 85721, USA}



\begin{abstract}
We present the first results from a new survey for high-redshift $(z\gtrsim5)$
gravitationally lensed quasars and close quasar pairs.
We carry out candidate selection based on the colors and shapes of objects in public imaging surveys,
then conduct follow-up observations to confirm the nature of high-priority candidates.
In this paper, we report the discoveries of J0025--0145 ($z=5.07$) which we identify as an  {intermediately-lensed quasar,
and J2329--0522 ($z=4.85$) which is a kpc-scale close quasar pair.}
 {The {\em Hubble Space Telescope (HST)} image of J0025--0145 shows a foreground lensing galaxy located $0\farcs6$ away from the quasar.
However, J0025--0145 does not exhibit multiple lensed images of the quasar, 
and we identify J0025--0145 as an intermediate lensing system (a lensing system that is not multiply imaged but has a significant magnification). 
The spectrum of J0025--0145 implies an extreme Eddington ratio if the quasar luminosity is intrinsic,
which could be explained by a large lensing magnification. 
The {\em HST} image of J0025--0145 also 
indicates a tentative detection of the quasar host galaxy in rest-frame UV, 
illustrating the power of lensing magnification and distortion
in studies of high-redshift quasar host galaxies.}
J2329--0522 consists of two resolved components with significantly different spectral properties, and a lack of lensing galaxy detection under sub-arcsecond seeing. We identify it as a close quasar pair, which is the highest confirmed kpc-scale quasar pair to date. 
We also report four lensed quasars and quasar pairs at $2<z<4$,
and discuss possible improvements to our survey strategy.
\end{abstract}



\section{Introduction} \label{sec:intro}

Gravitationally lensed quasars are rare and valuable objects
that enable important studies in extragalactic astronomy and cosmology.
Measuring the time lags between the lensed images of the background quasar
gives independent constraints on key cosmological parameters like the Hubble constant
\citep[e.g.,][]{wong19,shajib20}.
Microlensing events caused by stars in the lensing galaxy 
have been used to probe the structure of quasar accretion disks \citep[e.g.,][]{kochanek04,cornachione20}.
Analyzing the lensing model puts constraints on the properties
of dark matter \citep[e.g.,][]{gilman20}.
Measuring the absorption features {in the spectra of the lensed quasar images} 
probes the structure of the foreground galaxy's circumgalactic medium \citep[e.g.,][]{cashman21}.
With the help of lensing magnification, lensed quasars offer unique opportunities
to measure the properties of quasars with enhanced sensitivity \citep[e.g.,][]{yang19lens,yang22}
and investigate the small-scale structures in the quasar host galaxies 
\citep[e.g.,][]{bayliss17,paraficz18, yue21a}.

Physical pairs of quasars are another population of valuable objects
that provide critical information about the
evolution of supermassive black holes (SMBHs) and their host galaxies.
Quasar pairs with separations of $\gtrsim100$ kpc have been used to
measure the clustering of quasars up to $z\sim5$ \citep[e.g.,][]{mcgreer16,eft17}.
Kpc-scale quasar pairs are likely the results of ongoing galaxy mergers,
in which the SMBHs in both progenitor galaxies are ignited by the merging event
\citep[for a recent review, see][]{dr19}.
Comparison between detailed observations of close quasar pairs \citep[e.g.,][]{green10}
and zoom-in hydrodynamical simulations of galaxy mergers \citep[e.g.,][]{capelo17}
helps us to understand how galaxy mergers trigger quasar activities.
The statistics of quasar pairs (e.g., the fraction of close pairs among all quasars)
encode information about the coevolution of SMBHs, their host galaxies,
and their close environments \citep[e.g.,][]{steinborn16,volonteri16,shen22}.

Given the importance of these objects, 
there have been intensive efforts to search for
gravitationally lensed quasars 
\citep[e.g.,][]{inada12,more16,anguita18,treu18,lemon18,fan19,lemon19,lemon20}
and quasar pairs \citep[e.g.,][]{hennawi06,hennawi10,silverman20,tang21,chen22a}.
Most of these studies use public imaging surveys
such as the Sloan Digital Sky Survey 
\citep[SDSS;][]{sdss}
to select candidates that have colors and shapes
consistent with lensed quasars or quasar pairs,
then confirm the nature of the candidates with follow-up
spectroscopy and/or high-resolution imaging.
Interestingly, the observational features of arcsecond-scale quasar pairs 
are very similar to that of doubly-imaged lensed quasars
with faint lensing galaxies \citep[e.g.,][]{shen21}. 
As a result, surveys for lensed quasars naturally lead to discoveries
of close quasar pairs \citep[e.g.,][]{lemon22}
and vice versa \citep[e.g.,][]{chen22b}.

In the past few years,
the {\em Gaia} mission \citep{gaia} has revolutionized the
searches for lensed quasars and close (arcsecond-scale) quasar pairs.
{\em Gaia} is an all-sky imaging and spectroscopic survey 
that primarily aims at mapping the stars in the Milky Way.
Using a space-based telescope, {\em Gaia} delivers {an effective angular resolution}
of $\sim0\farcs4$ in Data Release (DR) 2 \citep{gaiadr2}.
The astrometric properties measured by {\em Gaia},
especially the parallax and proper motions of objects,
are useful in identifying lensed quasars and suppressing 
the contamination of galactic stars \citep[e.g.,][]{lemon17}.
\citet{lemon22} show that the number of known of lensed quasars
at redshift $z<1.5$ with separations $\Delta \theta>1''$ 
that are detected in the {\em Gaia} survey
is consistent with model predictions,
suggesting a high completeness for {\em Gaia}-based lens searches.

Despite the success of previous studies, to date
it is still unclear how to efficiently find 
high-redshift lensed quasars and quasar pairs.
{Although some lensed quasars at $z\sim4$
have been discovered using {\em Gaia} data \citep[e.g.,][]{desira22},}
most $z\gtrsim5$ quasars are {too faint to be detected} in the {\em Gaia} survey.
In addition, the relatively poor physical resolution of most observations 
makes it challenging to find kpc-scale quasar pairs at high redshifts.
So far, only one lensed quasar has been reported at $z>5$
\citep[][]{fan19}
and only one kpc-scale quasar pair has been confirmed at $z>3$
\citep[][]{tang21}.
These numbers are much lower than predictions from simulations 
\citep[e.g.,][]{om10,dr19}.

In order to draw a comprehensive picture of lensed quasars and quasar pairs,
we need to investigate how to efficiently identify
faint distant lensed quasars and quasar pairs from imaging surveys.
This task is urgent given the upcoming wide-area sky surveys like 
the Vera C. Rubin Observatory Legacy Survey for Space and Time \citep[LSST;][]{lsst}
and the {\em Euclid} survey \citep[][]{euclid}. In particular, LSST will deliver imaging that is several magnitudes
deeper than {\em Gaia}, meaning that {\em Gaia}-based survey methods
will be less useful in the LSST era.

The above considerations motivate us to design and carry out a new survey for 
lensed quasars and close quasar pairs,
with a focus on high-redshift ones $(z\gtrsim5)$.
We explore strategies that make the most use of current imaging surveys,
which can also be applied to upcoming sky surveys like LSST.
This paper reports the lensed quasars and close quasar pairs
discovered in the initial stage of this survey.

The layout of this paper is as follows.
We describe the data sets and the method we use for candidate selection in Section \ref{sec:survey}.
We describe how we confirm the nature of candidates with follow-up observations in Section \ref{sec:followup}.
{We then present the discoveries of an intermediately-lensed quasar at $z=5.07$ (J0025--0145, Section \ref{sec:J0025})
and a kpc-scale quasar pair at $z=4.85$ (J2329--0522, Section \ref{sec:J2329}).}
We also report a few lensed quasars and close quasar pairs at lower redshifts in Section \ref{sec:results:objects}.
We discuss future work to improve this survey in Section \ref{sec:discussion}.
We use a flat $\Lambda$CDM cosmology with 
$H_0=70\text{ km s}^{-1} \text{ Mpc}^{-1}$ and $\Omega_M=0.3$.

\section{Building the Candidate Sample} \label{sec:survey}

\begin{figure*}
    \centering
    \includegraphics[trim={0.5cm, 7.2cm, 2.4cm, 2cm}, clip, width=1\linewidth]{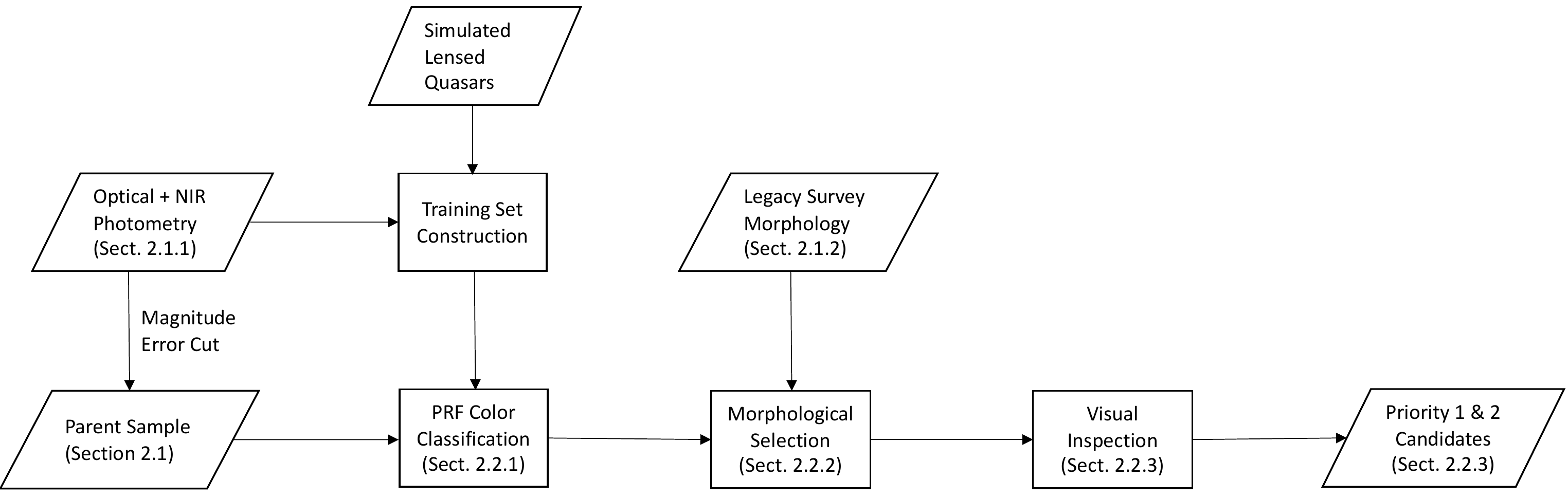}
    \caption{The flowchart of the candidate selection pipeline.}
    \label{fig:flowchart}
\end{figure*}

This section briefly summarizes the public data sets 
and the candidate selection method used in our survey.
 {Figure \ref{fig:flowchart}
presents a flowchart of the candidate selection pipeline.}

\subsection{Data Sets}

\subsubsection{The Parent Sample and Multi-band Photometry} \label{sec:survey:parent}

We construct our parent sample of candidates
using a number of optical and infrared sky surveys.
The public surveys we use in this work are
summarized in Table \ref{tbl:survey}.
Specifically, we start with objects in 
the Dark Energy Survey DR2  \citep[DES;][]{desdr2}
and the Pan-STARRS1 survey DR2 \citep[PS1;][]{ps1}. 
DES covers $\sim5,000$ deg$^2$ of high-galactic-latitude area
in the Southern Galactic Cap, providing deep imaging
in the $g, r, i, z$ and $Y-$bands.
PS1 covers sky areas at $\delta>-30^\circ$ with 
$g,r,i,z$ and $y$-band imaging.

\begin{deluxetable*}{c|cccc}
\label{tbl:survey}
\tablecaption{Information of Public Surveys Used in the Candidate Selection}
\tablewidth{0pt}
\tablehead{\colhead{Survey} & \colhead{Filters\tablenotemark{1}} & \colhead{PSF size\tablenotemark{2}} & \colhead{Depth\tablenotemark{3}} & \colhead{Reference} \\
\colhead{} & \colhead{} & \colhead{$('')$} & \colhead{(mag)} & \colhead{}}
\startdata
\hline
DES     & $g,r,i,z,Y$ & 1.11, 0.95, 0.88, 0.83, 0.90 & 25.4, 25.1, 24.5, 23.8, 22.4 & \citet{desdr2}\\
PS1     & $g,r,i,z,y$ & 1.31, 1.19, 1.11, 1.07, 1.02 & 23.3, 23.2, 23.1, 22.3, 21.3 & \citet{ps1} \\
VHS     & $J,K_s$ & 0.99, 0.91 & 20.8, 20.0 & \citet{vhs}\tablenotemark{4}\\
UKIDSS  & $J,H,K$ & $<1.2$ & 19.9, 18.6, 18.2 & \citet{ukidss} \\
UHS     & $J$ & $<1.2$ & 19.9 & \citet{uhs}\\
{\em WISE} & $W1,W2$ & 6, 6 & 18,17 & \citet{catwise} \\\hline
Legacy  & $g,r,z$ &  1.29, 1.18, 1.11 & 23.95, 23.54, 22.50 & \citet{legacy}\\
{\em Gaia} & $G$ & 0.4 & 21 & \citet{gaiadr2}
\enddata
\tablenotetext{1}{Magnitudes used in the color classification (except for the Legacy Survey and the {\em Gaia} survey).}
\tablenotetext{2}{The full-width half maximum (FWHM) of the PSF, in arcsecond.}
\tablenotetext{3}{The $5-\sigma$ depth for point sources. We use AB magnitudes for the $g,r,i,z$ and $y(Y)-$bands and Vega magnitudes for the $J, H, K(K_s), W1$ and $W2-$bands.}
\tablenotetext{4}{Also see https://www.eso.org/rm/api/v1/public/releaseDescriptions/144}
\tablecomments{We use the Legacy Survey to gather morphological information of objects,
and use {\em Gaia} to obtain parallaxes and proper motions of objects.
The other surveys are used to obtain the magnitudes of objects.}
\end{deluxetable*}

We obtain near-infrared (NIR) photometry   from 
the Vista Hemisphere Survey \citep[VHS;][]{vhs}, 
the UKIRT Infrared Deep Sky Survey \citep[UKIDSS;][]{ukidss}, 
and the UKIRT Hemisphere Survey \citep[UHS;][]{uhs}.
VHS covers most parts of the southern hemisphere in $Y, J, H$,
and $K_s$ bands.
UKIDSS covers $7,000\text{ deg}^2$ in the northern hemisphere 
with the $Z, Y, J, H$ and $K-$bands, 
and UHS observes the area at $0^\circ<\delta<60^\circ$ in $J$ and $K-$bands
that are not included in UKIDSS.
When we start the candidate selection, 
UKIDSS finished all the observations,
while UHS has only released the $J-$band photometry and 
VHS has only released the $J$ and $K_s-$band photometry for most sky areas.
As a result, the exact photometry available for an object
depends on its position on the sky.
We further obtain the {\em WISE} W1 and W2 magnitudes 
from the CatWISE2020 catalog \citep{catwise},
which is the deepest {\em WISE} data release available 
when the candidate selection was carried out.

Most lensed quasars and kpc-scale quasar pairs 
have separations of $\lesssim3''$ \citep[e.g.,][]{collett15}.
To include flux from all the components of these objects, 
we use aperture magnitudes with a diameter of $5''$ in our candidate selection,
with the exception of {\em WISE} W1 and W2 bands that 
 have PSF sizes of $\sim6''$.
We use a matching radius of $2''$ when cross-matching the survey catalogs.

We then construct the parent sample of objects  by selecting objects 
at high galactic latitudes with good photometry.
The exact criteria are
\begin{equation} \label{eq:parent}
\begin{gathered}
    |b|>30^\circ,\\
    m_z<20, \sigma_{m_i}<0.2, \sigma_{m_z}<0.1,\\
    \sigma_{m_X}<0.3, X \in [r,y(Y),J,H,K(K_s),W1,W2].
\end{gathered}
\end{equation}
Here, we do not have requirement on the $g-$band magnitude error,
as quasars at $z\gtrsim5$ have little or no flux at wavelengths 
bluer than the redshifted Ly$\alpha$ \citep[e.g.,][]{worseck14}.
This parent sample goes into the candidate selection pipeline described in 
Section \ref{sec:survey:candselect}.

We divide the parent sample into point sources and extended sources.
For objects in the DES field, we use $\texttt{spread\_model\_z}<0.005$ as the criterion
of point sources \citep[e.g.,][]{sn18};
for objects in the PS1 field, the criterion is $\text{\texttt{zPSFMag}}-\text{\texttt{zApMag}}<0.1$,
which is determined using the same approach as \citet{farrow14}.

\subsubsection{Morphological Information}

The morphological information of objects
is crucial in surveys of lensed quasars and quasar pairs
\citep[e.g.,][]{oguri06}. In this work,
we obtain morphological information from the 
DESI Legacy Imaging Survey \citep[hereafter the Legacy Survey;][]{legacy} DR8.
The Legacy Survey provides $g,r$ and $z-$band imaging for $\sim20,000\text{ deg}^2$
of sky areas at $|b|>18^\circ$, which is used in the target selection 
 of the Dark Energy Spectroscopy Instrument \citep[DESI;][]{desi}.
The $z-$band imaging has the sharpest PSF among all the three bands,
which has a full-width half maximum (FWHM) of $\sim1\farcs1$.
Starting from DR8, the Legacy Survey also includes 
the DES $g,r$, and $z-$band images, 
reduced using the same Legacy Survey photometric pipeline.

The Legacy Survey measures the magnitudes and the best-fit flux profiles of objects
using {\em Tractor} \citep{tractor}.
{\em Tractor} is a python-based image fitting and photometry tool. 
Briefly speaking, {\em Tractor} fits the images of an object 
as a point source and a S\'ersic profile,
then picks up the best-fit model based on the reduced $\chi^2$.
{\em Tractor} also generates residual images for the best-fit model.
The source detection and fitting pipeline can be executed iteratively,
i.e., new sources can be detected in the residual image of the previous iteration. 
In this way, the Legacy Survey can detect faint objects close to a bright object.

The best-fit image models and the residuals
provided by {\em Tractor} is useful in identifying 
objects with close companions.
Specifically, many lensed quasars and kpc-scale quasar pairs have 
separations of $\Delta \theta\lesssim1''$ \citep[e.g.,][]{yue22b,shen22}
and are blended in ground-based images.
When fitted as a PSF or a S\'ersic profile,
these objects will exhibit large image fitting residuals.
 {We thus obtain the residual images of objects and the corresponding reduced $\chi^2$
 from the Legacy Survey for candidate selection.}\footnote{ {Note that the Legacy Survey covers the whole DES footprint starting from DR8.
Although some objects in the PS1 survey are not covered by the Legacy Survey, these objects have low galactic latitudes $(|b|\lesssim20^\circ)$ and are not included in the Parent Sample (Section \ref{sec:survey:parent}}).}

\subsection{Candidate Selection} \label{sec:survey:candselect}

The candidate selection method in this work is based on the colors and shapes of objects.
Specifically, we first identify objects with colors consistent with a lensed quasar or a quasar pair,
then select candidates that have shapes suggestive of lensing structures
or close companions.

\subsubsection{Color Classification} \label{sec:survey:candselect:color}

For each object in the parent sample,
we use a probabilistic random forest \citep[PRF;][]{prf}
to predict the probability of the object to be 
a lensed quasar, a normal galaxy, and a galactic star.
Close quasar pairs have colors similar to lensed quasars
with faint deflector galaxies
and are thus covered by the class of lensed quasars.
PRFs are variations of Random Forests (RFs), 
a class of supervised machine learning algorithm
that is widely used in classification and regression.
Unlike RFs where the features of data points have no errors,
PRFs take the error of features into account by 
considering each data point as a probabilistic distribution.
PRFs are thus the ideal choice for classifying and fitting noisy data.

We construct training sets for galactic stars and normal galaxies
using point sources and extended sources from the parent sample.
Specifically, we select objects with magnitude errors 
smaller than 0.1 in all bands as training set objects.
We notice that these training sets might contain
objects that are not galactic stars or normal galaxies,
e.g., quasars and blended pairs of stars.
Nevertheless, these objects have little impact on the color classification,
as normal stars and galaxies dominate the training sets.

We generate training sets for lensed quasars
using simulated quasars and galaxies.
We do not use observed lensed quasars due to their limited numbers,
especially at high redshifts. 
Specifically, we use SIMQSO \citep[][]{simqso} to generate mock quasars at $z>4.5$
following the method in \citet{yue22b},
and use simulated galaxies from the JAGUAR mock galaxy catalog \citep[][]{jaguar}
to model the foreground deflector galaxies.
We estimate the velocity dispersion $\sigma$ of the simulated galaxies 
using their stellar mass and S\'ersic parameters
based on the relation in \citet{bezanson11}.
We then calculate the Einstein radius $\theta_E$
of all {pairs of mock quasars and simulated galaxies}
by assuming a Singular Isothermal Sphere (SIS) lensing model,
\begin{equation} \label{eq:thetaE}
    \theta_E=4\pi\left(\frac{\sigma}{c}\right)^2\frac{D_\text{ds}}{D_\text{s}}
\end{equation}
where $\sigma$ is the velocity dispersion of the galaxy,
$D_\text{s}$ is the angular diameter distance from the observer to the source,
and $D_\text{ds}$ is the angular diameter distance from the deflector to the source.
In SIS lensing models, the strong lensing cross section of 
a deflector galaxy is $\pi\theta_E^2$. Accordingly, 
we generate mock lensed quasars by
randomly drawing 1,000 simulated galaxies as deflectors for each mock quasar
using a weight of $\theta_E^2$.

For the purpose of color classification,
we only need the colors of the mock lensed quasars
instead of the accurate lensing models.
We thus randomly draw a magnification $\mu$ from the distribution $P(\mu)=8/\mu^3$
for each mock lensed quasar.
This distribution corresponds to SIS deflectors
and is a good approximation for galaxy-scale lensing systems \citep[e.g.,][]{yue22a}.
We then calculate the synthetic magnitudes of the lensing system by summing the
spectra of the deflector galaxy and the magnified spectra of the background quasar.

Because there are multiple optical and NIR imaging surveys
used to construct the parent sample, it is necessary to train
several PRFs to account for the differences in filter response curves
between the imaging surveys.
To be specific, we consider the following combinations of filters that cover
all objects in the parent sample:
\begin{itemize}
    \item PS1 $g,r,i,z,y$ + UHS $J$ + {\em WISE} W1, W2
    \item PS1 $g,r,i,z,y$ + UKIDSS $J, H, K$ + {\em WISE} W1, W2
    \item PS1 $g,r,i,z,y$ + VHS $J, K_s$ + {\em WISE} W1, W2
    \item DES $g,r,i,z,Y$ + VHS $J, K_s$ + {\em WISE} W1, W2
\end{itemize}
We then randomly draw $10^6$ objects from each training set
to train the PRFs.
We use flux ratios between other bands and the $z-$band
as features of the PRFs.
{We use the PRFs to predict the probabilities for each parent sample object
to be a star, a galaxy, or a lensed quasar}, and select objects that satisfy
\begin{equation} \label{eq:colorselect}
    P(\text{star})<0.1 \text{ and } P(\text{lensQSO})>0.3
\end{equation}
This criterion selects non-stellar objects
that have some chances to be a lensed quasar. 
We set a relatively low threshold in $P(\text{lensQSO})$ to
include deflector-dominated lensed quasars 
that may have high $P(\text{galaxy})$ values.
Objects selected by this criterion enter the morphological selection
described in the following subsection.

\subsubsection{Morphological Selection}

The second step of the candidate selection is to identify
objects that have shapes indicative of a possible lensing structure
or close companions.
In imaging surveys, lensed quasars and quasar pairs should either
be resolved into multiple components, or appear to be one object 
that cannot be well-described by regular flux profiles
(e.g., a PSF and a S\'ersic profile). Accordingly,
we match the objects that pass the color classification
to the Legacy Survey DR8 {\em Tractor} catalog using a matching radius of $4''$,
and select two samples of candidates:

(1) {\em Resolved candidates.} 
For objects that are resolved into multiple components in the Legacy Survey
(i.e., with more than one match in the {\em Tractor} catalog), 
we measure the smallest separation between the resolved components 
$(\Delta \theta_\text{min})$  and
select objects with $\Delta \theta_\text{min}<3''$ as candidates.
 {This separation limit is set to reduce the contamination 
from projected pairs of objects, as most lensed quasars and kpc-scale
quasar pairs have separations less than $3''$.}

(2) {\em Unresolved candidates.}
{We select objects with large $z-$band image fitting reduced $\chi^2$ ({\texttt{rchisq\_z}})
as candidates of small-separation (i.e., unresolved) lensed quasars and quasar pairs.}
Here we use the $z-$band image fitting residual 
because the $z-$band has a sharper PSF than the $g$ and the $r-$band.
We note that the distribution of {\texttt{rchisq\_z}} depends
on the magnitudes of objects, as the imperfectness of flux models
have higher statistical significance for brighter objects.
In this work, we adopt a selection criterion of
{\texttt{rchisq\_z}}$>1.55+0.05\times|{\texttt{mag\_z}}-20.5|^{4.5}$,
where {\texttt{mag\_z}} is the $z-$band magnitude in the Legacy Survey.
This criterion is tuned to include at least 90\% of the known lensed quasars
\footnote{Obtained from https://research.ast.cam.ac.uk/lensedquasars/}
\citep{lemon19} in the parent sample.
{We note that objects with close companions may also be selected
into this candidate sample, which allows us to include large-separation
lensed quasars that have one component poorly fitted by a single flux profile
(e.g., when the lensing galaxy is blended with one of the lensed quasar images).}

{Table \ref{tbl:ncand} lists the number of candidates that pass each step in the
candidate selection. The columns represent different combinations of photometric surveys,
as described in Section \ref{sec:survey:candselect:color}. 
About $0.1\%$ of the parent sample candidates pass the color selection and the morphological selection,
and the selection efficiency depends on the depth and the seeing of imaging surveys. 
We visually inspect the selected candidates to identify high-priority ones (Section \ref{sec:survey:candselect:vis}).
We note that the current candidate selection criteria (especially the color selection)
are designed to include as many candidates as possible while keeping 
visual inspection feasible. In the future,
the color selection and morphological selection will be improved by adopting
new data releases from sky surveys and updating the candidate selection methods,
and the visual inspection can be replaced by machine-learning-based methods.
More discussions about possible improvements to the candidate selection
can be found in Section \ref{sec:discussion}.}


\begin{deluxetable*}{c|cccc}
\label{tbl:ncand}
\tablecaption{Number of Candidates Selected in Each Step for Each Subset}
\tablewidth{0pt}
\tablehead{\colhead{} & {PS1+UKIDSS} & {PS1+UHS} & {PS1+VHS} & {DES+VHS}}
\startdata
\hline
Parent Sample & {12,240,526} & 22,613,325 & 27,100,453 & 29,417,942\\
Pass Color Selection & {52,810} & 54,290 & 158,168 & 78,616\\
Pass Morphological Selection\tablenotemark{1} & 17,492(r) + 8,837(u) & 11,462(r) + 9,185(u) & 19,993(r) + 7,610(u)& 11,728(r) + 5,136(u) \\\hline
Pass Visual Inspection\tablenotemark{2} & \multicolumn{4}{c}{88 (Priority 1) + 136 (Priority 2)} \\
\enddata
\tablenotetext{1}{``r" stands for resolved candidates and ``u" stands for unresolved candidates.}
\tablenotetext{2}{We do not distinguish the subsets of candidates when performing visual inspection.}
\tablecomments{Each column in this table correspond to a subset of candidates. 
The subsets are determined by the photometry used in color selection. 
For example, the subset ``PS1+UKIDSS" contains objects that have photometry from the PS1 and the UKIDSS survey.
See Section \ref{sec:survey:candselect:color} for details. 
Also note that {\em WISE} photometry is available for all the candidates.}
\end{deluxetable*}

\subsubsection{Visual Inspection} \label{sec:survey:candselect:vis}


We visually inspect the  {images and spectral energy distributions (SEDs) of the candidates}  
and subjectively classify the candidates into
Priority 1 (likely), Priority 2 (unsure), and
Priority 3 (not likely).
For resolved candidates,
we pay special attention to the fluxes and positions 
of the resolved components.
Objects that contain multiple point sources with similar colors
are assigned high priorities, where the point sources may be 
lensed images of the same quasar or two quasars at the same redshift.
For unresolved candidates, 
if a candidate exhibits the typical features of a quasar,
i.e., a power-law SED
and bumps in the SED suggesting broad emission lines, 
it is assigned a high priority.
{As shown in Table \ref{tbl:ncand},
the majority of the visually-inspected candidates are low-priority candidates,
which include irregular galaxies, galaxy mergers, pairs of galaxies, 
 pairs of point sources with star-like colors,
and pairs of a galaxy and a star-like point source.}

We also match the candidate list to the Million Quasar Catalog \citep[][]{milliquas2021}
to identify known quasars.
We assign priority 1 to known quasars that exhibit a lens-like structure
or have a companion with similar colors.
We further use the parallax and the proper motion
from the {\em Gaia} DR2 \citep{gaiadr2} to exclude galactic stars.
{Specifically, if an object has a non-zero parallax $\varpi$,
defined as
\begin{equation}
    \varpi >3\sigma_\varpi 
\end{equation}
or a non-zero proper motion $\mu$, defined as
\begin{equation}
    \mu=\sqrt{\mu_\alpha^2+\mu_\delta^2} >3\sqrt{\sigma_{\mu_\alpha}^2+\sigma_{\mu_\delta}^2}
\end{equation}
the object is rejected from the candidate list.}
We do not require a candidate to be detected in the {\em Gaia} survey.

Note that our candidate sample contains lensed quasars and quasar pairs at 
low redshifts $(z\lesssim5)$, though the primary targets of this project
are those at high redshifts. In this work,
candidates that exhibit features of low-redshift lensed quasars and quasar pairs
are also assigned high priorities.

The visual inspection returns 88 priority 1 candidates and 136 priority 2 candidates.
We then carry out follow-up imaging and spectroscopy for these candidates
to confirm their nature, 
as described in Section \ref{sec:followup}.
Limited by the telescope resources, we do not observe priority 3 candidates 
in this project.

We end this Section by briefly discussing the survey completeness.
{The completeness of color selection is a function of 
the redshifts and the SEDs of the quasar and, for lensed quasars,
the fluxes of the lensing galaxy and the lensing magnification.
The morphological selection relies on the image reduction pipeline
of the Legacy Survey and the {\em Tractor} algorithm, 
and quantifying their effects on the completeness function
is beyond the scope of this paper.}
The visual inspection further complicates the completeness function.
In addition, the completeness function will change  
as we improve the candidate selection method in the future 
(Section \ref{sec:discussion}).
We thus leave quantitative analyses of the completeness function
to subsequent papers and only provide rough estimates here.
Our morphology selection is similar to that of the SQLS,
which has a completeness of
$\sim50\%$ $(\sim90\%)$ for simulated lensed quasars with separation $\Delta \theta=0\farcs8$ $(1'')$  \citep{oguri06}.
Previous quasar surveys have reported a high completeness for 
RF-based color selection \citep[$\gtrsim95\%$; e.g.,][]{schindler18,khramtsov19}.
We thus expect that the completeness of our color classification and
morphological selection is $\gtrsim80\%$
for marginally-resolved $(\theta\gtrsim1'')$ lensed quasars.
The completeness of quasar pairs might be higher,
as there is no flux contamination from foreground lensing galaxies.

\section{Follow-up Observations} \label{sec:followup}

In semesters 2021A and 2021B, we successfully observed 66 priority 1 candidates
and 10 priority 2 candidates.
Generally speaking, we carry out imaging and spectroscopy for each candidate with sufficient spatial resolution,
where the candidate is resolved into multiple components.
The imaging gives the positions and fluxes of the resolved components,
and the spectroscopy reveals if these components are quasars.
Depending on the angular sizes of the candidates,
the observations may be carried out using ground-based telescopes
under good seeing conditions $(\lesssim0\farcs7)$,
using ground-based telescopes with adaptive optics, 
or using the {\em Hubble Space Telescope (HST)}.

We then classify all candidates into four categories
based on their images and spectra as follows
\citep[also see the discussion in, e.g.,][]{lemon22,chen22b}:
\begin{enumerate}
    \item {\em Lensed Quasars}: 
    Quadruply-lensed quasars are confirmed if the lensing structure is clearly resolved
    in the images. For doubly-lensed quasars, we require that the two lensed images are
    resolved and have similar emission line profiles, and that the lensing galaxy is detected
    in the image. 
    
    \item {\em Physical Quasar Pairs}:
    Physical quasar pairs are confirmed if the two quasars are resolved,
    have a line-of-sight velocity difference of $\Delta v<2000\text{ km s}^{-1}$ \citep[e.g.,][]{hennawi06}, 
    and show significant differences in their spectra. 
    We further require that no lensing galaxy 
    is detected in the field. We discuss the meaning of ``significant difference"
    below with more details.
    
    \item {\em Nearly Identical Quasars (NIQs)}:
    If a candidate contains two quasar images at the same redshift,
    but does not meet the criteria for a lensed quasar or a physical quasar pair,
    the candidate is classified as an NIQ \citep[e.g.,][]{anguita18}. 
    NIQs may be physical quasar pairs
    or doubly-lensed quasars with a faint deflector galaxy,
    and existing data cannot rule out either hypothesis.
    These objects are referred to as ``lensless twins" in \citet{schechter17}.
    
    \item {\em Contaminants}:
    The main contaminants in this project are emission line galaxies (ELGs),
     projected pairs of galaxies and stars, and projected pairs of stars. 
    Specifically, the narrow emission lines in ELGs generate features
    similar to the broad emission lines of quasars in broad-band photometry;
    late-type stars have long been recognized as the main contaminants 
    in surveys of high-redshift quasars due to their red colors \citep[e.g.,][]{yang19}.
    A few contaminants turn out to be projected pairs of a quasar and a star, {or a single quasar without evidence of strong lensing.}
\end{enumerate}

To help understand why we adopt the above criteria,
it is worth discussing the long-standing problem in 
distinguishing close quasar pairs and doubly-lensed quasars
\citep[e.g.,][]{shen21}.
If a candidate contains two quasar images at the same redshift
and no lensing galaxy is detected, 
we cannot rule out the possibility that the system is a physical quasar pair,
even if the two quasar images have nearly identical spectral profiles.
Meanwhile, for lensed quasars,
microlensing and differential reddening of the lensing galaxy 
can lead to small differences between the spectra of the lensed images of the same background quasar
\citep[e.g.,][]{sluse12}.
An apparent pair of quasars with subtle differences in spectral features
might be a doubly-lensed quasar with an undetected deflector
(e.g., a heavily obscured galaxy).
More discussions can be found in \citet{schechter17}, \citet{shen21} and \citet{yue21b}.

As such, we require the detection of the deflector galaxy to confirm a
doubly-lensed quasar. 
Low-redshift quasar pairs can be confirmed by detecting their merging host galaxies
\citep[e.g.,][]{silverman20,chen22c}.
At higher redshifts, it becomes difficult to detect the quasar host galaxies
even with  {\em HST} \citep[e.g.,][]{yue21b}.
In this case,
confirming a physical quasar pair requires significant differences
in spectral shapes that cannot be explained by microlensing and differential reddening.
Examples of such features include
large velocity offsets between emission lines, 
different broad absorption line (BAL) features, and
very different line widths and/or flux ratios.
It is still unclear how to confirm a quasar pair
if there are only subtle differences in the spectra of the two quasars.
In this work, we consider the features of each object individually,
which are presented in Section \ref{sec:J0025} to Section \ref{sec:results:objects}.

So far, we have discovered ten lensed quasars and quasar pairs in this survey. 
Among these objects, three lensed quasars 
(J0803+3908, J0918--0220, and J1408+0422) were independently reported in \citet{lemon22}, 
 and one NIQ was reported in our previous paper \citep[J2037--4537 at $z=5.66$;][]{yue21b}.
 {In this paper, we report the discoveries of 
an intermediately-lensed quasar at $z=5.07$ (J0025--0145, Section \ref{sec:J0025}) 
and a kpc-scale quasar pair at $z=4.85$ (J2329--0522, Section \ref{sec:J2329}).
We also mention four lensed quasars and quasar pairs at $z<4$ found in our survey in Section \ref{sec:results:objects}.}
Table \ref{tbl:obs} summarizes the observations of the lensed quasars and close quasar pairs, 
and Table \ref{tbl:obj} lists the properties of these objects.
The data are available upon request to the corresponding author.
Unless further specified, all the spectra are reduced using {\texttt{ PypeIt}}
\citep{pypeit}, and the PSF models of the images are generated using IRAF task
{\texttt {psf}} based on isolated stars in the field.

\begin{deluxetable*}{c|cccccc}
\label{tbl:obs}
\tablecaption{Summary of Observations}
\tablewidth{0pt}
\tablehead{\colhead{Object} & \colhead{Telescope} & \colhead{Instrument} &\colhead{Filter/Grating} & \colhead{$R$} & \colhead{Seeing} & \colhead{$t_\text{exp}$\tablenotemark{1}}}
\startdata
\hline
\multicolumn{7}{c}{Spectroscopy}\\\hline
J0025--0145 &  \makecell{Magellan/Clay\\Magellan/Clay\\Magellan/Baade} & \makecell{LDSS3\\LDSS3\\FIRE} & \makecell{VPH-all\\VPHBlue\\Echelle} & \makecell{645\\1425\\6000} & \makecell{$0\farcs7$\\$0\farcs7$\\$1\farcs0$} & \makecell{$1\times600$s\\$4\times900$s\\$1\times900$s}\\
J0402--4220 &  Magellan/Clay & LDSS3 & VPH-all & 645 & $0\farcs7$ & $3\times900$s\\
J0954--0022 &  Magellan/Clay & LDSS3 & VPH-all & 645 & $0\farcs7$ & $1\times600$s\\
J1051+0334  &  Magellan/Clay & LDSS3 & VPH-all & 645 & $0\farcs7$ & $3\times600$s\\
J1225+4831  &  LBT &LUCI1 & HKSpec & 645 & $0\farcs4$ & $20\times60$s\\
J2329--0522 &  \makecell{Magellan/Clay \\ Magellan/Baade } & \makecell{LDSS3\\FIRE} & \makecell{VPH-Red\\Echelle} & \makecell{1357\\6000} & \makecell{$0\farcs7$\\0\farcs7} & \makecell{$3\times1200$s\\$9\times909$s}\\\hline\hline
\multicolumn{7}{c}{Imaging}\\\hline
J0025--0145 &  {\em HST} & ACS/WFC & \makecell{F435W\\F606W} & - & \makecell{$0\farcs08$\\$0\farcs08$} & $4\times516$s\\
J0402--4220 &  Magellan/Clay & LDSS3 & $z$ & - & $0\farcs7$ & $1\times60$s\\
J0954--0022 &  Magellan/Clay & LDSS3 & $z$ & - & $0\farcs7$ & $1\times60$s\\
J1051+0334  &  Magellan/Clay & LDSS3 & $z$ & - & $0\farcs7$ & $1\times60$s\\
J1225+4831  &  LBT &LUCI1 & $K$ & - & $0\farcs35$ & $2\times60$s\\
J2329--0522 &  Magellan/Clay & LDSS3 & $z$ & - & $0\farcs7$ & $1\times30$s\\\hline\hline
\enddata
\tablenotetext{1}{The total exposure time, in the form of (number of exposures) $\times$ (length of a single exposure).}
\end{deluxetable*}

\begin{deluxetable*}{ccccc|cccc}[!t]
\label{tbl:obj}
\tablecaption{Properties of the Lensed Quasars and Quasar Pairs}
\tablewidth{0pt}
\tablehead{\colhead{Object} & \colhead{Redshift} & \colhead{$\Delta\theta$\tablenotemark{1}} & \colhead{$N_{gaia}$\tablenotemark{2}} &\colhead{Comment} & \colhead{Component\tablenotemark{3}} & \colhead{RA} & \colhead{Dec} &\colhead{Magnitude}\\
\colhead{} & \colhead{} & \colhead{} & \colhead{} & \colhead{} & \colhead{} & \colhead{} & \colhead{} & \colhead{}}
\startdata
\hline
J0025--0145 &  5.07 & - & 1 & lensed quasar & Q & 00:25:26.83 & --01:45:32.48 & $m_\text{F606W}=20.782\pm0.001$  \\
            &       &   &   &               & G & 00:25:26.79 & --01:45:32.51 & $m_\text{F606W}=21.34\pm0.03$ \\\hline
J0402--4220 & 2.88  & $1\farcs27$ & 1 & lensed quasar & Q1 & 04:02:22.16 & --42:20:52.82 & $m_z=18.948\pm0.002$ \\
            &       &             &   &               & Q2 & 04:02:22.06 & --42:20:53.44 & $m_z=20.44\pm0.02$\\
            &       &             &   &               &  G & 04:02:22.08 & --42:20:53.40 & $m_z=20.60\pm0.03$ \\\hline
J0954--0022 & 2.27  & $1\farcs21$ & 1 & lensed quasar & Q1 & 09:54:06.97 & --00:22:25.08 & $m_z=19.92\pm0.01$ \\
            &       &             &   &               & Q2 & 09:54:06.96 & --00:22:26.28 & $m_z=21.06\pm0.04$ \\
            &       &             &   &               &  G & - & - & - \\\hline
J1051+0334  & 2.40  & $0\farcs99$ & 0 & NIQ  & Q1 & 10:51:32.43 & +03:34:17.71 & $m_z=19.76\pm0.01$ \\
            &       &             &   &      & Q2 & 10:51:32.37 & +03:34:17.59 & $m_z=19.82\pm0.01$ \\\hline
J1225+4831  & 3.09  & $0\farcs89$ & 2 & NIQ  & Q1 & 12:25:18.62 & +48:31:16.15 & $m_{K_s}=15.77\pm0.01$ \\
            &       &             &   &      & Q2 & 12:25:18.63 & +48:31:17.03 & $m_{K_s}=16.88\pm0.01$ \\\hline
J2329--0522 & 4.85  & $1\farcs55$ & 0 & quasar pair & Q1 & 23:29:15.07 & --05:22:35.88 & $m_z=19.95\pm0.02$ \\
            &       &             &   &             & Q2 & 23:29:15.18 & --05:22:35.96 & $m_z=20.80\pm0.04$ \\\hline
\enddata
\tablenotetext{1}{The separation between the two quasar components.}
\tablenotetext{2}{The Number of {\em Gaia} objects matched with a radius of $2''$.}
\tablenotetext{3}{Components Q1 and Q2 are quasar components, and component G is the lensing galaxy for lensed quasars.}
\tablecomments{Coordinates and magnitudes are from the best-fit $galfit$ model.
The F606W and the $z-$band magnitudes are AB magnitudes,
and the $K_s-$band magnitudes are Vega magnitudes. 
The information about the lensing galaxy in J0954-0022 is not included as the 
parameters do not converge when fitting. See text for details about image fitting.}
\end{deluxetable*}

\section{J0025--0145: an Intermediately-Lensed Quasar at $z=5.07$} \label{sec:J0025}

\begin{figure*}
    \centering
    \includegraphics[width=0.6\textwidth]{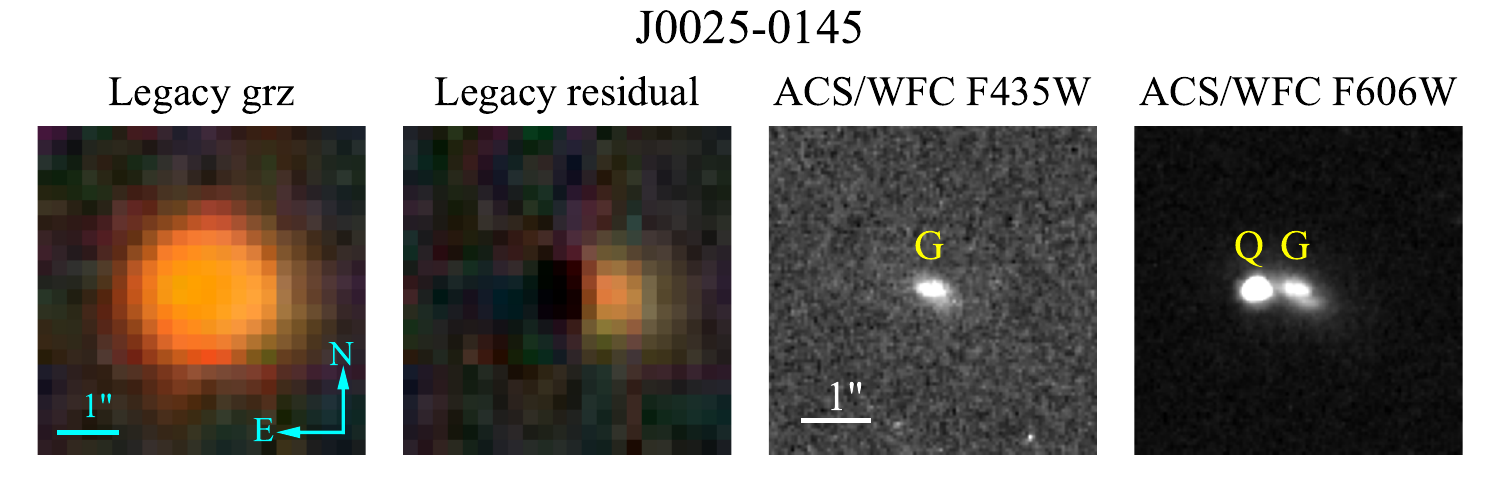}
    \includegraphics[width=0.7\textwidth]{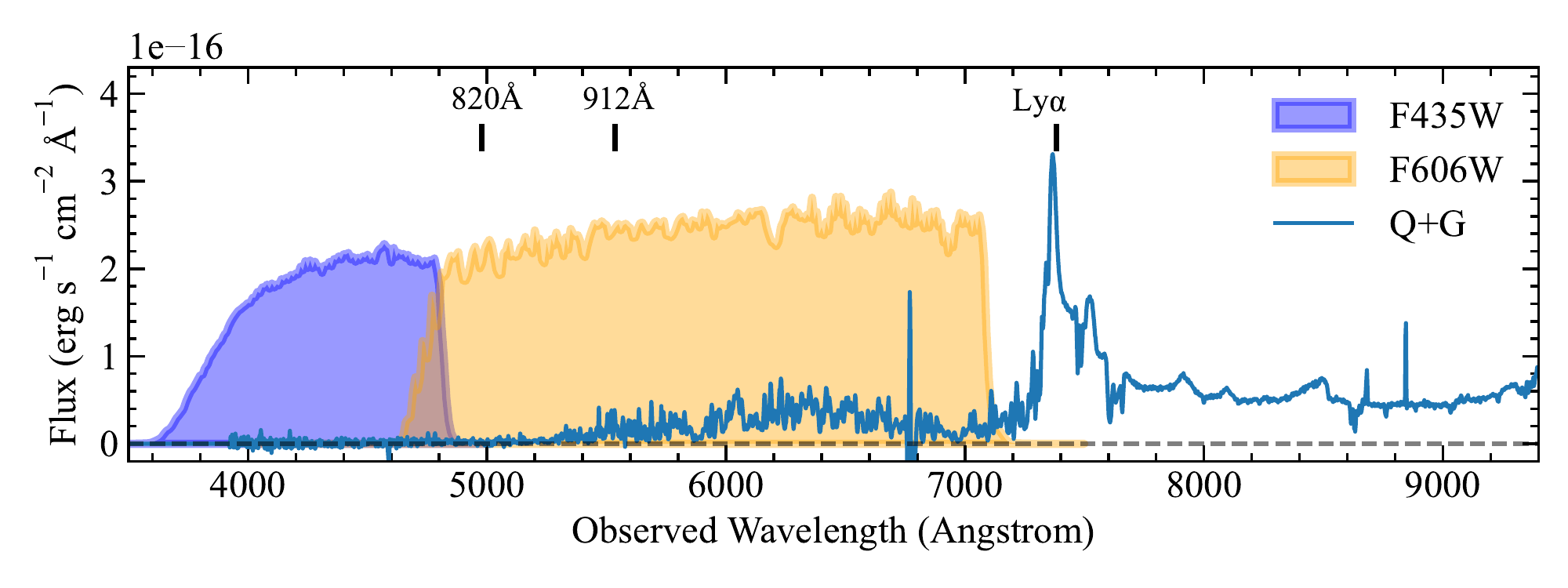}
    \caption{{\em Top}: the images of J0025--0145. From left to right: the
    Legacy Survey $grz$, the Legacy Survey residual,
    {\em HST} ACS/WFC F435W, {\em HST} ACS/WFC F606W.
    J0025--0145 is modeled as a point source in the Legacy Survey, 
    while the residual image indicates a nearby companion.
    The {\em HST} images show the existence of a foreground galaxy.
    The separation between the quasar and the galaxy is $0\farcs60$.
    {\em Bottom}: the LDSS3 spectra of J0025--0145.
    This spectrum is a combination of a VPH-Blue spectrum $(\lambda<5500\text{\AA})$
    and a VPH-All spectrum $(\lambda>5500\text{\AA})$.
    We also plot the response curve of the {\em HST} filters and mark
    the rest-frame wavelengths.
    The background quasar has no flux at rest-frame $\lambda<820\text{\AA}$,
    and the F435W image only captures the flux from the foreground lensing galaxy.
    The F606W image shows both the quasar and the lensing galaxy.}
    \label{fig:J0025}
\end{figure*}

J0025--0145 was initially reported in \citet{wang16}
as a luminous quasar with $M_{1450}=-28.63$.
The upper panel of Figure \ref{fig:J0025} shows the images of J0025--0145.
The object is modeled as a point source in the Legacy Survey DR8,
and the residual image
indicates a close companion next to the quasar. 
J0025--0145 was observed by {\em HST} Advanced Camera for Surveys
Wide Field Camera (ACS/WFC) in the F435W and F606W bands
(Proposal \#16460, PI: Yue).
In the F606W image, J0025--0145 is resolved into a point source
(i.e., the quasar) and an extended source, separated by $0\farcs60$.
The F435W only detects the extended source.

Note that objects at $z>5$ have essentially no flux
at rest-frame $\lambda<820\text{\AA}$ 
due to the absorption of the IGM \citep[e.g.,][]{worseck14}.
The lower panel of Figure \ref{fig:J0025} shows the spectrum of J0025--0145 
taken by the Low Dispersion Survey Spectrograph\footnote{https://www.lco.cl/?epkb\_post\_type\_1=index-2} (LDSS3)
on the Magellan/Clay telescope,
where we also plot the response curves of the ACS/WFC filters.
Since the extended source is detected in the F435W filter,
the object must be a foreground galaxy.
The small separation between the quasar and the foreground galaxy
suggests that the background quasar must be magnified by the galaxy.
Interestingly, there is only one point source in the F606W image, 
meaning that the background quasar is not multiply-imaged.
 {We thus report J0025--0145 as an intermediate lensing system,
i.e., lensing systems that do not have multiple images but
have a significant magnification, in contrast with weak and strong lensing \citep[e.g.,][]{mason15}.}

\begin{figure}
    \centering
    \includegraphics[width=1\columnwidth]{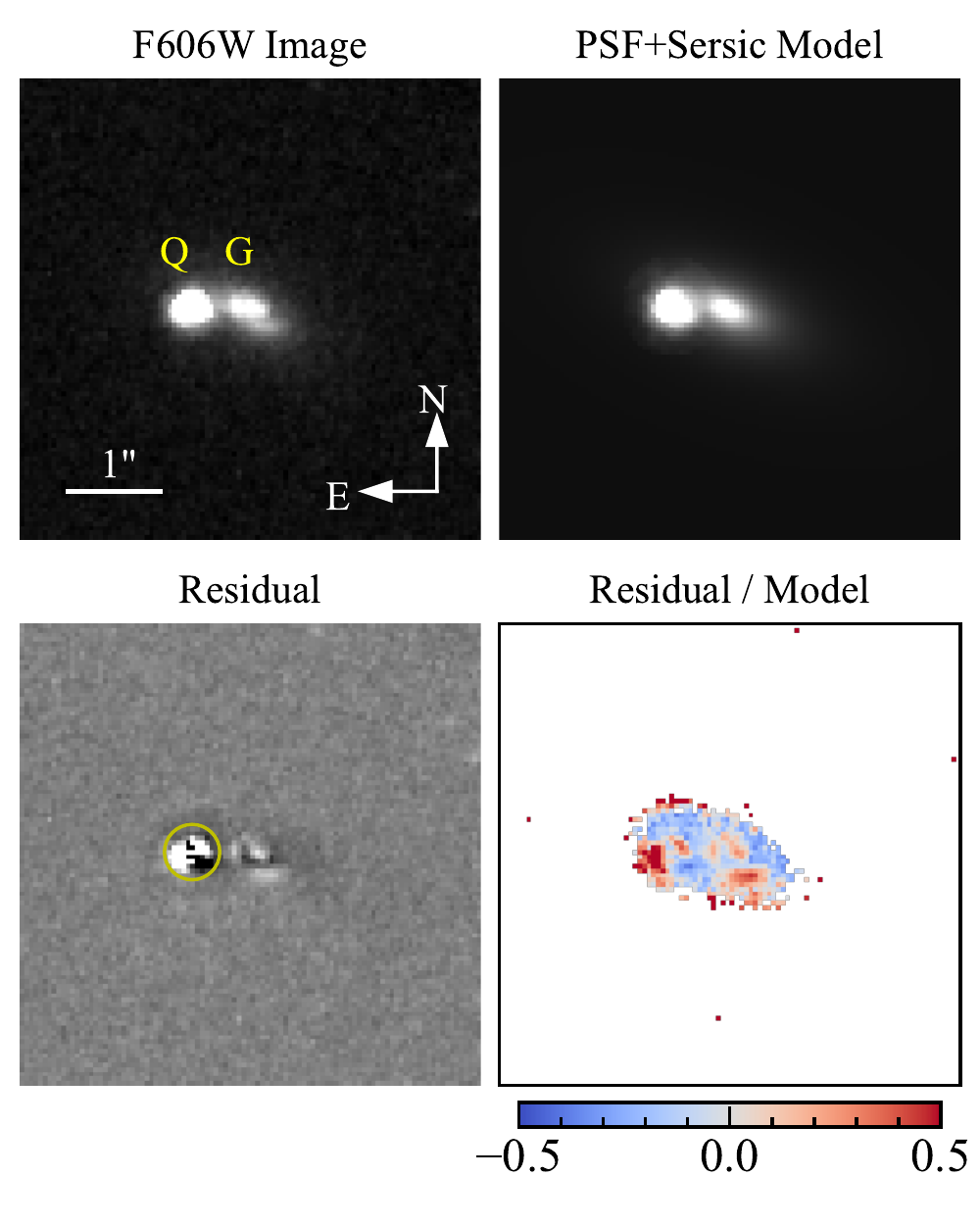}
    \caption{The {\em galfit} model of J0025--0145. 
    {\em Top left}: the {\em HST} ACS/WFC F606W image;
    {\em Top right}: the model image (one PSF and one S\'ersic profile);
    {\em Bottom left}: the residual image;
    {\em Bottom right}: the relative residual map.
    There is no evidence for a second lensed image of the background quasar.
    The deflector galaxy has a high ellipticity $(e=0.53)$
    and exhibit irregular structures. The quasar shows positive flux in the 
    residual image, which cannot be explained by PSF inaccuracies
    and thus indicates a detection of the quasar host galaxy (see text for details).
    The residual flux in the yellow aperture (diameter $0\farcs6$)
    is $m_\text{F606W}=24.2$.}
    \label{fig:J0025galfit}
\end{figure}

To further investigate the structure of J0025--0145, we fit the F606W image
by a PSF plus a S\'ersic profile {using {\em galfit}
\citep{galfit}}.
The result is presented in Figure \ref{fig:J0025galfit}. 
The foreground galaxy has a high ellipticity of $e=1-b/a=0.53$.
Deflectors of such high ellipticities can generate large magnifications
$(\mu\gg2)$ without producing multiple images of the source \citep[e.g.,][]{keeton05}.
Also note that the residual image shows irregular structures in the galaxy,
 suggesting a complex mass profile for the deflector galaxy.

 {
Another noticeable feature is the positive flux residual around the quasar.
The lower right panel of Figure \ref{fig:J0025galfit} shows the relative difference
between the model and the real image.
The relative difference reaches $\gtrsim50\%$ for the regions with positive
residual flux, which is much higher than the PSF variances in 
ACS/WFC images \citep[$\lesssim3\%$ for individual pixels; e.g.,][]{acspsf}.
Using an aperture of $0\farcs6$ centered at the quasar's position 
(marked by the solid yellow circle in Figure \ref{fig:J0025galfit}),
we estimate the residual flux to be $m_\text{F606W}=24.2$, 
i.e., about 4.4\% of the total quasar flux.
In comparison, the $5-\sigma$ magnitude limit for this aperture
is $m_\text{F606W}=25.8$.}

 {The most plausible scenario is that the positive flux 
comes from the emission of the quasar host galaxy.
Lensing magnification increases the angular size of the host galaxy,
making it easier to get detected.
Moreover, some regions in the galaxy 
might have large magnifications up to infinity.}
Detecting the rest-frame UV emission from quasar host galaxies at $z\gtrsim5$
has never been successful with {\em HST} and is very difficult even with 
{\em the James Webb Space Telescope (JWST)} \citep[e.g., ][]{marshall20,ding22}. 
J0025--0145 illustrates the power of lensing magnification
in the studies of high-redshift quasar host galaxies.

{We note that naked-cusp lenses can generate
multiply-imaged lensing systems with infinitely small lensing separation
\citep[e.g.,][]{fujimoto20}. 
We thus consider the possibility that 
J0025--0145 is a naked-cusp lens with a tiny lensing separation.
Specifically, we use 
a Singular Isothermal Ellipsoid (SIE) deflector
plus an external shear to model the lensing potential, 
where we fix the ellipticity of the SIE to be $e=0.53$.
Note that we are not able to fit the {\em HST} image
due to the limited spatial resolution;
instead, we tune the model parameters 
to test if we can reproduce systems that
resemble the features of J0025--0145.}

{We find that, in order to generate compact lensing systems 
with lensing separations $\Delta\theta\lesssim0\farcs1$,
we need a large external shear $(\gamma\gtrsim0.1)$
and fine-tuning of the source position.
In addition, such models always generate positive flux residuals
at the north or south side of the quasar, 
which is inconsistent with the {\em HST} images 
where the flux residual lies on the east side
(Figure \ref{fig:J0025galfit}).
We thus conclude that the naked-cusp lensing model
is disfavored by existing data.}

With only one image of the background quasar, 
we are not able to build the lensing model 
and reconstruct the source plane emission for J0025--0145.
Instead, we use the apparent Eddington ratio $(\lambda_\text{Edd})$ 
of J0025--0145 to estimate its lensing magnification. 
Specifically, the Eddington ratio follows the relation
$\lambda_\text{Edd} \propto L_\text{bol}/M_\text{BH}$,
where $L_\text{bol}$ is the bolometric luminosity of the quasar.
The SMBH mass, $M_\text{BH}$, is usually measured using broad emission line widths 
and continuum luminosities \citep[e.g.,][]{vo09}:
\begin{equation}\label{eq:mbh}
    M_\text{BH} = 10^\text{zp} \left(\frac{\text{FWHM}}{1000 \text{ km s}^{-1}}\right)^2 \left(\frac{\lambda L_\lambda}{10^{44}\text{ erg s}^{-1}}\right)^{0.5}
\end{equation}
The scaling factor, $10^\text{zp}$, depends on the exact line used in the measurement.
For lensed quasars, lensing magnification changes the luminosities $(L_\text{bol}$ and $ \lambda L_\lambda)$
but does not influence the line width. Combining the above relations
gives $\lambda_\text{Edd}\propto \mu^{0.5}$, i.e.,
the {\em observed} Eddington ratio is higher than the {\em intrinsic} value 
if the lensing magnification is not corrected.

\begin{figure*}
    \centering
    \includegraphics[width=0.5\textwidth]{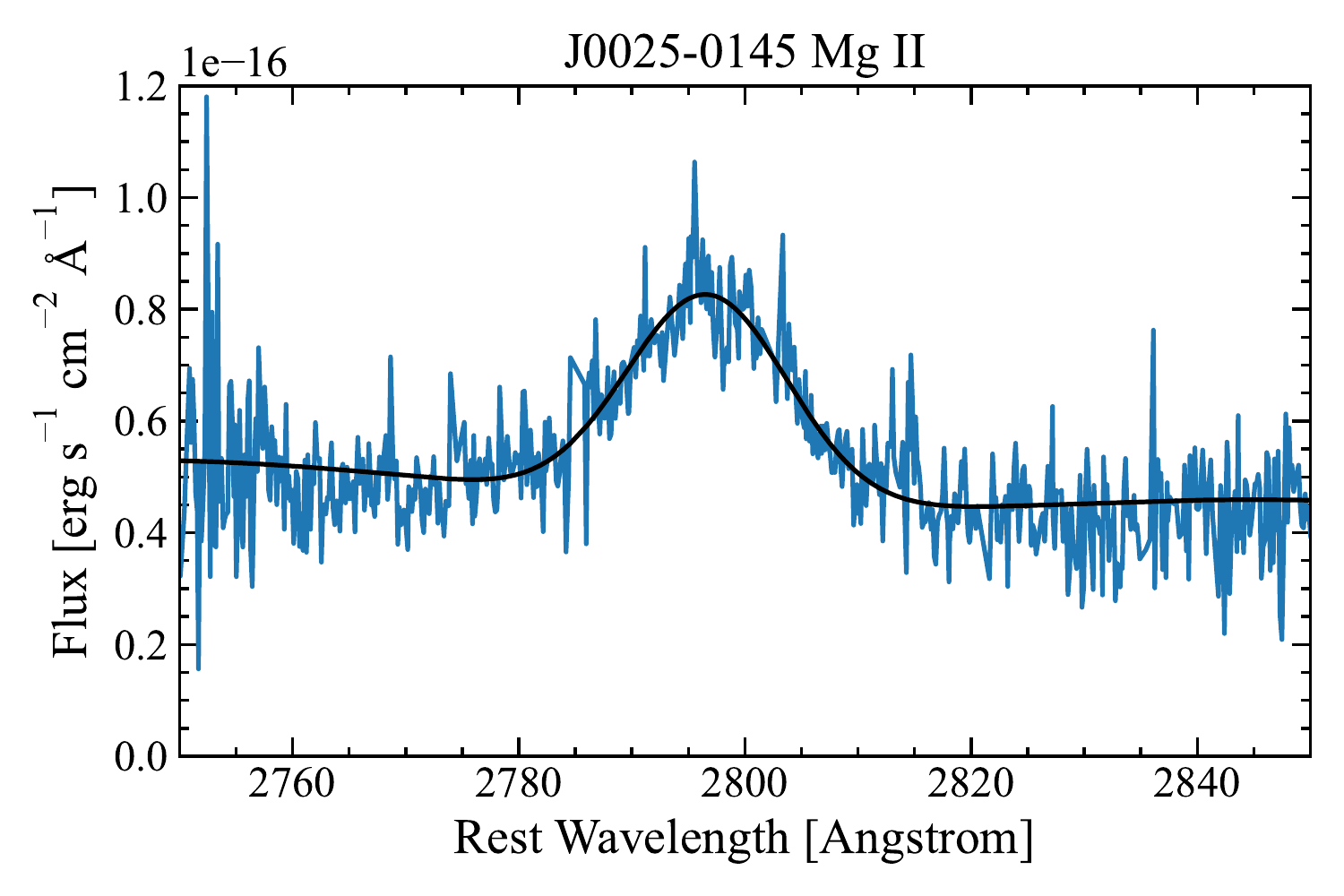}
    \includegraphics[width=0.4\textwidth]{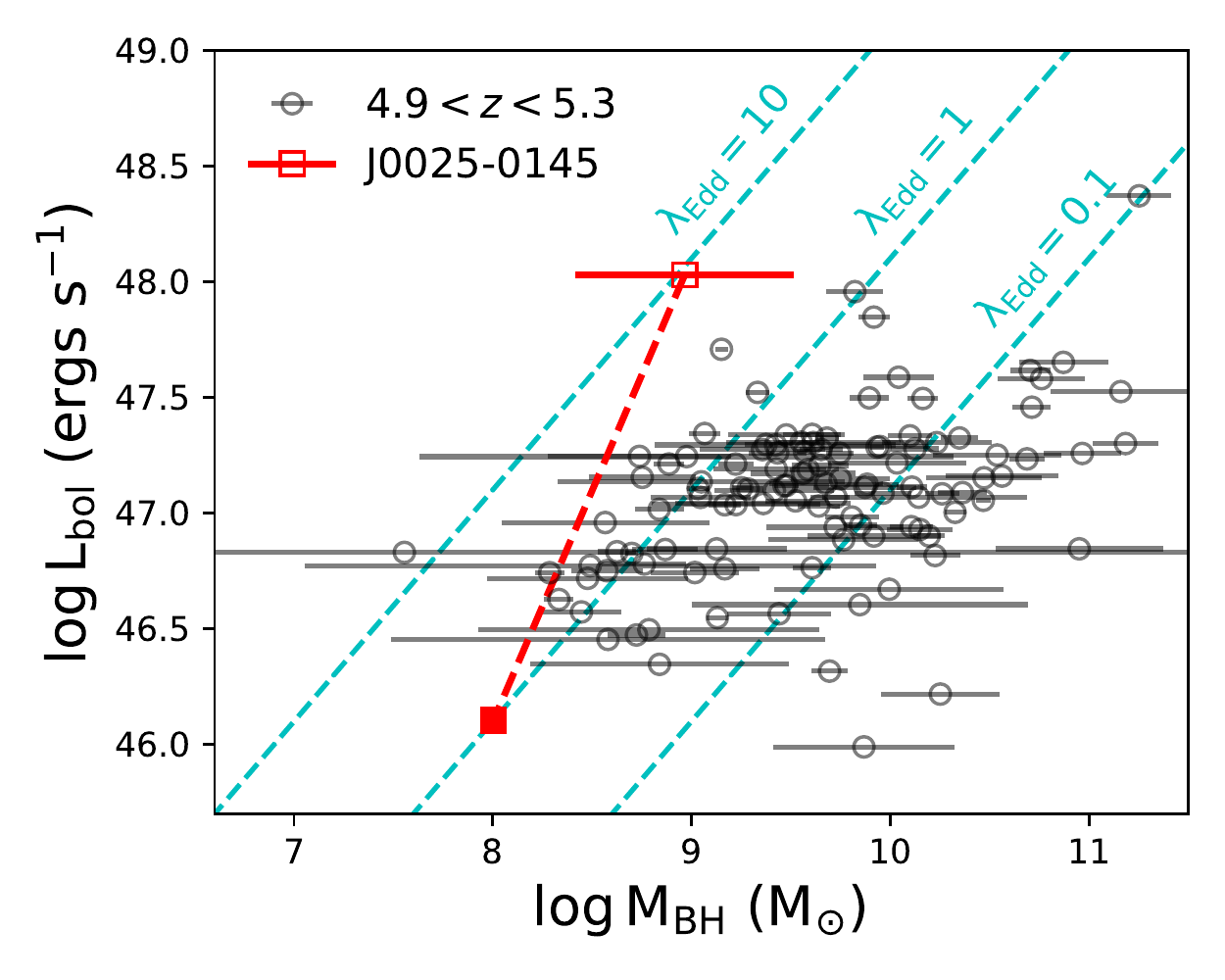}
    \caption{{\em Left}: the {\mgii} line of J0025--0145, 
    taken by the Magellan/Baade FIRE spectrograph. We fit the region around the 
    {\mgii} line as a power law plus a Gaussian profile and a series of {\feii} lines.
    The best-fit model gives $\text{FWHM}_{\text{\mgii}}=1798\pm37\text{ km s}^{-1}$,
    using which we calculate the SMBH mass of J0025--0145 to be $M_\text{BH}=9.24\times10^8M_\odot$
    (before correcting for the magnification effect).
    See text for details. {\em Right}: the $M_\text{BH}-L_\text{bol}$ distribution
    of quasars. 
    The red open square marks the apparent values (i.e., without correcting 
    for the lensing magnification) of J0025--0145, and the
    gray circles represents SDSS DR16 quasars at $4.9<z<5.3$ from \citet{dr16q}.
    Cyan dashed lines mark constant Eddington ratios. Most quasars have Eddington ratios
    smaller than or consistent with one, while J0025--0145 has a large apparent 
    Eddington ratio of $\lambda_\text{Edd}=9$. 
    The red dashed line illustrates how lensing magnification moves a quasar 
    in this diagram.
    A magnification of $\mu=84$ is needed to move J0025--0145 onto the 
    line of $\lambda_\text{Edd}=1$ (the filled red square).}
    \label{fig:J0025_Redd}
\end{figure*}

We use the {\mgii} emission line to measure the SMBH mass for J0025--0145.
The left panel of Figure \ref{fig:J0025_Redd} 
shows the {\mgii} line of J0025--0145
taken by the Folded-port InfraRed Echelle\footnote{http://web.mit.edu/\~{}rsimcoe/www/FIRE/} (FIRE) on the Magellan/Baade telescope.
We fit the region around the {\mgii} line using a power-law continuum plus a single 
{\mgii} line and a series of {\feii} lines.
We use the {\feii} line template from \citet{tsuzuki06} 
and use Gaussian functions to describe the profiles of emission lines.
The best-fit model gives $\text{FWHM}_{\text{\mgii}}=1798\pm37\text{ km s}^{-1}$
and $L_{3000}=1.56\times10^{47}\text{ erg s}^{-1}$.
Using the relation in \citet{vo09}, we estimate the black hole mass to be
$M_\text{BH}=9.24\times10^8M_\odot$ without correcting the magnification effect.
The error of $M_\text{BH}$ is about 0.55 dex and is dominated by 
the scatters of the empirical relation (Equation \ref{eq:mbh}).
We estimate the bolometric luminosity using the bolometric correction suggested by \citet{runnoe12}, which gives $L_\text{bol}=8.26\times10^{47}\text{ erg s}^{-1}$.

The right panel of Figure \ref{fig:J0025_Redd}
shows the distribution of SMBH masses and bolometric luminosities 
of quasars at $4.9<z<5.3$ from the SDSS DR16 quasar catalog \citep{dr16q}.
The median Eddington ratio of these quasars is 0.4, and
most quasars have Eddington ratios consistent with or smaller than one.
In contrast, 
J0025--0145 has a high apparent Eddington ratio of $\lambda_\text{Edd}=9$
before correcting for lensing magnification.
Again, the error of $\lambda_\text{Edd}$ is 0.55 dex and is dominated by the 
scatters of Equation \ref{eq:mbh}.
Assuming that the quasar has an intrinsic Eddington ratio of $\lambda_\text{Edd}=1$
(i.e., Eddington-limit accretion),
we need a magnification of $\mu=84$.
Even if J0025--0145 is accreting at a super-Eddington rate of $\lambda_\text{Edd}\approx3$,
we still get a magnification of $\mu\approx9$.

We notice that the estimation above is subject to several systematic uncertainties,
in particular the errors in the SMBH mass measurement and in the bolometric correction.
Nevertheless, the result indicates that J0025--0145 has a high magnification of $\mu\gg1$.
Future observations with the Atacama Large Millimeter/Submillimeter Array
will characterize the lensed arc of the quasar host galaxy,
enabling accurate lens modeling of this system.
If the high magnification is confirmed by the lensing model,
J0025--0145 will offer unique opportunities to investigate
a distant quasar host galaxy with high spatial resolution.

\section{J2329--0522: a Kpc-scale Quasar Pair at $z=4.85$} \label{sec:J2329}

\begin{figure*}
    \centering
    \includegraphics[width=0.6\textwidth]{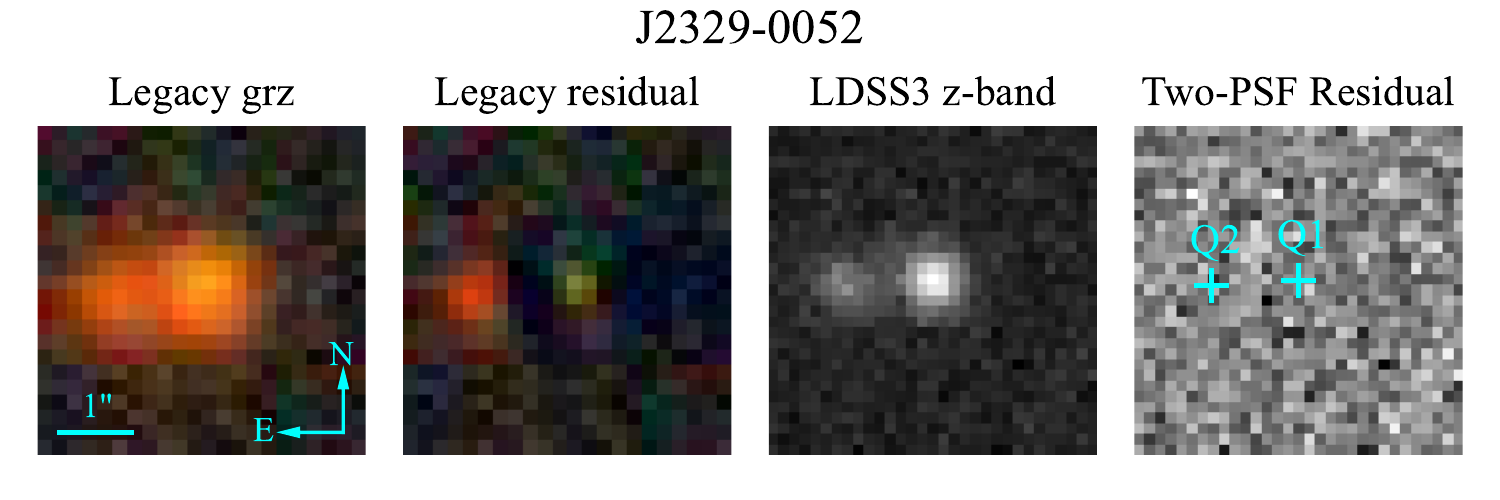}
    \includegraphics[width=0.7\textwidth]{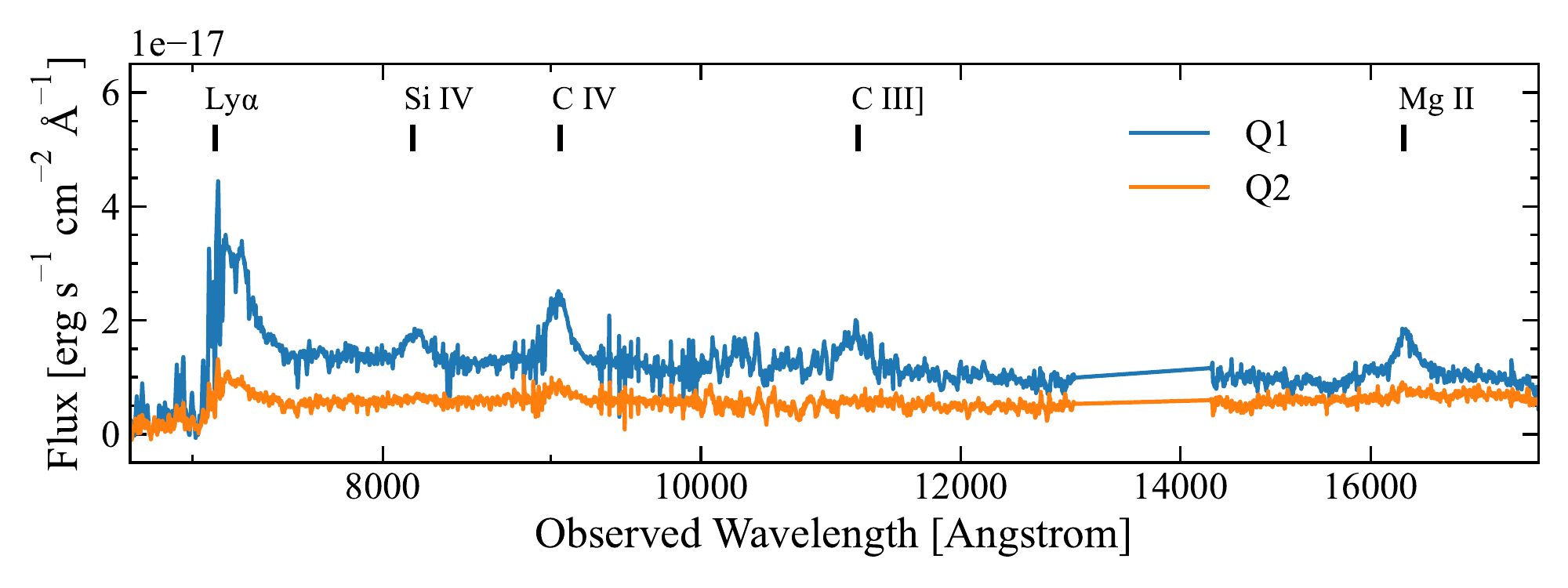}
    \caption{{\em Top}: the images of J2329--0522. From left to right: the
    Legacy Survey $grz$ image, the Legacy Survey residual, the LDSS3 $z-$band image,
    the two-PSF residual of the LDSS3 image. 
    J2329--0522 is modeled as a de Vaucouleurs profile in the Legacy Survey DR8.
    The two-PSF residual of the LDSS3 image shows no evidence of a foreground lensing galaxy.
    {\em Bottom}: the spectra of J2329--0522. These spectra are combinations of 
    LDSS3 VPH-Red spectra $(\lambda<1\mu\text{m})$ and FIRE Echelle spectra $(\lambda>1\mu\text{m})$. The emission line profiles of the two quasars
    show significant differences, and we classify this object as a physical quasar pair
    (see text for details).}
    \label{fig:J2329}
\end{figure*}

Figure \ref{fig:J2329} shows the images and spectra of J2329--0522.
In the Legacy Survey DR8, J2329--0522 is modeled as a de Vaucouleurs profile,
and the large image fitting residual suggests that the flux model is incorrect.
J2329--0522 was observed using LDSS3 in both imaging mode and spectroscopy mode under a seeing of $\sim0\farcs7$.
The LDSS3 $z-$band image is well-described by 
two point sources, showing no signs of a lensing galaxy in the field.
The two point sources are separated by $1\farcs55$.
The lower panel of Figure \ref{fig:J2329} shows the spectra of the two point sources,
{both of which exhibit features of quasars at $z=4.85$}.
The NIR part $(\lambda>1\mu\text{m})$ of the spectra was taken with FIRE.
The two components have different spectral shapes, 
with the fainter component (Q2) being redder.
The broad emission line profiles of the two components also show
significant differences. 
In particular, Q1 has a prominent {\ciii} line, which is nearly
absent in the spectrum of Q2.
The {\civ} profiles of the two components also look different.

\begin{figure*}
    \centering
    \includegraphics[width=1\textwidth]{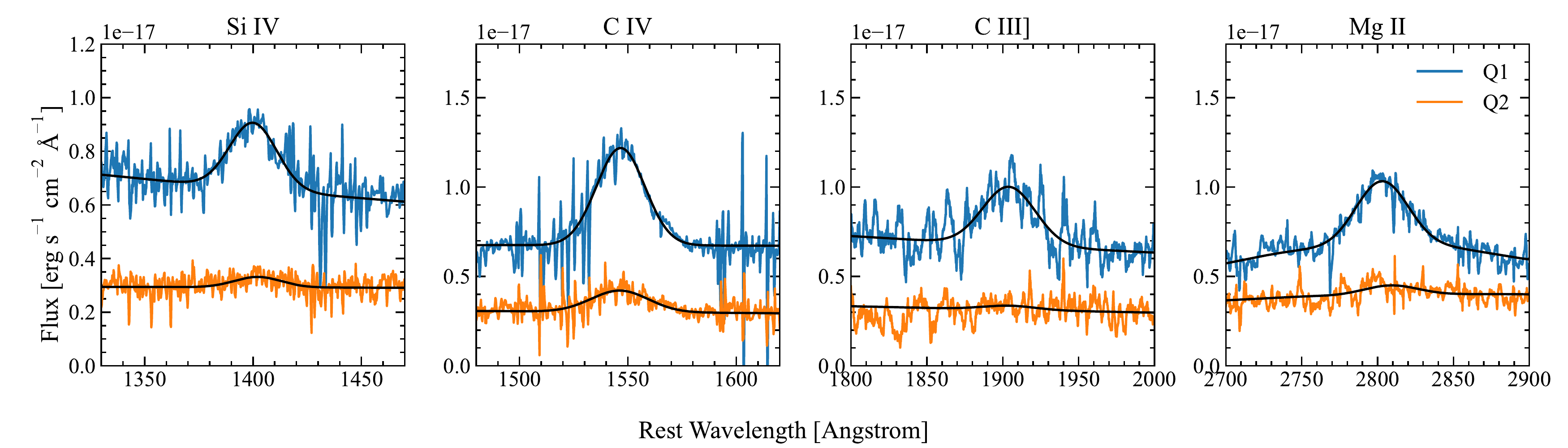}
    \caption{The {\siiv}, {\civ}, {\ciii}, and {\mgii} lines of J2329--0522. We fit the lines
    as a power-law continuum and a Gaussian profile emission line,
    except for the {\mgii} line where we add a series of {\feii} lines.
    The best-fit models are demonstrated by the black lines, and 
    the best-fit parameters are listed in Table \ref{tbl:J2329lines}. 
    These emission lines show vastly different EW ratios between the two quasars,
    which is difficult to be explained by microlensing. 
    In particular, 
    the {\ciii} line {is prominent in Q1 but is
    only marginally detected in Q2.} 
    In addition, the two quasars show a $2.8\sigma$ difference in the {\civ} FWHM.
    These features suggest that J2329--0522 is a physical pair of quasars
    instead of a doubly-lensed quasar.}
    \label{fig:J2329lines}
\end{figure*}

To further investigate the possibility of J2329--0522 being
a doubly-lensed quasar, we fit the {\siiv}, {\civ}, {\ciii} lines
using a power-law continuum plus a Gaussian profile.
We also fit the {\mgii} lines using the method described in Section \ref{sec:J0025}.
Figure \ref{fig:J2329lines} presents the result, and Table \ref{tbl:J2329lines}
summarizes the best-fit parameters. 
The {\mgii} and the {\siiv} redshifts of Q2 are higher than 
that of Q1,
while Q2 has a lower {\civ} redshift than Q1. 
There is a $2.8\sigma$ difference between the
FWHM of the {\civ} lines of the two components.
The {\ciii} line of Q2 is only marginally detected, and the {\ciii}
equivalent width (EW) of Q1 is about six times larger than that of Q2.
Meanwhile, the EWs of the two {\civ} lines only differ by $\sim50\%$.
 {
It is implausible that the FWHM and EW differences between the two quasars 
in J2329--0522 are results of microlensing.}
For instance, \citet{sluse12} report the microlensing effect of 17 lensed quasars, 
where the microlensing magnification of different broad emission lines are similar.
Also, note that differential reddening has no impact on the FWHM and EW of emission lines.

We thus conclude that  J2329--0522 is a physical quasar pair.
The projected separation is 9.88 kpc at $z=4.85$, and
J2329--0522 is the first quasar pair with projected separation $\Delta d<10$ kpc 
confirmed at $z>3.2$.
The mass ratio of the two SMBHs is close to unity,
which is similar to J2037--4537 (an NIQ at $z=5.66$) reported in \citet{yue21b}\footnote{\citet{yue21b} report J2037--4537 as an NIQ
(or a ``candidate" quasar pair), as the two components in J2037--4537 show similar emission line profiles and
the lensing scenario cannot be fully excluded based on current data. However, \citet{yue21b} 
rule out lensing galaxies with $m_\text{F850LP}<26.7$,
and show that J2037--4537 cannot be explained by lensing galaxies with
regular mass and dust distribution.} 
and is consistent with the prediction of hydrodynamical simulations
\citep[e.g.,][]{capelo17}.
The small separation between the two quasars suggests that the two quasar host galaxies
are in the progress of merging.
According to the analysis in \citet[][]{yue21b},
the clustering of quasars cannot explain 
the existence of kpc-scale quasar pairs at $z\sim5$.
As such, the two quasars in J2329--0522
are likely triggered by the ongoing galaxy merger.

\begin{deluxetable}{ccc}
\label{tbl:J2329lines}
\tablecaption{Emission Line Properties of J2329--0522}
\tablewidth{0pt}
\tablehead{\colhead{Component} & \colhead{Q1} & \colhead{Q2}}
\startdata
\hline
\multicolumn{3}{c}{{\siiv} line}\\\hline
Redshift & $4.8643\pm 0.0011$ & $4.8725 \pm 0.0062$ \\
FWHM $(\text{km s}^{-1})$& $5506\pm156$ & $5421\pm887$ \\
EW {(\AA)} & $60\pm2$ & $21\pm4$\\\hline
\multicolumn{3}{c}{{\civ} line}\\\hline
Redshift & $4.8410\pm0.0005$ & $4.8396\pm0.0023$ \\
FWHM $(\text{km s}^{-1})$& $5099\pm65$ & $6034\pm332$ \\
EW {(\AA)} & $132\pm2$ & $77\pm6$\\\hline
\multicolumn{3}{c}{{\ciii} line}\\\hline
Redshift & $4.8363 \pm 0.0030$ & [4.8363] \\
FWHM $(\text{km s}^{-1})$& $6449\pm416$ & [6449] \\
EW {(\AA)} & $123\pm10$ & $18\pm4$\\\hline
\multicolumn{3}{c}{{\mgii} line}\\\hline
Redshift & $4.857 \pm 0.001$ & $4.868\pm0.005$ \\
FWHM $(\text{km s}^{-1})$& $4549\pm103$ & $4586\pm580$ \\
EW {(\AA)} & $246\pm5$ & $46\pm6$\\\hline
\multicolumn{3}{c}{Other Properties}\\\hline
$\log (M_\text{BH, MgII}/M_\odot)$ & 9.49 & 9.42 \\
$M_{1450}$ & -26.1 & -25.1\\
$M_{3000}$ & -27.5 & -27.1\\\hline
\enddata
\tablecomments{
The fit does not converge for the {\ciii} line of Q2 if we keep all parameters free.
We thus fix the FWHM and the central wavelength
to the best-fit values of Q1 when fitting the {\ciii} line of Q2.}
\end{deluxetable}

\section{Low-redshift Lensed Quasars and Quasar Pairs} \label{sec:results:objects}

In this Section,  we present the low-redshift $(z<4)$
lensed quasars and quasar pairs found in our survey. 

\subsection{J0402--4220: a lensed quasar at $z=2.88$}

\begin{figure*}
    \centering
    \includegraphics[width=0.6\textwidth]{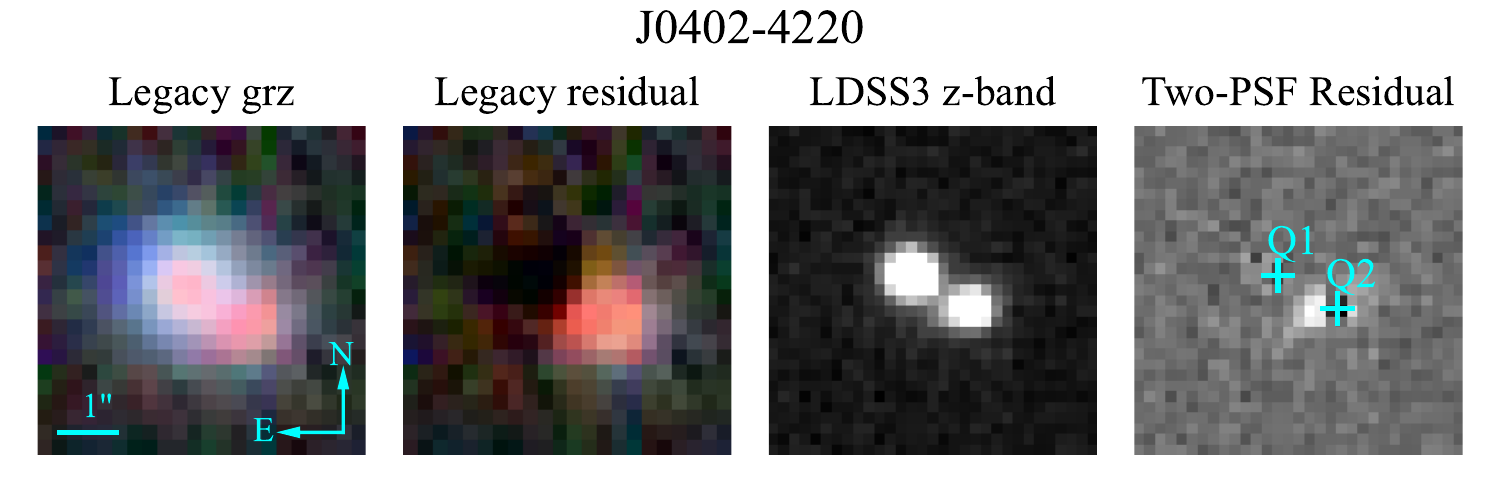}
    \includegraphics[width=0.7\textwidth]{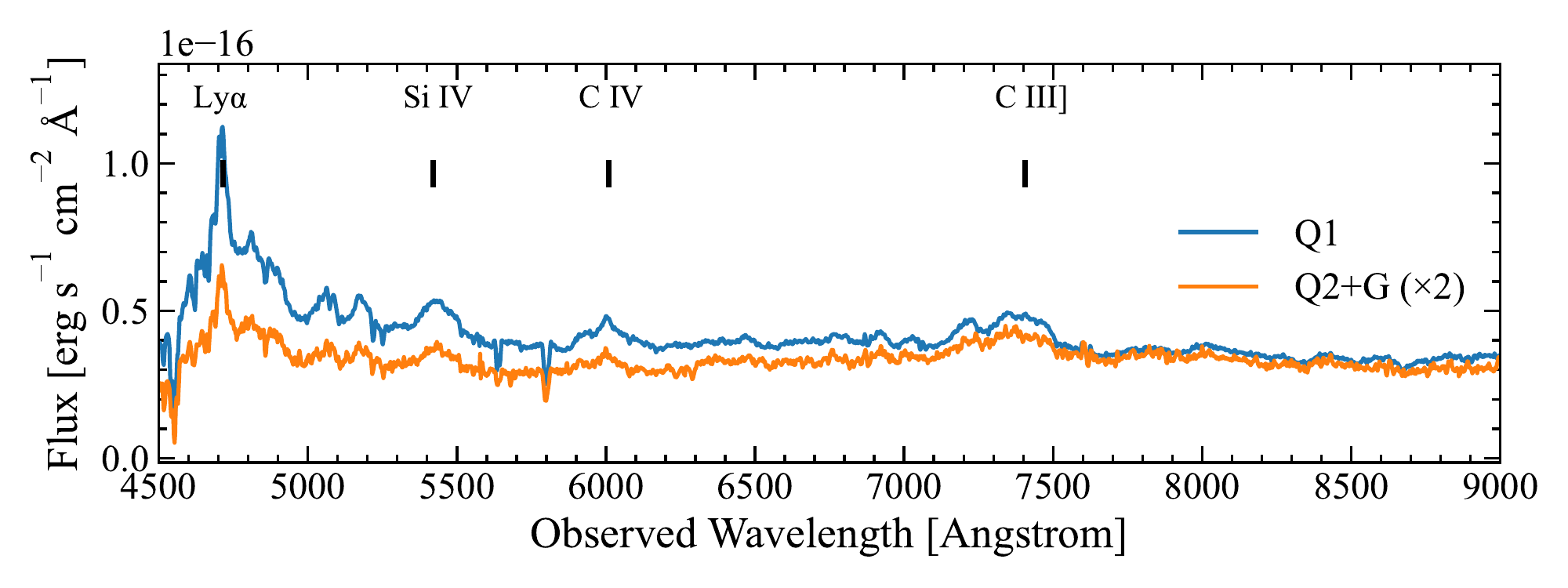}
    \caption{{\em Top}: the images of J0402--4220. 
    From left to right: the
    Legacy Survey $grz$ image, the Legacy Survey residual image, 
    the LDSS3 $z-$band image, the two-PSF residual of the LDSS3 image.
    J0402-4220 is modeled as a point source in the Legacy Survey DR8.
    The two-PSF residual of the LDSS3 image shows another source 
    between the two point sources, i.e., the lensing galaxy.
    {\em Bottom}: the LDSS3 spectra of J0402-4220, 
    clearly showing that the
    two components are two quasars at the same redshift, $z=2.88$.
    The spectrum of the fainter component (Q2+G) is scaled by a factor of two.}
    \label{fig:J0402}
\end{figure*}

Figure \ref{fig:J0402} shows the images and spectra of J0402--4220.
J0402--4220 is modeled as a point source in the Legacy Survey DR8,
and the residual image indicates the existence of a close companion.
J0402--4220 was observed using LDSS3
in both imaging and spectroscopic mode.
Under a seeing of $\sim0\farcs7$, J0402--4220 is clearly resolved into two components
separated by $1\farcs27$.
The two traces exhibit broad emission lines at the same redshift, $z=2.88$,
and the line profiles are similar. This feature suggests that J0402--4220 is 
a doubly lensed quasar.

To test the strong-lensing scenario,
we fit the LDSS3 $z-$band image as two PSFs.
The residual image clearly shows a lensing galaxy
lying between the two PSF components, which
 confirms J0402--4220 as a lensed quasar.
We also fit the image as two PSFs plus a de Vaucouleurs profile, and
the best-fit de Vaucouleurs profile has a half-light radius of $R_e=0\farcs37$ 
and an axis ratio of $q=0.23$.
The small radius and the extreme axis ratio indicate that the 
lensing galaxy is only marginally resolved in the image.
Since the ellipticity of the deflector might be inaccurate,
we do not carry out lens modeling for J0402--4220 in this paper.
{J0402--4220 was also reported by \citet{lemon20}
as an inconclusive candidate, where this object was modeled as
a quasar blended with a galaxy without decisive features of strong lensing.}

\subsection{J0954--0022: a lensed quasar at $z=2.27$}

\begin{figure*}
    \centering
    \includegraphics[width=0.6\textwidth]{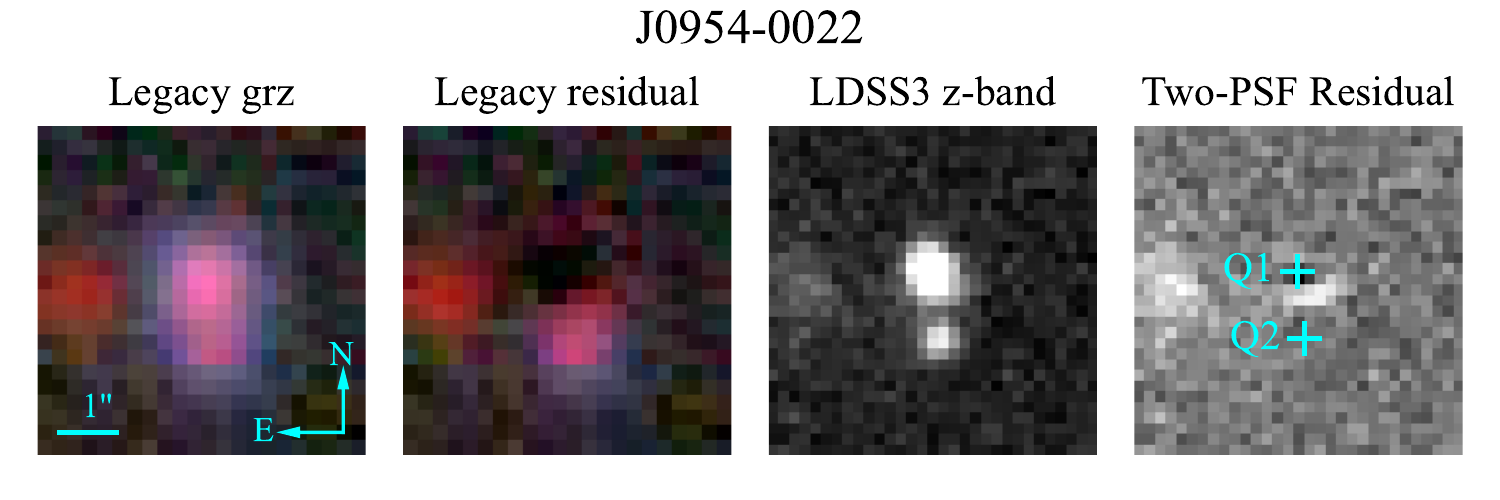}
    \includegraphics[width=0.7\textwidth]{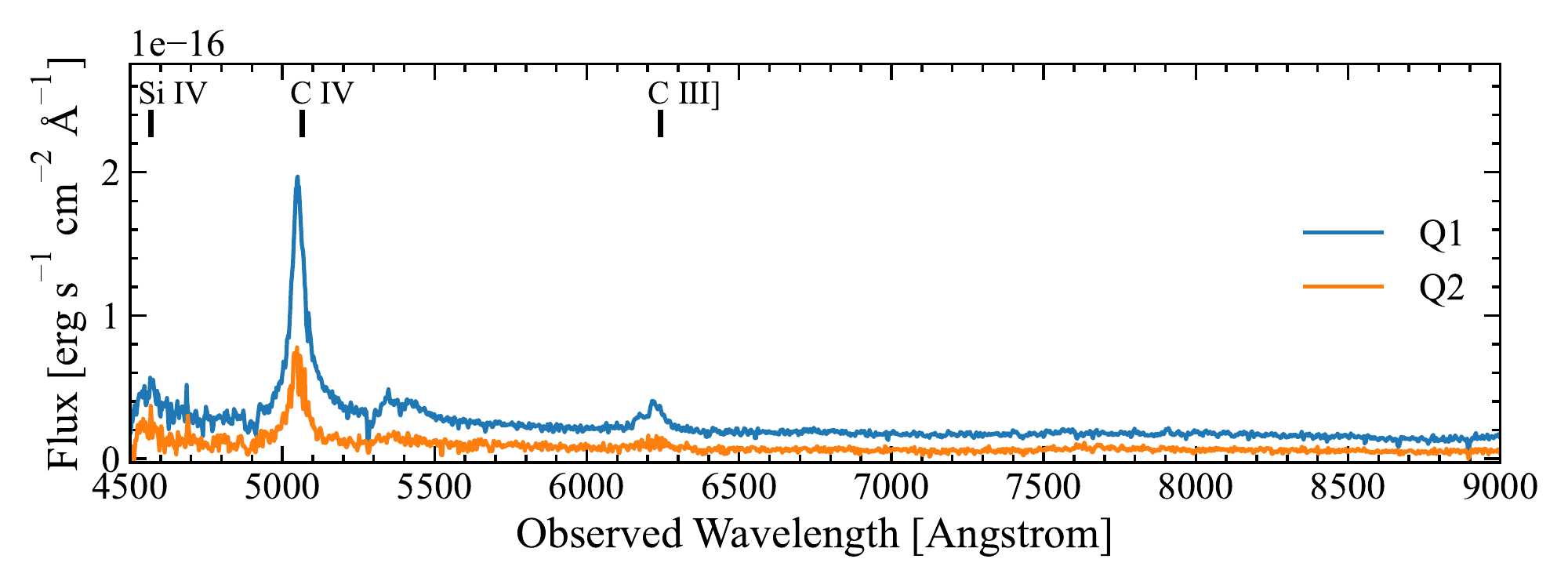}
    \caption{Same as Figure \ref{fig:J0402}, but for J0954--0022.
    J0954-0022 is modeled as a single point source in the Legacy Survey DR8.
    The two-PSF residual of the Legacy Survey image shows the existence of a lensing galaxy, and the spectra of both components exhibit features of a quasar at $z=2.27$.}
    \label{fig:J0954}
\end{figure*}

J0954--0022 (Figure \ref{fig:J0954}) was initially reported by \citet{croom01} 
as a luminous quasar.
The Legacy Survey image indicates that J0954--0022 consists of
two components with similar colors.
J0954--0022 was observed using LDSS3 under a seeing of $\sim0\farcs7$.
In the LDSS3 $z-$band image, J0954--0022 is resolved into two components
with a separation of $1\farcs21$.
The broad emission line profiles of the two components are nearly identical.

We fit the LDSS3 image as two PSFs, 
and the fitting residual show a faint lensing galaxy 
lying between the two PSFs.
We thus confirm J0954--0022 as a lensed quasar.
We also try to add a de Vaucouleurs profile or an exponential profile
to the model, but the fitting does not converge.
We argue that the current image does not have sufficient
spatial resolution and/or depth to characterize the foreground lensing galaxy,
and we do not perform lens modeling for J0954--0022.

\subsection{J1051+0334: {a lensed quasar} at $z=2.40$}

\begin{figure*}
    \centering
    \includegraphics[width=0.6\textwidth]{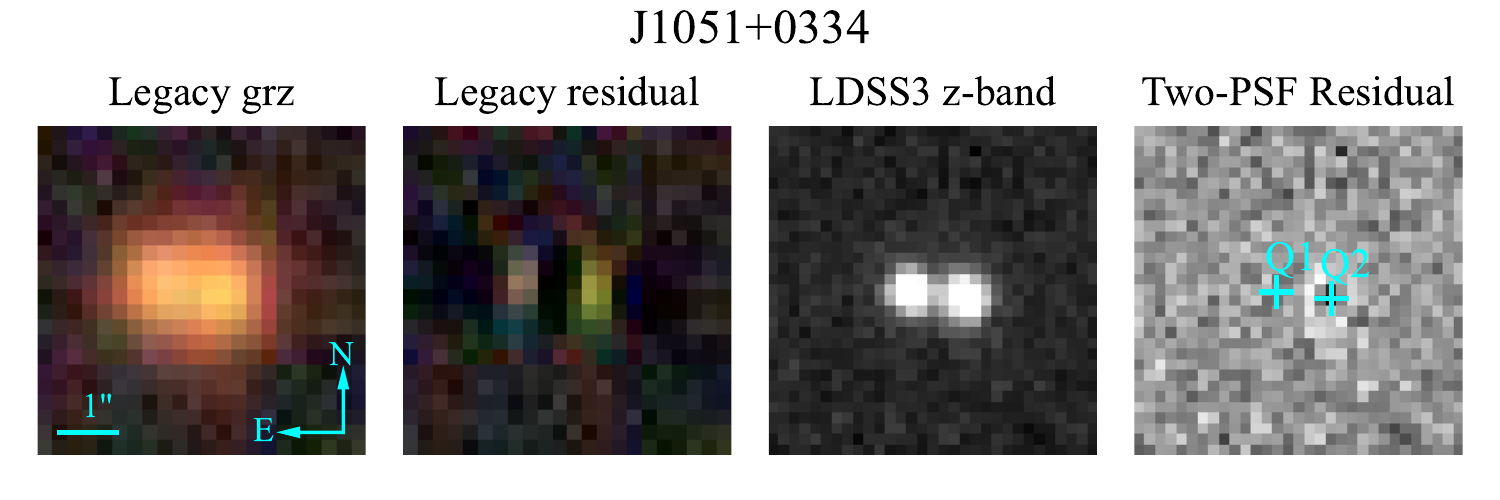}
    \includegraphics[width=0.7\textwidth]{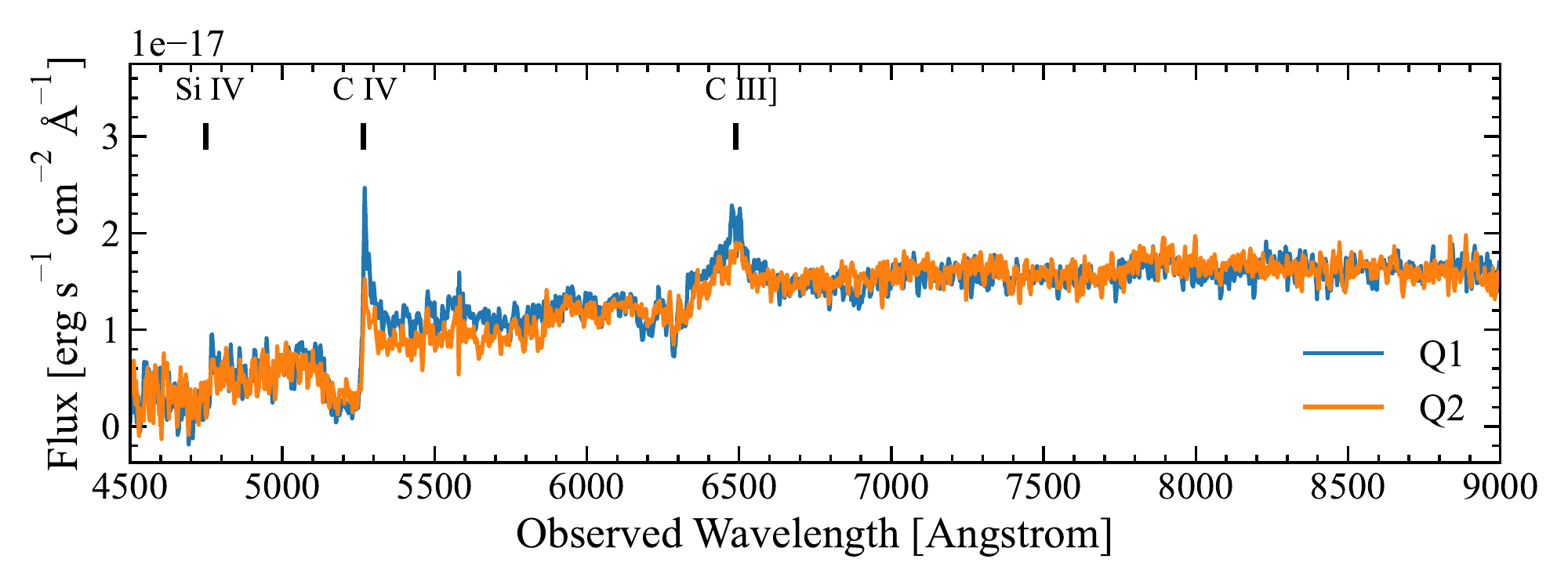}
    \caption{Same as Figure \ref{fig:J0402}, but for J1051+0334. 
    J1051+0334 is modeled as a S\'ersic profile in the Legacy Survey DR8.
    The spectra of both components exhibit features of a BAL quasar at $z=2.40$. The emission line shapes of the two components are similar, while Q1 has larger emission line fluxes than Q2.
    {Although we do not see clear evidence of a lensing galaxy in the two-PSF residual of the LDSS3 image,
    the nearly identical BAL features suggest that J1051+0334 
    must be a lensed quasar.}}
    \label{fig:J1051}
\end{figure*}

Figure \ref{fig:J1051} shows the images and spectra of J1051+0334.
J1051+0334 was observed using LDSS3 under a seeing of $\sim0\farcs6$.
In the LDSS3 $z-$band image, J1051+0334 is resolved into two components
with a separation of $0\farcs99$.
The spectroscopy shows that 
both components are heavily reddened BAL quasars at $z=2.40$.
The emission line profiles and the BAL features of the two components are similar,
while quasar Q1 shows stronger emission lines than quasar Q2.

The LDSS3 $z-$band image of J1051+0334 can be well-fitted by two PSFs,
and no foreground lensing galaxy is detected.
{However, it is extremely unlikely to have a close pair of BAL quasars
exhibiting spectra so similar to each other.}
The difference in the emission line strengths 
can be explained by  microlensing effect on the broad line region.
We thus classify this object as {a lensed quasar.
A similar case is reported in \citet{hut20}.}

\subsection{J1225+4831: an NIQ at $z=3.09$}

\begin{figure*}
    \centering
    \includegraphics[width=0.6\textwidth]{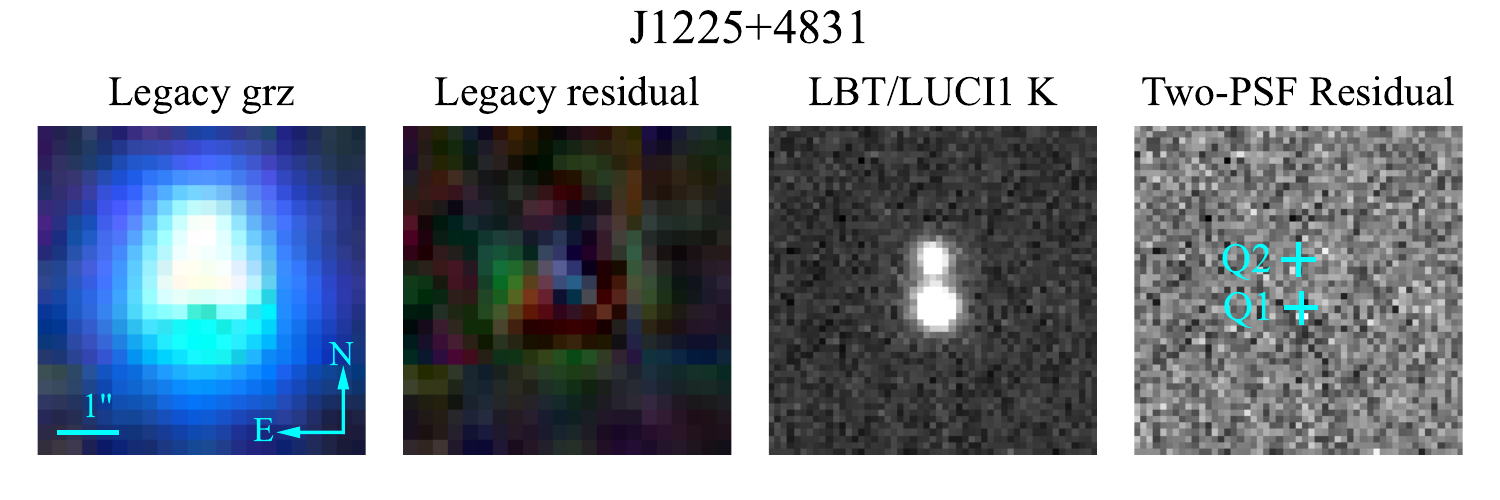}
    \includegraphics[width=0.4\textwidth]{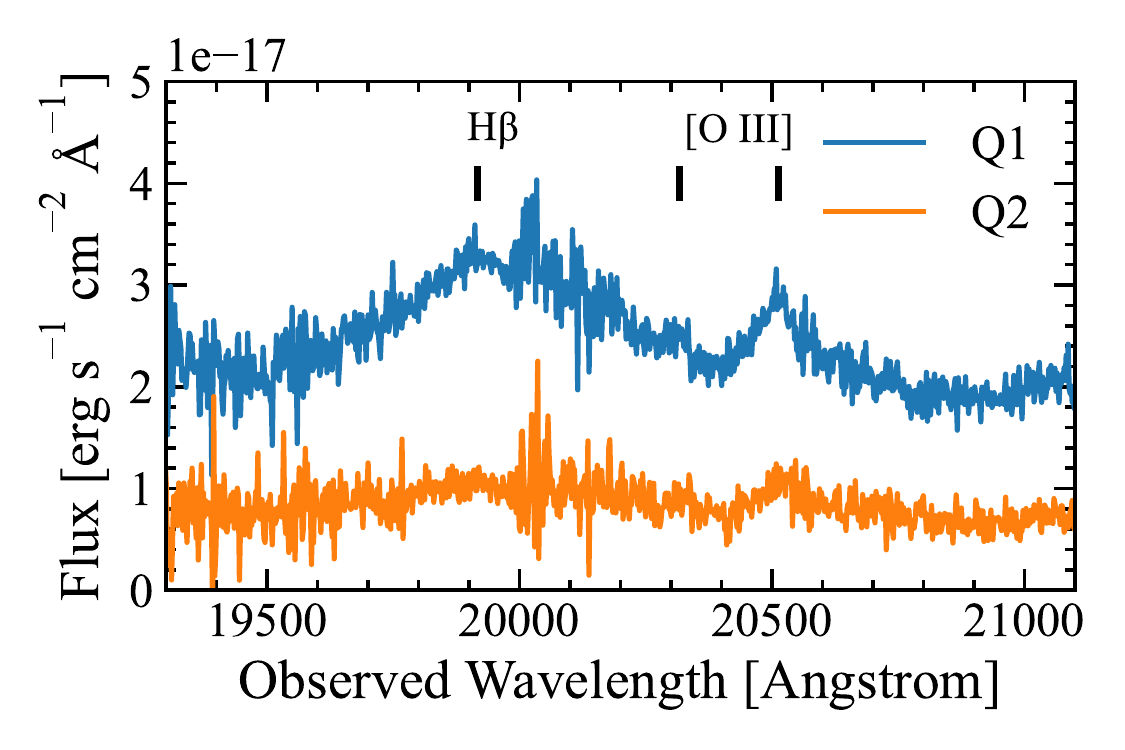}
    \caption{{\em Top}: the images of J1225+4831. From left to right: the
    Legacy Survey $grz$, the Legacy Survey residual, the LBT/LUCI1 $K-$band image,
    the two-PSF residual of the LUCI image. 
    J1225+4831 is modeled as two PSFs in the Legacy Survey DR8.
    The $K-$band image has a PSF FWHM of
    $0\farcs35$, and is well-fitted by two PSFs. {\em Bottom}:
    the LBT/LUCI1 HKspec spectrum of the two components, showing that
    both components are a quasar at $z=3.09$.}
    \label{fig:J1225}
\end{figure*}

J1225+4831 was initially reported in \citet{schindler18} 
as a luminous quasar {and was selected as a grade A candidate 
of lensed quasars and quasar pairs in \citet{dawes22}}.
J1225+4831 was observed using the Large Binocular Telescope (LBT)
LUCI1 \footnote{https://scienceops.lbto.org/luci/} in both imaging and spectroscopic mode.
We enabled the Enhance Seeing Mode (ESM) in the observing runs,
which delivered a PSF size of $\sim0\farcs35$.
The images and spectra are shown in Figure \ref{fig:J1225}.
In the LUCI1 $K-$band image, J1225+4831 is clearly resolved into 
two point sources separated by $0\farcs89$.
The LUCI1 spectra show that both components 
exhibit features of  a quasar at $z=3.095$.

As a side effect of the ESM observation,
the PSF of the LUCI1 image depends on the position on the image.
Consequently, we cannot use field stars to construct the PSF model.
To test if J1225+4831 can be well-described by two point sources,
we fit the image as two Moffat profiles, where we force
the two components to have the same profile.
The residual is shown in Figure  \ref{fig:J1225}, 
which does not exhibit signs of a lensing galaxy.

We notice that the line profiles of the two components in J1225+4831
are different; specifically, the H$\beta$ peak of Q1
is higher than the {\oiii} peak, 
while for Q2, the peaks of the two lines are close.
This difference can be explained by 
 microlensing effect on the broad line region.
We thus classify this object as an NIQ.

{
J1225+4831 is an extremely luminous quasar
with $m_r=17.9$ (PS1).
We thus test the possibility that the high luminosity of J1225+4831
is a result of lensing magnification.
Following the argument in Section \ref{sec:J0025},
if J1225+4831 is a lensed quasar, the {\em apparent} Eddington ratio
estimated using the total spectrum (i.e., the sum of components A and B)
will be $\mu^{0.5}$ times the intrinsic Eddington ratio.
We thus fit the total spectrum 
as a power law plus three Gaussian profiles, 
which represent the H$\beta$, the {\oiii}4959{\AA}, 
and the{\oiii}5007{\AA} line, respectively.
We fix the flux ratio of the {\oiii}5007{\AA} line
and the {\oiii}4959{\AA} line to be 3 
and require the two lines to have the same redshift and FWHM.
This fitting gives $\text{FWHM}_{\text{H}\beta}=7130\text{ km s}^{-1}$ 
and $L_{5100\text{\AA}}=4.56\times10^{46}\text{ erg s}^{-1}$.
Using the SMBH mass estimator given by \citet{vp06} 
and the bolometric correction given by \citet{runnoe12},
we estimate the {\em apparent} Eddington ratio of J1225+4831 to be 0.3. 
This {\em apparent} Eddington ratio is not indicative of a large
lensing magnification.}


\section{Discussions and Future Work} \label{sec:discussion}


 {
In this work, we present six lensed quasars and quasar pairs discovered in 
the initial stage of our survey.
In addition to these objects,
we have also independently found three lensed quasars that are reported earlier in \citet{lemon22}
and one NIQ that is reported in \citet{yue21b}.
We thus estimate the current survey efficiency to be $10/76\approx13\%$.
All these objects are Priority 1 candidates;
if we count Priority 1 candidates only, the efficiency is $10/66\approx15\%$. 
Recent {\em Gaia-}based surveys for lensed quasar and close quasar pairs have efficiencies of $\sim30\%$ \citep[e.g.,][]{lemon20,chen22a}.
Since our survey includes faint candidates that are not detected in {\em Gaia}, a selection efficiency of $\sim15\%$ is reasonable.}

The candidate selection method of this work relies on 
the colors and shapes of objects, which is similar to the approach of
the SQLS \citep[e.g.,][]{oguri06, inada12}.
There are two major differences between the SQLS and this work.
First, the parent sample of SQLS consists of spectroscopically confirmed SDSS quasars, 
while this work uses color-selected quasar candidates.
By applying relatively loose color selection criteria, 
we can include lensed quasars with bright deflector galaxies.
Second, the morphological selection in the SQLS identifies quasars that 
cannot be well-described by PSFs,
while in this work, we select objects that can be well-fitted by 
neither PSFs nor S\'ersic profiles. 
The more strict morphological selection of this work 
reduces the contamination from regular galaxies,
resulting in a viable number of candidates for visual inspection.


Recent searches for lensed quasars and close quasar pairs 
heavily rely on the {\em Gaia} survey,
which offers high spatial resolution $(0\farcs4)$
and high-precision parallax and proper motion measurements.
Techniques used in these studies include
(1) searching for quasars that have multiple matches in the {\em Gaia} catalog
\citep[e.g.,][]{agnello17,ducourant18};
(2) selecting quasars that have abnormal astrometry \citep[e.g.,][]{lemon17,chen22a};
(3) selecting quasars that show differences between {\em Gaia} photometry and ground-based
photometry \citep[e.g.,][]{lemon17}.

 {
Since the primary targets of our survey are quasar pairs and lensed quasars at $z\gtrsim5$, 
in this work, we do not require candidates to be detected in {\em Gaia}.
The reason is that {\em Gaia} mainly covers the optical wavelengths,
while high-redshift quasars have little flux at the blue side of Ly$\alpha$ due to IGM absorption.
Candidate selection methods based on object colors and morphology
help us to find lensed quasars and quasar pairs beyond the 
flux limit of {\em Gaia}.
Indeed, three quasar pairs and NIQs discovered 
in our survey (J1051+0334, J2329--0522, and J2037--4537)
 are not detected in {\em Gaia} DR3.}

 {In the future, we will adopt new data releases from public surveys
(e.g., the Legacy Survey DR9 and the {\em Gaia} survey DR3)
and improve the candidate selection method.
In particular, some candidates are resolved into multiple components
in the Legacy Survey, 
though the survey catalog incorrectly describes the candidates as single-component objects.
J2329--0522 (Figure \ref{fig:J2329}), J0402--4220 (Figure \ref{fig:J0402}), J0954--0022 (Figure \ref{fig:J0954}), and J1051+0334 (Figure \ref{fig:J1051})
are all examples of such cases.
It is thus useful to reprocess the images of candidates using optimal source deblending algorithms,
{which will provide the flux of the resolved components.
We note that one major source of contaminants
is projected pairs of objects 
(e.g., galaxy+galaxy, galaxy+star, star+star, and quasar+star).
Reprocessing the images from public surveys will help us to 
get accurate SEDs of the resolved components 
and to exclude projected pairs of objects in the color selection.}

 {Another critical improvement to make 
is to develop automatic algorithms in place of visual inspection. 
For example, neural networks have been found useful in finding lensed galaxies \citep[e.g.,][]{cmudeeplens}
as well as quadruply-lensed quasars \citep[e.g.,][]{akh22}.
However, the application of deep learning in searching for high-redshift  lensed quasars and quasar pairs has been poorly explored.
Developing these algorithms is necessary to handle the large candidate sample
from the LSST and the {\em Euclid} survey, as well as to estimate the
survey completeness and correctly analyze the statistics of 
 lensed quasars and quasar pairs found in this survey.}

\section{Summary} \label{sec:summary}

We present the first results from a new survey for high-redshift lensed quasars and close quasar pairs.
The candidate selection is based on broad-band photometry  
 from a number of public imaging surveys and
 morphological information from the Legacy Survey DR8.
We use PRF classifiers to identify objects that have colors 
consistent with a lensed quasar or a quasar pair.
We further select objects that have close companions
or have large image fitting residuals in the Legacy Survey 
as candidates.
We visually inspect the candidates and carry out
follow-up imaging and spectroscopy for high-priority ones.

In semesters 2021A and 2021B, we observed 66 priority 1 candidates
and 10 priority 2 candidates, from which we
independently discovered ten lensed quasars and close quasar pairs.
Three out of ten objects discovered in this survey are not detected in {\em Gaia}.
In this paper, 
we highlight J0025--0145 at $z=5.07$ (an intermediately-lensed quasar)
and J2329--0522 at $z=4.85$ (the most distant close quasar pair reported so far).
 {The {\em HST} images of J0025--0145 indicate a tentative detection of the quasar host galaxy,
with the help of lensing magnification and distortion.}
We also mention four lensed quasars and NIQs at $z<4$
that are not reported by previous studies.

In the future, we will improve our survey by adopting new data releases from public sky surveys and updating the candidate selection method.
 {We will also develop automatic tools (e.g., deep-learning-based algorithms)
in place of the visual inspection.}
Upcoming wide-area sky surveys like LSST and {\em Euclid}
will provide unprecedented depth and spatial resolution
and probe compact lensing systems and sub-kpc quasar pairs,
which are largely unexplored so far \citep[e.g.,][]{chen22a}.
The methods used in this work can be adapted and applied in
these future imaging surveys.

\acknowledgments
We thank the referee for the valuable comments.
MY, XF and JY acknowledge supports by NSF grants AST 19-08284.
MY and XF acknowledges supports by HST-GO-16460 grant from the Space Telescope Science Institute. 
FW thanks the support provided by NASA through the NASA Hubble Fellowship grant 
\#HST-HF2-51448.001-A awarded by the Space Telescope Science Institute, 
which is operated by the Association of Universities for Research in Astronomy, Incorporated, 
under NASA contract NAS5-26555.

This research is based on observations made with the NASA/ESA Hubble Space Telescope 
obtained from the Space Telescope Science Institute, 
which is operated by the Association of Universities for Research in Astronomy, Inc., 
under NASA contract NAS 5–26555. These observations are associated with program GO-16460.
Some of the data presented in this paper were obtained from the Mikulski Archive for Space Telescopes (MAST)
at the Space Telescope Science Institute. 
The specific observations analyzed can be accessed via
\dataset[doi:10.17909/nmc1-1g12]{http://dx.doi.org/10.17909/nmc1-1g12}.
This paper includes data gathered with the 6.5 meter Magellan Telescopes located at Las Campanas Observatory, Chile.
This research used the facilities of the Italian Center for Astronomical Archive (IA2) operated by INAF at the Astronomical Observatory of Trieste.

%

\vspace{5mm}
\facilities{{\em{HST}}(ACS/WFC), Magellan/Clay(LDSS3), Magellan/Baade(FIRE), LBT(LUCI1)}

\software{NumPy \citep{numpy1,numpy2},
        Astropy \citep{2013A&A...558A..33A,2018AJ....156..123A},  SIMQSO \citep{simqso},
         galfit \citep{galfit}}



\bibliography{sample631}{}

\begin{thebibliography}{}
\expandafter\ifx\csname natexlab\endcsname\relax\def\natexlab#1{#1}\fi
\providecommand{\url}[1]{\href{#1}{#1}}
\providecommand{\dodoi}[1]{doi:~\href{http://doi.org/#1}{\nolinkurl{#1}}}
\providecommand{\doeprint}[1]{\href{http://ascl.net/#1}{\nolinkurl{http://ascl.net/#1}}}
\providecommand{\doarXiv}[1]{\href{https://arxiv.org/abs/#1}{\nolinkurl{https://arxiv.org/abs/#1}}}

\bibitem[{{Agnello}(2017)}]{agnello17}
{Agnello}, A. 2017, \mnras, 471, 2013, \dodoi{10.1093/mnras/stx1650}

\bibitem[{{Akhazhanov} {et~al.}(2022){Akhazhanov}, {More}, {Amini}, {Hazlett},
  {Treu}, {Birrer}, {Shajib}, {Liao}, {Lemon}, {Agnello}, {Nord}, {Aguena},
  {Allam}, {Andrade-Oliveira}, {Annis}, {Brooks}, {Buckley-Geer}, {Burke},
  {Carnero Rosell}, {Carrasco Kind}, {Carretero}, {Choi}, {Conselice},
  {Costanzi}, {da Costa}, {Pereira}, {De Vicente}, {Desai}, {Dietrich}, {Doel},
  {Everett}, {Ferrero}, {Finley}, {Flaugher}, {Frieman}, {Garc{\'\i}a-Bellido},
  {Gerdes}, {Gruen}, {Gruendl}, {Gschwend}, {Gutierrez}, {Hinton}, {Hollowood},
  {Honscheid}, {James}, {Kim}, {Kuehn}, {Kuropatkin}, {Lahav}, {Lima}, {Lin},
  {Maia}, {March}, {Menanteau}, {Miquel}, {Morgan}, {Palmese},
  {Paz-Chinch{\'o}n}, {Pieres}, {Plazas Malag{\'o}n}, {Sanchez}, {Scarpine},
  {Serrano}, {Sevilla-Noarbe}, {Smith}, {Soares-Santos}, {Suchyta}, {Swanson},
  {Tarle}, {To}, {Varga}, {Weller}, \& {(DES Collaboration)}}]{akh22}
{Akhazhanov}, A., {More}, A., {Amini}, A., {et~al.} 2022, \mnras, 513, 2407,
  \dodoi{10.1093/mnras/stac925}

\bibitem[{{Anguita} {et~al.}(2018){Anguita}, {Schechter}, {Kuropatkin},
  {Morgan}, {Ostrovski}, {Abramson}, {Agnello}, {Apostolovski}, {Fassnacht},
  {Hsueh}, {Motta}, {Rojas}, {Rusu}, {Treu}, {Williams}, {Auger},
  {Buckley-Geer}, {Lin}, {McMahon}, {Abbott}, {Allam}, {Annis}, {Bernstein},
  {Bertin}, {Brooks}, {Burke}, {Carnero Rosell}, {Carrasco-Kind}, {Carretero},
  {Cunha}, {D'Andrea}, {De Vicente}, {DePoy}, {Desai}, {Diehl}, {Doel},
  {Flaugher}, {Garc{\'\i}a-Bellido}, {Gerdes}, {Gruen}, {Gruendl}, {Gschwend},
  {Hartley}, {Hollowood}, {Honscheid}, {James}, {Kuehn}, {Lima}, {Maia},
  {Miquel}, {Plazas}, {Sanchez}, {Scarpine}, {Smith}, {Soares-Santos},
  {Sobreira}, {Suchyta}, {Tarle}, \& {Walker}}]{anguita18}
{Anguita}, T., {Schechter}, P.~L., {Kuropatkin}, N., {et~al.} 2018, \mnras,
  480, 5017, \dodoi{10.1093/mnras/sty2172}

\bibitem[{{Astropy Collaboration} {et~al.}(2013){Astropy Collaboration},
  {Robitaille}, {Tollerud}, {Greenfield}, {Droettboom}, {Bray}, {Aldcroft},
  {Davis}, {Ginsburg}, {Price-Whelan}, {Kerzendorf}, {Conley}, {Crighton},
  {Barbary}, {Muna}, {Ferguson}, {Grollier}, {Parikh}, {Nair}, {Unther},
  {Deil}, {Woillez}, {Conseil}, {Kramer}, {Turner}, {Singer}, {Fox}, {Weaver},
  {Zabalza}, {Edwards}, {Azalee Bostroem}, {Burke}, {Casey}, {Crawford},
  {Dencheva}, {Ely}, {Jenness}, {Labrie}, {Lim}, {Pierfederici}, {Pontzen},
  {Ptak}, {Refsdal}, {Servillat}, \& {Streicher}}]{2013A&A...558A..33A}
{Astropy Collaboration}, {Robitaille}, T.~P., {Tollerud}, E.~J., {et~al.} 2013,
  \aap, 558, A33, \dodoi{10.1051/0004-6361/201322068}

\bibitem[{{Astropy Collaboration} {et~al.}(2018){Astropy Collaboration},
  {Price-Whelan}, {Sip{\H{o}}cz}, {G{\"u}nther}, {Lim}, {Crawford}, {Conseil},
  {Shupe}, {Craig}, {Dencheva}, {Ginsburg}, {VanderPlas}, {Bradley},
  {P{\'e}rez-Su{\'a}rez}, {de Val-Borro}, {Aldcroft}, {Cruz}, {Robitaille},
  {Tollerud}, {Ardelean}, {Babej}, {Bach}, {Bachetti}, {Bakanov}, {Bamford},
  {Barentsen}, {Barmby}, {Baumbach}, {Berry}, {Biscani}, {Boquien}, {Bostroem},
  {Bouma}, {Brammer}, {Bray}, {Breytenbach}, {Buddelmeijer}, {Burke},
  {Calderone}, {Cano Rodr{\'\i}guez}, {Cara}, {Cardoso}, {Cheedella}, {Copin},
  {Corrales}, {Crichton}, {D'Avella}, {Deil}, {Depagne}, {Dietrich}, {Donath},
  {Droettboom}, {Earl}, {Erben}, {Fabbro}, {Ferreira}, {Finethy}, {Fox},
  {Garrison}, {Gibbons}, {Goldstein}, {Gommers}, {Greco}, {Greenfield},
  {Groener}, {Grollier}, {Hagen}, {Hirst}, {Homeier}, {Horton}, {Hosseinzadeh},
  {Hu}, {Hunkeler}, {Ivezi{\'c}}, {Jain}, {Jenness}, {Kanarek}, {Kendrew},
  {Kern}, {Kerzendorf}, {Khvalko}, {King}, {Kirkby}, {Kulkarni}, {Kumar},
  {Lee}, {Lenz}, {Littlefair}, {Ma}, {Macleod}, {Mastropietro}, {McCully},
  {Montagnac}, {Morris}, {Mueller}, {Mumford}, {Muna}, {Murphy}, {Nelson},
  {Nguyen}, {Ninan}, {N{\"o}the}, {Ogaz}, {Oh}, {Parejko}, {Parley}, {Pascual},
  {Patil}, {Patil}, {Plunkett}, {Prochaska}, {Rastogi}, {Reddy Janga},
  {Sabater}, {Sakurikar}, {Seifert}, {Sherbert}, {Sherwood-Taylor}, {Shih},
  {Sick}, {Silbiger}, {Singanamalla}, {Singer}, {Sladen}, {Sooley},
  {Sornarajah}, {Streicher}, {Teuben}, {Thomas}, {Tremblay}, {Turner},
  {Terr{\'o}n}, {van Kerkwijk}, {de la Vega}, {Watkins}, {Weaver}, {Whitmore},
  {Woillez}, {Zabalza}, \& {Astropy Contributors}}]{2018AJ....156..123A}
{Astropy Collaboration}, {Price-Whelan}, A.~M., {Sip{\H{o}}cz}, B.~M., {et~al.}
  2018, \aj, 156, 123, \dodoi{10.3847/1538-3881/aabc4f}

\bibitem[{{Bayliss} {et~al.}(2017){Bayliss}, {Sharon}, {Acharyya}, {Gladders},
  {Rigby}, {Bian}, {Bordoloi}, {Runnoe}, {Dahle}, {Kewley}, {Florian},
  {Johnson}, \& {Paterno-Mahler}}]{bayliss17}
{Bayliss}, M.~B., {Sharon}, K., {Acharyya}, A., {et~al.} 2017, \apjl, 845, L14,
  \dodoi{10.3847/2041-8213/aa831a}

\bibitem[{{Bellini} {et~al.}(2018){Bellini}, {Anderson}, \& {Grogin}}]{acspsf}
{Bellini}, A., {Anderson}, J., \& {Grogin}, N.~A. 2018, {Focus-diverse,
  empirical PSF models for the ACS/WFC}, Instrument Science Report ACS 2018-8

\bibitem[{{Bezanson} {et~al.}(2011){Bezanson}, {van Dokkum}, {Franx},
  {Brammer}, {Brinchmann}, {Kriek}, {Labb{\'e}}, {Quadri}, {Rix}, {van de
  Sande}, {Whitaker}, \& {Williams}}]{bezanson11}
{Bezanson}, R., {van Dokkum}, P.~G., {Franx}, M., {et~al.} 2011, \apjl, 737,
  L31, \dodoi{10.1088/2041-8205/737/2/L31}

\bibitem[{{Capelo} {et~al.}(2017){Capelo}, {Dotti}, {Volonteri}, {Mayer},
  {Bellovary}, \& {Shen}}]{capelo17}
{Capelo}, P.~R., {Dotti}, M., {Volonteri}, M., {et~al.} 2017, \mnras, 469,
  4437, \dodoi{10.1093/mnras/stx1067}

\bibitem[{{Cashman} {et~al.}(2021){Cashman}, {Kulkarni}, \&
  {Lopez}}]{cashman21}
{Cashman}, F.~H., {Kulkarni}, V.~P., \& {Lopez}, S. 2021, \aj, 161, 90,
  \dodoi{10.3847/1538-3881/abd2b0}

\bibitem[{{Chambers} {et~al.}(2016){Chambers}, {Magnier}, {Metcalfe},
  {Flewelling}, {Huber}, {Waters}, {Denneau}, {Draper}, {Farrow}, {Finkbeiner},
  {Holmberg}, {Koppenhoefer}, {Price}, {Rest}, {Saglia}, {Schlafly}, {Smartt},
  {Sweeney}, {Wainscoat}, {Burgett}, {Chastel}, {Grav}, {Heasley}, {Hodapp},
  {Jedicke}, {Kaiser}, {Kudritzki}, {Luppino}, {Lupton}, {Monet}, {Morgan},
  {Onaka}, {Shiao}, {Stubbs}, {Tonry}, {White}, {Ba{\~n}ados}, {Bell},
  {Bender}, {Bernard}, {Boegner}, {Boffi}, {Botticella}, {Calamida},
  {Casertano}, {Chen}, {Chen}, {Cole}, {Deacon}, {Frenk}, {Fitzsimmons},
  {Gezari}, {Gibbs}, {Goessl}, {Goggia}, {Gourgue}, {Goldman}, {Grant},
  {Grebel}, {Hambly}, {Hasinger}, {Heavens}, {Heckman}, {Henderson}, {Henning},
  {Holman}, {Hopp}, {Ip}, {Isani}, {Jackson}, {Keyes}, {Koekemoer}, {Kotak},
  {Le}, {Liska}, {Long}, {Lucey}, {Liu}, {Martin}, {Masci}, {McLean}, {Mindel},
  {Misra}, {Morganson}, {Murphy}, {Obaika}, {Narayan}, {Nieto-Santisteban},
  {Norberg}, {Peacock}, {Pier}, {Postman}, {Primak}, {Rae}, {Rai}, {Riess},
  {Riffeser}, {Rix}, {R{\"o}ser}, {Russel}, {Rutz}, {Schilbach}, {Schultz},
  {Scolnic}, {Strolger}, {Szalay}, {Seitz}, {Small}, {Smith}, {Soderblom},
  {Taylor}, {Thomson}, {Taylor}, {Thakar}, {Thiel}, {Thilker}, {Unger},
  {Urata}, {Valenti}, {Wagner}, {Walder}, {Walter}, {Watters}, {Werner},
  {Wood-Vasey}, \& {Wyse}}]{ps1}
{Chambers}, K.~C., {Magnier}, E.~A., {Metcalfe}, N., {et~al.} 2016, arXiv
  e-prints, arXiv:1612.05560.
\newblock \doarXiv{1612.05560}

\bibitem[{{Chen}(2021)}]{chen22b}
{Chen}, Y.-C. 2021, arXiv e-prints, arXiv:2109.06881.
\newblock \doarXiv{2109.06881}

\bibitem[{{Chen} {et~al.}(2022{\natexlab{a}}){Chen}, {Hwang}, {Shen}, {Liu},
  {Zakamska}, {Yang}, \& {Li}}]{chen22a}
{Chen}, Y.-C., {Hwang}, H.-C., {Shen}, Y., {et~al.} 2022{\natexlab{a}}, \apj,
  925, 162, \dodoi{10.3847/1538-4357/ac401b}

\bibitem[{{Chen} {et~al.}(2022{\natexlab{b}}){Chen}, {Liu}, {Foord}, {Shen},
  {Chen}, {Holgado}, {Di Matteo}, {Oguri}, {Hwang}, \& {Zakamska}}]{chen22c}
{Chen}, Y.-C., {Liu}, X., {Foord}, A., {et~al.} 2022{\natexlab{b}}, arXiv
  e-prints, arXiv:2209.11249.
\newblock \doarXiv{2209.11249}

\bibitem[{{Collett}(2015)}]{collett15}
{Collett}, T.~E. 2015, \apj, 811, 20, \dodoi{10.1088/0004-637X/811/1/20}

\bibitem[{{Cornachione} {et~al.}(2020){Cornachione}, {Morgan}, {Millon},
  {Bentz}, {Courbin}, {Bonvin}, \& {Falco}}]{cornachione20}
{Cornachione}, M.~A., {Morgan}, C.~W., {Millon}, M., {et~al.} 2020, \apj, 895,
  125, \dodoi{10.3847/1538-4357/ab557a}

\bibitem[{{Croom} {et~al.}(2001){Croom}, {Smith}, {Boyle}, {Shanks}, {Loaring},
  {Miller}, \& {Lewis}}]{croom01}
{Croom}, S.~M., {Smith}, R.~J., {Boyle}, B.~J., {et~al.} 2001, \mnras, 322,
  L29, \dodoi{10.1046/j.1365-8711.2001.04474.x}

\bibitem[{{Dawes} {et~al.}(2022){Dawes}, {Storfer}, {Huang}, {Aldering}, {Dey},
  \& {Schlegel}}]{dawes22}
{Dawes}, C., {Storfer}, C., {Huang}, X., {et~al.} 2022, arXiv e-prints,
  arXiv:2208.06356.
\newblock \doarXiv{2208.06356}

\bibitem[{{De Rosa} {et~al.}(2019){De Rosa}, {Vignali}, {Bogdanovi{\'c}},
  {Capelo}, {Charisi}, {Dotti}, {Husemann}, {Lusso}, {Mayer}, {Paragi},
  {Runnoe}, {Sesana}, {Steinborn}, {Bianchi}, {Colpi}, {del Valle}, {Frey},
  {Gab{\'a}nyi}, {Giustini}, {Guainazzi}, {Haiman}, {Herrera Ruiz},
  {Herrero-Illana}, {Iwasawa}, {Komossa}, {Lena}, {Loiseau}, {Perez-Torres},
  {Piconcelli}, \& {Volonteri}}]{dr19}
{De Rosa}, A., {Vignali}, C., {Bogdanovi{\'c}}, T., {et~al.} 2019, \nar, 86,
  101525, \dodoi{10.1016/j.newar.2020.101525}

\bibitem[{{DES Collaboration} {et~al.}(2021){DES Collaboration}, {Abbott},
  {Adamow}, {Aguena}, {Allam}, {Amon}, {Annis}, {Avila}, {Bacon}, {Banerji},
  {Bechtol}, {Becker}, {Bernstein}, {Bertin}, {Bhargava}, {Bridle}, {Brooks},
  {Burke}, {Carnero Rosell}, {Carrasco Kind}, {Carretero}, {Castander},
  {Cawthon}, {Chang}, {Choi}, {Conselice}, {Costanzi}, {Crocce}, {da Costa},
  {Davis}, {De Vicente}, {DeRose}, {Desai}, {Diehl}, {Dietrich},
  {Drlica-Wagner}, {Eckert}, {Elvin-Poole}, {Everett}, {Evrard}, {Ferrero},
  {Fert{\'e}}, {Flaugher}, {Fosalba}, {Friedel}, {Frieman},
  {Garc{\'\i}a-Bellido}, {Gelman}, {Gerdes}, {Giannantonio}, {Gill}, {Gruen},
  {Gruendl}, {Gschwend}, {Gutierrez}, {Hartley}, {Hinton}, {Hollowood},
  {Honscheid}, {Huterer}, {James}, {Jeltema}, {Johnson}, {Kent}, {Kron},
  {Kuehn}, {Kuropatkin}, {Lahav}, {Li}, {Lidman}, {Lin}, {MacCrann}, {Maia},
  {Manning}, {March}, {Marshall}, {Martini}, {Melchior}, {Menanteau}, {Miquel},
  {Morgan}, {Myles}, {Neilsen}, {Ogando}, {Palmese}, {Paz-Chinch{\'o}n},
  {Petravick}, {Pieres}, {Plazas}, {Pond}, {Rodriguez-Monroy}, {Romer},
  {Roodman}, {Rykoff}, {Sako}, {Sanchez}, {Santiago}, {Serrano},
  {Sevilla-Noarbe}, {Allyn. Smith}, {Smith}, {Soares-Santos}, {Suchyta},
  {Swanson}, {Tarle}, {Thomas}, {To}, {Tremblay}, {Troxel}, {Tucker}, {Turner},
  {Varga}, {Walker}, {Wechsler}, {Weller}, {Wester}, {Wilkinson}, {Yanny},
  {Zhang}, {Nikutta}, {Fitzpatrick}, {Jacques}, {Scott}, {Olsen}, {Huang},
  {Herrera}, {Juneau}, {Nidever}, {Weaver}, {Adean}, {Correia}, {de Freitas},
  {Freitas}, {Singulani}, \& {Vila-Verde}}]{desdr2}
{DES Collaboration}, {Abbott}, T.~M.~C., {Adamow}, M., {et~al.} 2021, arXiv
  e-prints, arXiv:2101.05765.
\newblock \doarXiv{2101.05765}

\bibitem[{{DESI Collaboration} {et~al.}(2016){DESI Collaboration}, {Aghamousa},
  {Aguilar}, {Ahlen}, {Alam}, {Allen}, {Allende Prieto}, {Annis}, {Bailey},
  {Balland}, {Ballester}, {Baltay}, {Beaufore}, {Bebek}, {Beers}, {Bell},
  {Bernal}, {Besuner}, {Beutler}, {Blake}, {Bleuler}, {Blomqvist}, {Blum},
  {Bolton}, {Briceno}, {Brooks}, {Brownstein}, {Buckley-Geer}, {Burden},
  {Burtin}, {Busca}, {Cahn}, {Cai}, {Cardiel-Sas}, {Carlberg}, {Carton},
  {Casas}, {Castander}, {Cervantes-Cota}, {Claybaugh}, {Close}, {Coker},
  {Cole}, {Comparat}, {Cooper}, {Cousinou}, {Crocce}, {Cuby}, {Cunningham},
  {Davis}, {Dawson}, {de la Macorra}, {De Vicente}, {Delubac}, {Derwent},
  {Dey}, {Dhungana}, {Ding}, {Doel}, {Duan}, {Ealet}, {Edelstein},
  {Eftekharzadeh}, {Eisenstein}, {Elliott}, {Escoffier}, {Evatt}, {Fagrelius},
  {Fan}, {Fanning}, {Farahi}, {Farihi}, {Favole}, {Feng}, {Fernandez},
  {Findlay}, {Finkbeiner}, {Fitzpatrick}, {Flaugher}, {Flender}, {Font-Ribera},
  {Forero-Romero}, {Fosalba}, {Frenk}, {Fumagalli}, {Gaensicke}, {Gallo},
  {Garcia-Bellido}, {Gaztanaga}, {Pietro Gentile Fusillo}, {Gerard},
  {Gershkovich}, {Giannantonio}, {Gillet}, {Gonzalez-de-Rivera},
  {Gonzalez-Perez}, {Gott}, {Graur}, {Gutierrez}, {Guy}, {Habib}, {Heetderks},
  {Heetderks}, {Heitmann}, {Hellwing}, {Herrera}, {Ho}, {Holland}, {Honscheid},
  {Huff}, {Hutchinson}, {Huterer}, {Hwang}, {Illa Laguna}, {Ishikawa},
  {Jacobs}, {Jeffrey}, {Jelinsky}, {Jennings}, {Jiang}, {Jimenez}, {Johnson},
  {Joyce}, {Jullo}, {Juneau}, {Kama}, {Karcher}, {Karkar}, {Kehoe}, {Kennamer},
  {Kent}, {Kilbinger}, {Kim}, {Kirkby}, {Kisner}, {Kitanidis}, {Kneib},
  {Koposov}, {Kovacs}, {Koyama}, {Kremin}, {Kron}, {Kronig}, {Kueter-Young},
  {Lacey}, {Lafever}, {Lahav}, {Lambert}, {Lampton}, {Landriau}, {Lang},
  {Lauer}, {Le Goff}, {Le Guillou}, {Le Van Suu}, {Lee}, {Lee}, {Leitner},
  {Lesser}, {Levi}, {L'Huillier}, {Li}, {Liang}, {Lin}, {Linder}, {Loebman},
  {Luki{\'c}}, {Ma}, {MacCrann}, {Magneville}, {Makarem}, {Manera}, {Manser},
  {Marshall}, {Martini}, {Massey}, {Matheson}, {McCauley}, {McDonald},
  {McGreer}, {Meisner}, {Metcalfe}, {Miller}, {Miquel}, {Moustakas}, {Myers},
  {Naik}, {Newman}, {Nichol}, {Nicola}, {Nicolati da Costa}, {Nie}, {Niz},
  {Norberg}, {Nord}, {Norman}, {Nugent}, {O'Brien}, {Oh}, {Olsen}, {Padilla},
  {Padmanabhan}, {Padmanabhan}, {Palanque-Delabrouille}, {Palmese},
  {Pappalardo}, {P{\^a}ris}, {Park}, {Patej}, {Peacock}, {Peiris}, {Peng},
  {Percival}, {Perruchot}, {Pieri}, {Pogge}, {Pollack}, {Poppett}, {Prada},
  {Prakash}, {Probst}, {Rabinowitz}, {Raichoor}, {Ree}, {Refregier}, {Regal},
  {Reid}, {Reil}, {Rezaie}, {Rockosi}, {Roe}, {Ronayette}, {Roodman}, {Ross},
  {Ross}, {Rossi}, {Rozo}, {Ruhlmann-Kleider}, {Rykoff}, {Sabiu}, {Samushia},
  {Sanchez}, {Sanchez}, {Schlegel}, {Schneider}, {Schubnell}, {Secroun},
  {Seljak}, {Seo}, {Serrano}, {Shafieloo}, {Shan}, {Sharples}, {Sholl},
  {Shourt}, {Silber}, {Silva}, {Sirk}, {Slosar}, {Smith}, {Smoot}, {Som},
  {Song}, {Sprayberry}, {Staten}, {Stefanik}, {Tarle}, {Sien Tie}, {Tinker},
  {Tojeiro}, {Valdes}, {Valenzuela}, {Valluri}, {Vargas-Magana}, {Verde},
  {Walker}, {Wang}, {Wang}, {Weaver}, {Weaverdyck}, {Wechsler}, {Weinberg},
  {White}, {Yang}, {Yeche}, {Zhang}, {Zhao}, {Zheng}, {Zhou}, {Zhou}, {Zhu},
  {Zou}, \& {Zu}}]{desi}
{DESI Collaboration}, {Aghamousa}, A., {Aguilar}, J., {et~al.} 2016, arXiv
  e-prints, arXiv:1611.00036.
\newblock \doarXiv{1611.00036}

\bibitem[{{Desira} {et~al.}(2022){Desira}, {Shu}, {Auger}, {McMahon}, {Lemon},
  {Anguita}, \& {Neira}}]{desira22}
{Desira}, C., {Shu}, Y., {Auger}, M.~W., {et~al.} 2022, \mnras, 509, 738,
  \dodoi{10.1093/mnras/stab2960}

\bibitem[{{Dey} {et~al.}(2019){Dey}, {Schlegel}, {Lang}, {Blum}, {Burleigh},
  {Fan}, {Findlay}, {Finkbeiner}, {Herrera}, {Juneau}, {Landriau}, {Levi},
  {McGreer}, {Meisner}, {Myers}, {Moustakas}, {Nugent}, {Patej}, {Schlafly},
  {Walker}, {Valdes}, {Weaver}, {Y{\`e}che}, {Zou}, {Zhou}, {Abareshi},
  {Abbott}, {Abolfathi}, {Aguilera}, {Alam}, {Allen}, {Alvarez}, {Annis},
  {Ansarinejad}, {Aubert}, {Beechert}, {Bell}, {BenZvi}, {Beutler}, {Bielby},
  {Bolton}, {Brice{\~n}o}, {Buckley-Geer}, {Butler}, {Calamida}, {Carlberg},
  {Carter}, {Casas}, {Castander}, {Choi}, {Comparat}, {Cukanovaite}, {Delubac},
  {DeVries}, {Dey}, {Dhungana}, {Dickinson}, {Ding}, {Donaldson}, {Duan},
  {Duckworth}, {Eftekharzadeh}, {Eisenstein}, {Etourneau}, {Fagrelius},
  {Farihi}, {Fitzpatrick}, {Font-Ribera}, {Fulmer}, {G{\"a}nsicke},
  {Gaztanaga}, {George}, {Gerdes}, {Gontcho}, {Gorgoni}, {Green}, {Guy},
  {Harmer}, {Hernandez}, {Honscheid}, {Huang}, {James}, {Jannuzi}, {Jiang},
  {Joyce}, {Karcher}, {Karkar}, {Kehoe}, {Kneib}, {Kueter-Young}, {Lan},
  {Lauer}, {Le Guillou}, {Le Van Suu}, {Lee}, {Lesser}, {Perreault Levasseur},
  {Li}, {Mann}, {Marshall}, {Mart{\'\i}nez-V{\'a}zquez}, {Martini}, {du Mas des
  Bourboux}, {McManus}, {Meier}, {M{\'e}nard}, {Metcalfe},
  {Mu{\~n}oz-Guti{\'e}rrez}, {Najita}, {Napier}, {Narayan}, {Newman}, {Nie},
  {Nord}, {Norman}, {Olsen}, {Paat}, {Palanque-Delabrouille}, {Peng},
  {Poppett}, {Poremba}, {Prakash}, {Rabinowitz}, {Raichoor}, {Rezaie},
  {Robertson}, {Roe}, {Ross}, {Ross}, {Rudnick}, {Safonova}, {Saha},
  {S{\'a}nchez}, {Savary}, {Schweiker}, {Scott}, {Seo}, {Shan}, {Silva},
  {Slepian}, {Soto}, {Sprayberry}, {Staten}, {Stillman}, {Stupak}, {Summers},
  {Sien Tie}, {Tirado}, {Vargas-Maga{\~n}a}, {Vivas}, {Wechsler}, {Williams},
  {Yang}, {Yang}, {Yapici}, {Zaritsky}, {Zenteno}, {Zhang}, {Zhang}, {Zhou}, \&
  {Zhou}}]{legacy}
{Dey}, A., {Schlegel}, D.~J., {Lang}, D., {et~al.} 2019, \aj, 157, 168,
  \dodoi{10.3847/1538-3881/ab089d}

\bibitem[{{Ding} {et~al.}(2022){Ding}, {Silverman}, \& {Onoue}}]{ding22}
{Ding}, X., {Silverman}, J.~D., \& {Onoue}, M. 2022, arXiv e-prints,
  arXiv:2209.03359.
\newblock \doarXiv{2209.03359}

\bibitem[{{Ducourant} {et~al.}(2018){Ducourant}, {Wertz}, {Krone-Martins},
  {Teixeira}, {Le Campion}, {Galluccio}, {Kl{\"u}ter}, {Delchambre}, {Surdej},
  {Mignard}, {Wambsganss}, {Bastian}, {Graham}, {Djorgovski}, \&
  {Slezak}}]{ducourant18}
{Ducourant}, C., {Wertz}, O., {Krone-Martins}, A., {et~al.} 2018, \aap, 618,
  A56, \dodoi{10.1051/0004-6361/201833480}

\bibitem[{{Dye} {et~al.}(2018){Dye}, {Lawrence}, {Read}, {Fan}, {Kerr},
  {Varricatt}, {Furnell}, {Edge}, {Irwin}, {Hambly}, {Lucas}, {Almaini},
  {Chambers}, {Green}, {Hewett}, {Liu}, {McGreer}, {Best}, {Zhang}, {Sutorius},
  {Froebrich}, {Magnier}, {Hasinger}, {Lederer}, {Bold}, \& {Tedds}}]{uhs}
{Dye}, S., {Lawrence}, A., {Read}, M.~A., {et~al.} 2018, \mnras, 473, 5113,
  \dodoi{10.1093/mnras/stx2622}

\bibitem[{{Eftekharzadeh} {et~al.}(2017){Eftekharzadeh}, {Myers}, {Hennawi},
  {Djorgovski}, {Richards}, {Mahabal}, \& {Graham}}]{eft17}
{Eftekharzadeh}, S., {Myers}, A.~D., {Hennawi}, J.~F., {et~al.} 2017, \mnras,
  468, 77, \dodoi{10.1093/mnras/stx412}

\bibitem[{{Fan} {et~al.}(2019){Fan}, {Wang}, {Yang}, {Keeton}, {Yue},
  {Zabludoff}, {Bian}, {Bonaglia}, {Georgiev}, {Hennawi}, {Li}, {McGreer},
  {Naidu}, {Pacucci}, {Rabien}, {Thompson}, {Venemans}, {Walter}, {Wang}, \&
  {Wu}}]{fan19}
{Fan}, X., {Wang}, F., {Yang}, J., {et~al.} 2019, \apjl, 870, L11,
  \dodoi{10.3847/2041-8213/aaeffe}

\bibitem[{{Farrow} {et~al.}(2014){Farrow}, {Cole}, {Metcalfe}, {Draper},
  {Norberg}, {Foucaud}, {Burgett}, {Chambers}, {Kaiser}, {Kudritzki},
  {Magnier}, {Price}, {Tonry}, \& {Waters}}]{farrow14}
{Farrow}, D.~J., {Cole}, S., {Metcalfe}, N., {et~al.} 2014, \mnras, 437, 748,
  \dodoi{10.1093/mnras/stt1933}

\bibitem[{{Flesch}(2021)}]{milliquas2021}
{Flesch}, E.~W. 2021, arXiv e-prints, arXiv:2105.12985.
\newblock \doarXiv{2105.12985}

\bibitem[{{Fujimoto} {et~al.}(2020){Fujimoto}, {Oguri}, {Nagao}, {Izumi}, \&
  {Ouchi}}]{fujimoto20}
{Fujimoto}, S., {Oguri}, M., {Nagao}, T., {Izumi}, T., \& {Ouchi}, M. 2020,
  \apj, 891, 64, \dodoi{10.3847/1538-4357/ab718c}

\bibitem[{{Gaia Collaboration} {et~al.}(2016){Gaia Collaboration}, {Prusti},
  {de Bruijne}, {Brown}, {Vallenari}, {Babusiaux}, {Bailer-Jones}, {Bastian},
  {Biermann}, {Evans}, {Eyer}, {Jansen}, {Jordi}, {Klioner}, {Lammers},
  {Lindegren}, {Luri}, {Mignard}, {Milligan}, {Panem}, {Poinsignon},
  {Pourbaix}, {Randich}, {Sarri}, {Sartoretti}, {Siddiqui}, {Soubiran},
  {Valette}, {van Leeuwen}, {Walton}, {Aerts}, {Arenou}, {Cropper}, {Drimmel},
  {H{\o}g}, {Katz}, {Lattanzi}, {O'Mullane}, {Grebel}, {Holland}, {Huc},
  {Passot}, {Bramante}, {Cacciari}, {Casta{\~n}eda}, {Chaoul}, {Cheek}, {De
  Angeli}, {Fabricius}, {Guerra}, {Hern{\'a}ndez}, {Jean-Antoine-Piccolo},
  {Masana}, {Messineo}, {Mowlavi}, {Nienartowicz}, {Ord{\'o}{\~n}ez-Blanco},
  {Panuzzo}, {Portell}, {Richards}, {Riello}, {Seabroke}, {Tanga},
  {Th{\'e}venin}, {Torra}, {Els}, {Gracia-Abril}, {Comoretto},
  {Garcia-Reinaldos}, {Lock}, {Mercier}, {Altmann}, {Andrae}, {Astraatmadja},
  {Bellas-Velidis}, {Benson}, {Berthier}, {Blomme}, {Busso}, {Carry},
  {Cellino}, {Clementini}, {Cowell}, {Creevey}, {Cuypers}, {Davidson}, {De
  Ridder}, {de Torres}, {Delchambre}, {Dell'Oro}, {Ducourant}, {Fr{\'e}mat},
  {Garc{\'\i}a-Torres}, {Gosset}, {Halbwachs}, {Hambly}, {Harrison}, {Hauser},
  {Hestroffer}, {Hodgkin}, {Huckle}, {Hutton}, {Jasniewicz}, {Jordan},
  {Kontizas}, {Korn}, {Lanzafame}, {Manteiga}, {Moitinho}, {Muinonen},
  {Osinde}, {Pancino}, {Pauwels}, {Petit}, {Recio-Blanco}, {Robin}, {Sarro},
  {Siopis}, {Smith}, {Smith}, {Sozzetti}, {Thuillot}, {van Reeven}, {Viala},
  {Abbas}, {Abreu Aramburu}, {Accart}, {Aguado}, {Allan}, {Allasia},
  {Altavilla}, {{\'A}lvarez}, {Alves}, {Anderson}, {Andrei}, {Anglada Varela},
  {Antiche}, {Antoja}, {Ant{\'o}n}, {Arcay}, {Atzei}, {Ayache}, {Bach},
  {Baker}, {Balaguer-N{\'u}{\~n}ez}, {Barache}, {Barata}, {Barbier}, {Barblan},
  {Baroni}, {Barrado y Navascu{\'e}s}, {Barros}, {Barstow}, {Becciani},
  {Bellazzini}, {Bellei}, {Bello Garc{\'\i}a}, {Belokurov}, {Bendjoya},
  {Berihuete}, {Bianchi}, {Bienaym{\'e}}, {Billebaud}, {Blagorodnova},
  {Blanco-Cuaresma}, {Boch}, {Bombrun}, {Borrachero}, {Bouquillon}, {Bourda},
  {Bouy}, {Bragaglia}, {Breddels}, {Brouillet}, {Br{\"u}semeister},
  {Bucciarelli}, {Budnik}, {Burgess}, {Burgon}, {Burlacu}, {Busonero}, {Buzzi},
  {Caffau}, {Cambras}, {Campbell}, {Cancelliere}, {Cantat-Gaudin}, {Carlucci},
  {Carrasco}, {Castellani}, {Charlot}, {Charnas}, {Charvet}, {Chassat},
  {Chiavassa}, {Clotet}, {Cocozza}, {Collins}, {Collins}, {Costigan}, {Crifo},
  {Cross}, {Crosta}, {Crowley}, {Dafonte}, {Damerdji}, {Dapergolas}, {David},
  {David}, {De Cat}, {de Felice}, {de Laverny}, {De Luise}, {De March}, {de
  Martino}, {de Souza}, {Debosscher}, {del Pozo}, {Delbo}, {Delgado},
  {Delgado}, {di Marco}, {Di Matteo}, {Diakite}, {Distefano}, {Dolding}, {Dos
  Anjos}, {Drazinos}, {Dur{\'a}n}, {Dzigan}, {Ecale}, {Edvardsson}, {Enke},
  {Erdmann}, {Escolar}, {Espina}, {Evans}, {Eynard Bontemps}, {Fabre},
  {Fabrizio}, {Faigler}, {Falc{\~a}o}, {Farr{\`a}s Casas}, {Faye}, {Federici},
  {Fedorets}, {Fern{\'a}ndez-Hern{\'a}ndez}, {Fernique}, {Fienga}, {Figueras},
  {Filippi}, {Findeisen}, {Fonti}, {Fouesneau}, {Fraile}, {Fraser}, {Fuchs},
  {Furnell}, {Gai}, {Galleti}, {Galluccio}, {Garabato}, {Garc{\'\i}a-Sedano},
  {Gar{\'e}}, {Garofalo}, {Garralda}, {Gavras}, {Gerssen}, {Geyer}, {Gilmore},
  {Girona}, {Giuffrida}, {Gomes}, {Gonz{\'a}lez-Marcos},
  {Gonz{\'a}lez-N{\'u}{\~n}ez}, {Gonz{\'a}lez-Vidal}, {Granvik}, {Guerrier},
  {Guillout}, {Guiraud}, {G{\'u}rpide}, {Guti{\'e}rrez-S{\'a}nchez}, {Guy},
  {Haigron}, {Hatzidimitriou}, {Haywood}, {Heiter}, {Helmi}, {Hobbs},
  {Hofmann}, {Holl}, {Holland}, {Hunt}, {Hypki}, {Icardi}, {Irwin}, {Jevardat
  de Fombelle}, {Jofr{\'e}}, {Jonker}, {Jorissen}, {Julbe}, {Karampelas},
  {Kochoska}, {Kohley}, {Kolenberg}, {Kontizas}, {Koposov}, {Kordopatis},
  {Koubsky}, {Kowalczyk}, {Krone-Martins}, {Kudryashova}, {Kull}, {Bachchan},
  {Lacoste-Seris}, {Lanza}, {Lavigne}, {Le Poncin-Lafitte}, {Lebreton},
  {Lebzelter}, {Leccia}, {Leclerc}, {Lecoeur-Taibi}, {Lemaitre}, {Lenhardt},
  {Leroux}, {Liao}, {Licata}, {Lindstr{\o}m}, {Lister}, {Livanou}, {Lobel},
  {L{\"o}ffler}, {L{\'o}pez}, {Lopez-Lozano}, {Lorenz}, {Loureiro},
  {MacDonald}, {Magalh{\~a}es Fernandes}, {Managau}, {Mann}, {Mantelet},
  {Marchal}, {Marchant}, {Marconi}, {Marie}, {Marinoni}, {Marrese},
  {Marschalk{\'o}}, {Marshall}, {Mart{\'\i}n-Fleitas}, {Martino}, {Mary},
  {Matijevi{\v{c}}}, {Mazeh}, {McMillan}, {Messina}, {Mestre}, {Michalik},
  {Millar}, {Miranda}, {Molina}, {Molinaro}, {Molinaro}, {Moln{\'a}r},
  {Moniez}, {Montegriffo}, {Monteiro}, {Mor}, {Mora}, {Morbidelli}, {Morel},
  {Morgenthaler}, {Morley}, {Morris}, {Mulone}, {Muraveva}, {Musella},
  {Narbonne}, {Nelemans}, {Nicastro}, {Noval}, {Ord{\'e}novic},
  {Ordieres-Mer{\'e}}, {Osborne}, {Pagani}, {Pagano}, {Pailler}, {Palacin},
  {Palaversa}, {Parsons}, {Paulsen}, {Pecoraro}, {Pedrosa}, {Pentik{\"a}inen},
  {Pereira}, {Pichon}, {Piersimoni}, {Pineau}, {Plachy}, {Plum}, {Poujoulet},
  {Pr{\v{s}}a}, {Pulone}, {Ragaini}, {Rago}, {Rambaux}, {Ramos-Lerate},
  {Ranalli}, {Rauw}, {Read}, {Regibo}, {Renk}, {Reyl{\'e}}, {Ribeiro},
  {Rimoldini}, {Ripepi}, {Riva}, {Rixon}, {Roelens}, {Romero-G{\'o}mez},
  {Rowell}, {Royer}, {Rudolph}, {Ruiz-Dern}, {Sadowski}, {Sagrist{\`a}
  Sell{\'e}s}, {Sahlmann}, {Salgado}, {Salguero}, {Sarasso}, {Savietto},
  {Schnorhk}, {Schultheis}, {Sciacca}, {Segol}, {Segovia}, {Segransan},
  {Serpell}, {Shih}, {Smareglia}, {Smart}, {Smith}, {Solano}, {Solitro},
  {Sordo}, {Soria Nieto}, {Souchay}, {Spagna}, {Spoto}, {Stampa}, {Steele},
  {Steidelm{\"u}ller}, {Stephenson}, {Stoev}, {Suess}, {S{\"u}veges}, {Surdej},
  {Szabados}, {Szegedi-Elek}, {Tapiador}, {Taris}, {Tauran}, {Taylor},
  {Teixeira}, {Terrett}, {Tingley}, {Trager}, {Turon}, {Ulla}, {Utrilla},
  {Valentini}, {van Elteren}, {Van Hemelryck}, {van Leeuwen}, {Varadi},
  {Vecchiato}, {Veljanoski}, {Via}, {Vicente}, {Vogt}, {Voss}, {Votruba},
  {Voutsinas}, {Walmsley}, {Weiler}, {Weingrill}, {Werner}, {Wevers},
  {Whitehead}, {Wyrzykowski}, {Yoldas}, {{\v{Z}}erjal}, {Zucker}, {Zurbach},
  {Zwitter}, {Alecu}, {Allen}, {Allende Prieto}, {Amorim},
  {Anglada-Escud{\'e}}, {Arsenijevic}, {Azaz}, {Balm}, {Beck}, {Bernstein},
  {Bigot}, {Bijaoui}, {Blasco}, {Bonfigli}, {Bono}, {Boudreault}, {Bressan},
  {Brown}, {Brunet}, {Bunclark}, {Buonanno}, {Butkevich}, {Carret}, {Carrion},
  {Chemin}, {Ch{\'e}reau}, {Corcione}, {Darmigny}, {de Boer}, {de Teodoro}, {de
  Zeeuw}, {Delle Luche}, {Domingues}, {Dubath}, {Fodor}, {Fr{\'e}zouls},
  {Fries}, {Fustes}, {Fyfe}, {Gallardo}, {Gallegos}, {Gardiol}, {Gebran},
  {Gomboc}, {G{\'o}mez}, {Grux}, {Gueguen}, {Heyrovsky}, {Hoar}, {Iannicola},
  {Isasi Parache}, {Janotto}, {Joliet}, {Jonckheere}, {Keil}, {Kim},
  {Klagyivik}, {Klar}, {Knude}, {Kochukhov}, {Kolka}, {Kos}, {Kutka}, {Lainey},
  {LeBouquin}, {Liu}, {Loreggia}, {Makarov}, {Marseille}, {Martayan},
  {Martinez-Rubi}, {Massart}, {Meynadier}, {Mignot}, {Munari}, {Nguyen},
  {Nordlander}, {Ocvirk}, {O'Flaherty}, {Olias Sanz}, {Ortiz}, {Osorio},
  {Oszkiewicz}, {Ouzounis}, {Palmer}, {Park}, {Pasquato}, {Peltzer}, {Peralta},
  {P{\'e}turaud}, {Pieniluoma}, {Pigozzi}, {Poels}, {Prat}, {Prod'homme},
  {Raison}, {Rebordao}, {Risquez}, {Rocca-Volmerange}, {Rosen}, {Ruiz-Fuertes},
  {Russo}, {Sembay}, {Serraller Vizcaino}, {Short}, {Siebert}, {Silva},
  {Sinachopoulos}, {Slezak}, {Soffel}, {Sosnowska}, {Strai{\v{z}}ys}, {ter
  Linden}, {Terrell}, {Theil}, {Tiede}, {Troisi}, {Tsalmantza}, {Tur},
  {Vaccari}, {Vachier}, {Valles}, {Van Hamme}, {Veltz}, {Virtanen}, {Wallut},
  {Wichmann}, {Wilkinson}, {Ziaeepour}, \& {Zschocke}}]{gaia}
{Gaia Collaboration}, {Prusti}, T., {de Bruijne}, J.~H.~J., {et~al.} 2016,
  \aap, 595, A1, \dodoi{10.1051/0004-6361/201629272}

\bibitem[{{Gaia Collaboration} {et~al.}(2018){Gaia Collaboration}, {Brown},
  {Vallenari}, {Prusti}, {de Bruijne}, {Babusiaux}, {Bailer-Jones}, {Biermann},
  {Evans}, {Eyer}, {Jansen}, {Jordi}, {Klioner}, {Lammers}, {Lindegren},
  {Luri}, {Mignard}, {Panem}, {Pourbaix}, {Randich}, {Sartoretti}, {Siddiqui},
  {Soubiran}, {van Leeuwen}, {Walton}, {Arenou}, {Bastian}, {Cropper},
  {Drimmel}, {Katz}, {Lattanzi}, {Bakker}, {Cacciari}, {Casta{\~n}eda},
  {Chaoul}, {Cheek}, {De Angeli}, {Fabricius}, {Guerra}, {Holl}, {Masana},
  {Messineo}, {Mowlavi}, {Nienartowicz}, {Panuzzo}, {Portell}, {Riello},
  {Seabroke}, {Tanga}, {Th{\'e}venin}, {Gracia-Abril}, {Comoretto},
  {Garcia-Reinaldos}, {Teyssier}, {Altmann}, {Andrae}, {Audard},
  {Bellas-Velidis}, {Benson}, {Berthier}, {Blomme}, {Burgess}, {Busso},
  {Carry}, {Cellino}, {Clementini}, {Clotet}, {Creevey}, {Davidson}, {De
  Ridder}, {Delchambre}, {Dell'Oro}, {Ducourant},
  {Fern{\'a}ndez-Hern{\'a}ndez}, {Fouesneau}, {Fr{\'e}mat}, {Galluccio},
  {Garc{\'\i}a-Torres}, {Gonz{\'a}lez-N{\'u}{\~n}ez}, {Gonz{\'a}lez-Vidal},
  {Gosset}, {Guy}, {Halbwachs}, {Hambly}, {Harrison}, {Hern{\'a}ndez},
  {Hestroffer}, {Hodgkin}, {Hutton}, {Jasniewicz}, {Jean-Antoine-Piccolo},
  {Jordan}, {Korn}, {Krone-Martins}, {Lanzafame}, {Lebzelter}, {L{\"o}ffler},
  {Manteiga}, {Marrese}, {Mart{\'\i}n-Fleitas}, {Moitinho}, {Mora}, {Muinonen},
  {Osinde}, {Pancino}, {Pauwels}, {Petit}, {Recio-Blanco}, {Richards},
  {Rimoldini}, {Robin}, {Sarro}, {Siopis}, {Smith}, {Sozzetti}, {S{\"u}veges},
  {Torra}, {van Reeven}, {Abbas}, {Abreu Aramburu}, {Accart}, {Aerts},
  {Altavilla}, {{\'A}lvarez}, {Alvarez}, {Alves}, {Anderson}, {Andrei},
  {Anglada Varela}, {Antiche}, {Antoja}, {Arcay}, {Astraatmadja}, {Bach},
  {Baker}, {Balaguer-N{\'u}{\~n}ez}, {Balm}, {Barache}, {Barata}, {Barbato},
  {Barblan}, {Barklem}, {Barrado}, {Barros}, {Barstow}, {Bartholom{\'e}
  Mu{\~n}oz}, {Bassilana}, {Becciani}, {Bellazzini}, {Berihuete}, {Bertone},
  {Bianchi}, {Bienaym{\'e}}, {Blanco-Cuaresma}, {Boch}, {Boeche}, {Bombrun},
  {Borrachero}, {Bossini}, {Bouquillon}, {Bourda}, {Bragaglia}, {Bramante},
  {Breddels}, {Bressan}, {Brouillet}, {Br{\"u}semeister}, {Brugaletta},
  {Bucciarelli}, {Burlacu}, {Busonero}, {Butkevich}, {Buzzi}, {Caffau},
  {Cancelliere}, {Cannizzaro}, {Cantat-Gaudin}, {Carballo}, {Carlucci},
  {Carrasco}, {Casamiquela}, {Castellani}, {Castro-Ginard}, {Charlot},
  {Chemin}, {Chiavassa}, {Cocozza}, {Costigan}, {Cowell}, {Crifo}, {Crosta},
  {Crowley}, {Cuypers}, {Dafonte}, {Damerdji}, {Dapergolas}, {David}, {David},
  {de Laverny}, {De Luise}, {De March}, {de Martino}, {de Souza}, {de Torres},
  {Debosscher}, {del Pozo}, {Delbo}, {Delgado}, {Delgado}, {Di Matteo},
  {Diakite}, {Diener}, {Distefano}, {Dolding}, {Drazinos}, {Dur{\'a}n},
  {Edvardsson}, {Enke}, {Eriksson}, {Esquej}, {Eynard Bontemps}, {Fabre},
  {Fabrizio}, {Faigler}, {Falc{\~a}o}, {Farr{\`a}s Casas}, {Federici},
  {Fedorets}, {Fernique}, {Figueras}, {Filippi}, {Findeisen}, {Fonti},
  {Fraile}, {Fraser}, {Fr{\'e}zouls}, {Gai}, {Galleti}, {Garabato},
  {Garc{\'\i}a-Sedano}, {Garofalo}, {Garralda}, {Gavel}, {Gavras}, {Gerssen},
  {Geyer}, {Giacobbe}, {Gilmore}, {Girona}, {Giuffrida}, {Glass}, {Gomes},
  {Granvik}, {Gueguen}, {Guerrier}, {Guiraud}, {Guti{\'e}rrez-S{\'a}nchez},
  {Haigron}, {Hatzidimitriou}, {Hauser}, {Haywood}, {Heiter}, {Helmi}, {Heu},
  {Hilger}, {Hobbs}, {Hofmann}, {Holland}, {Huckle}, {Hypki}, {Icardi},
  {Jan{\ss}en}, {Jevardat de Fombelle}, {Jonker}, {Juh{\'a}sz}, {Julbe},
  {Karampelas}, {Kewley}, {Klar}, {Kochoska}, {Kohley}, {Kolenberg},
  {Kontizas}, {Kontizas}, {Koposov}, {Kordopatis}, {Kostrzewa-Rutkowska},
  {Koubsky}, {Lambert}, {Lanza}, {Lasne}, {Lavigne}, {Le Fustec}, {Le
  Poncin-Lafitte}, {Lebreton}, {Leccia}, {Leclerc}, {Lecoeur-Taibi},
  {Lenhardt}, {Leroux}, {Liao}, {Licata}, {Lindstr{\o}m}, {Lister}, {Livanou},
  {Lobel}, {L{\'o}pez}, {Managau}, {Mann}, {Mantelet}, {Marchal}, {Marchant},
  {Marconi}, {Marinoni}, {Marschalk{\'o}}, {Marshall}, {Martino}, {Marton},
  {Mary}, {Massari}, {Matijevi{\v{c}}}, {Mazeh}, {McMillan}, {Messina},
  {Michalik}, {Millar}, {Molina}, {Molinaro}, {Moln{\'a}r}, {Montegriffo},
  {Mor}, {Morbidelli}, {Morel}, {Morris}, {Mulone}, {Muraveva}, {Musella},
  {Nelemans}, {Nicastro}, {Noval}, {O'Mullane}, {Ord{\'e}novic},
  {Ord{\'o}{\~n}ez-Blanco}, {Osborne}, {Pagani}, {Pagano}, {Pailler},
  {Palacin}, {Palaversa}, {Panahi}, {Pawlak}, {Piersimoni}, {Pineau}, {Plachy},
  {Plum}, {Poggio}, {Poujoulet}, {Pr{\v{s}}a}, {Pulone}, {Racero}, {Ragaini},
  {Rambaux}, {Ramos-Lerate}, {Regibo}, {Reyl{\'e}}, {Riclet}, {Ripepi}, {Riva},
  {Rivard}, {Rixon}, {Roegiers}, {Roelens}, {Romero-G{\'o}mez}, {Rowell},
  {Royer}, {Ruiz-Dern}, {Sadowski}, {Sagrist{\`a} Sell{\'e}s}, {Sahlmann},
  {Salgado}, {Salguero}, {Sanna}, {Santana-Ros}, {Sarasso}, {Savietto},
  {Schultheis}, {Sciacca}, {Segol}, {Segovia}, {S{\'e}gransan}, {Shih},
  {Siltala}, {Silva}, {Smart}, {Smith}, {Solano}, {Solitro}, {Sordo}, {Soria
  Nieto}, {Souchay}, {Spagna}, {Spoto}, {Stampa}, {Steele},
  {Steidelm{\"u}ller}, {Stephenson}, {Stoev}, {Suess}, {Surdej}, {Szabados},
  {Szegedi-Elek}, {Tapiador}, {Taris}, {Tauran}, {Taylor}, {Teixeira},
  {Terrett}, {Teyssandier}, {Thuillot}, {Titarenko}, {Torra Clotet}, {Turon},
  {Ulla}, {Utrilla}, {Uzzi}, {Vaillant}, {Valentini}, {Valette}, {van Elteren},
  {Van Hemelryck}, {van Leeuwen}, {Vaschetto}, {Vecchiato}, {Veljanoski},
  {Viala}, {Vicente}, {Vogt}, {von Essen}, {Voss}, {Votruba}, {Voutsinas},
  {Walmsley}, {Weiler}, {Wertz}, {Wevers}, {Wyrzykowski}, {Yoldas},
  {{\v{Z}}erjal}, {Ziaeepour}, {Zorec}, {Zschocke}, {Zucker}, {Zurbach}, \&
  {Zwitter}}]{gaiadr2}
{Gaia Collaboration}, {Brown}, A.~G.~A., {Vallenari}, A., {et~al.} 2018, \aap,
  616, A1, \dodoi{10.1051/0004-6361/201833051}

\bibitem[{{Gilman} {et~al.}(2020){Gilman}, {Birrer}, {Nierenberg}, {Treu},
  {Du}, \& {Benson}}]{gilman20}
{Gilman}, D., {Birrer}, S., {Nierenberg}, A., {et~al.} 2020, \mnras, 491, 6077,
  \dodoi{10.1093/mnras/stz3480}

\bibitem[{{Green} {et~al.}(2010){Green}, {Myers}, {Barkhouse}, {Mulchaey},
  {Bennert}, {Cox}, \& {Aldcroft}}]{green10}
{Green}, P.~J., {Myers}, A.~D., {Barkhouse}, W.~A., {et~al.} 2010, \apj, 710,
  1578, \dodoi{10.1088/0004-637X/710/2/1578}

\bibitem[{{Harris} {et~al.}(2020){Harris}, {Millman}, {van der Walt},
  {Gommers}, {Virtanen}, {Cournapeau}, {Wieser}, {Taylor}, {Berg}, {Smith},
  {Kern}, {Picus}, {Hoyer}, {van Kerkwijk}, {Brett}, {Haldane}, {del R{\'\i}o},
  {Wiebe}, {Peterson}, {G{\'e}rard-Marchant}, {Sheppard}, {Reddy}, {Weckesser},
  {Abbasi}, {Gohlke}, \& {Oliphant}}]{numpy2}
{Harris}, C.~R., {Millman}, K.~J., {van der Walt}, S.~J., {et~al.} 2020, \nat,
  585, 357, \dodoi{10.1038/s41586-020-2649-2}

\bibitem[{{Hennawi} {et~al.}(2006){Hennawi}, {Strauss}, {Oguri}, {Inada},
  {Richards}, {Pindor}, {Schneider}, {Becker}, {Gregg}, {Hall}, {Johnston},
  {Fan}, {Burles}, {Schlegel}, {Gunn}, {Lupton}, {Bahcall}, {Brunner}, \&
  {Brinkmann}}]{hennawi06}
{Hennawi}, J.~F., {Strauss}, M.~A., {Oguri}, M., {et~al.} 2006, \aj, 131, 1,
  \dodoi{10.1086/498235}

\bibitem[{{Hennawi} {et~al.}(2010){Hennawi}, {Myers}, {Shen}, {Strauss},
  {Djorgovski}, {Fan}, {Glikman}, {Mahabal}, {Martin}, {Richards}, {Schneider},
  \& {Shankar}}]{hennawi10}
{Hennawi}, J.~F., {Myers}, A.~D., {Shen}, Y., {et~al.} 2010, \apj, 719, 1672,
  \dodoi{10.1088/0004-637X/719/2/1672}

\bibitem[{{Hutsem{\'e}kers} {et~al.}(2020){Hutsem{\'e}kers}, {Sluse}, \&
  {Kumar}}]{hut20}
{Hutsem{\'e}kers}, D., {Sluse}, D., \& {Kumar}, P. 2020, \aap, 633, A101,
  \dodoi{10.1051/0004-6361/201936973}

\bibitem[{{Inada} {et~al.}(2012){Inada}, {Oguri}, {Shin}, {Kayo}, {Strauss},
  {Morokuma}, {Rusu}, {Fukugita}, {Kochanek}, {Richards}, {Schneider}, {York},
  {Bahcall}, {Frieman}, {Hall}, \& {White}}]{inada12}
{Inada}, N., {Oguri}, M., {Shin}, M.-S., {et~al.} 2012, \aj, 143, 119,
  \dodoi{10.1088/0004-6256/143/5/119}

\bibitem[{{Ivezi{\'c}} {et~al.}(2019){Ivezi{\'c}}, {Kahn}, {Tyson}, {Abel},
  {Acosta}, {Allsman}, {Alonso}, {AlSayyad}, {Anderson}, {Andrew}, {Angel},
  {Angeli}, {Ansari}, {Antilogus}, {Araujo}, {Armstrong}, {Arndt}, {Astier},
  {Aubourg}, {Auza}, {Axelrod}, {Bard}, {Barr}, {Barrau}, {Bartlett}, {Bauer},
  {Bauman}, {Baumont}, {Bechtol}, {Bechtol}, {Becker}, {Becla}, {Beldica},
  {Bellavia}, {Bianco}, {Biswas}, {Blanc}, {Blazek}, {Blandford}, {Bloom},
  {Bogart}, {Bond}, {Booth}, {Borgland}, {Borne}, {Bosch}, {Boutigny},
  {Brackett}, {Bradshaw}, {Brandt}, {Brown}, {Bullock}, {Burchat}, {Burke},
  {Cagnoli}, {Calabrese}, {Callahan}, {Callen}, {Carlin}, {Carlson},
  {Chandrasekharan}, {Charles-Emerson}, {Chesley}, {Cheu}, {Chiang}, {Chiang},
  {Chirino}, {Chow}, {Ciardi}, {Claver}, {Cohen-Tanugi}, {Cockrum}, {Coles},
  {Connolly}, {Cook}, {Cooray}, {Covey}, {Cribbs}, {Cui}, {Cutri}, {Daly},
  {Daniel}, {Daruich}, {Daubard}, {Daues}, {Dawson}, {Delgado}, {Dellapenna},
  {de Peyster}, {de Val-Borro}, {Digel}, {Doherty}, {Dubois},
  {Dubois-Felsmann}, {Durech}, {Economou}, {Eifler}, {Eracleous}, {Emmons},
  {Fausti Neto}, {Ferguson}, {Figueroa}, {Fisher-Levine}, {Focke}, {Foss},
  {Frank}, {Freemon}, {Gangler}, {Gawiser}, {Geary}, {Gee}, {Geha}, {Gessner},
  {Gibson}, {Gilmore}, {Glanzman}, {Glick}, {Goldina}, {Goldstein}, {Goodenow},
  {Graham}, {Gressler}, {Gris}, {Guy}, {Guyonnet}, {Haller}, {Harris},
  {Hascall}, {Haupt}, {Hernandez}, {Herrmann}, {Hileman}, {Hoblitt}, {Hodgson},
  {Hogan}, {Howard}, {Huang}, {Huffer}, {Ingraham}, {Innes}, {Jacoby}, {Jain},
  {Jammes}, {Jee}, {Jenness}, {Jernigan}, {Jevremovi{\'c}}, {Johns}, {Johnson},
  {Johnson}, {Jones}, {Juramy-Gilles}, {Juri{\'c}}, {Kalirai}, {Kallivayalil},
  {Kalmbach}, {Kantor}, {Karst}, {Kasliwal}, {Kelly}, {Kessler}, {Kinnison},
  {Kirkby}, {Knox}, {Kotov}, {Krabbendam}, {Krughoff}, {Kub{\'a}nek},
  {Kuczewski}, {Kulkarni}, {Ku}, {Kurita}, {Lage}, {Lambert}, {Lange},
  {Langton}, {Le Guillou}, {Levine}, {Liang}, {Lim}, {Lintott}, {Long},
  {Lopez}, {Lotz}, {Lupton}, {Lust}, {MacArthur}, {Mahabal}, {Mandelbaum},
  {Markiewicz}, {Marsh}, {Marshall}, {Marshall}, {May}, {McKercher}, {McQueen},
  {Meyers}, {Migliore}, {Miller}, {Mills}, {Miraval}, {Moeyens}, {Moolekamp},
  {Monet}, {Moniez}, {Monkewitz}, {Montgomery}, {Morrison}, {Mueller},
  {Muller}, {Mu{\~n}oz Arancibia}, {Neill}, {Newbry}, {Nief}, {Nomerotski},
  {Nordby}, {O'Connor}, {Oliver}, {Olivier}, {Olsen}, {O'Mullane}, {Ortiz},
  {Osier}, {Owen}, {Pain}, {Palecek}, {Parejko}, {Parsons}, {Pease},
  {Peterson}, {Peterson}, {Petravick}, {Libby Petrick}, {Petry},
  {Pierfederici}, {Pietrowicz}, {Pike}, {Pinto}, {Plante}, {Plate}, {Plutchak},
  {Price}, {Prouza}, {Radeka}, {Rajagopal}, {Rasmussen}, {Regnault}, {Reil},
  {Reiss}, {Reuter}, {Ridgway}, {Riot}, {Ritz}, {Robinson}, {Roby}, {Roodman},
  {Rosing}, {Roucelle}, {Rumore}, {Russo}, {Saha}, {Sassolas}, {Schalk},
  {Schellart}, {Schindler}, {Schmidt}, {Schneider}, {Schneider}, {Schoening},
  {Schumacher}, {Schwamb}, {Sebag}, {Selvy}, {Sembroski}, {Seppala}, {Serio},
  {Serrano}, {Shaw}, {Shipsey}, {Sick}, {Silvestri}, {Slater}, {Smith},
  {Smith}, {Sobhani}, {Soldahl}, {Storrie-Lombardi}, {Stover}, {Strauss},
  {Street}, {Stubbs}, {Sullivan}, {Sweeney}, {Swinbank}, {Szalay}, {Takacs},
  {Tether}, {Thaler}, {Thayer}, {Thomas}, {Thornton}, {Thukral}, {Tice},
  {Trilling}, {Turri}, {Van Berg}, {Vanden Berk}, {Vetter}, {Virieux},
  {Vucina}, {Wahl}, {Walkowicz}, {Walsh}, {Walter}, {Wang}, {Wang}, {Warner},
  {Wiecha}, {Willman}, {Winters}, {Wittman}, {Wolff}, {Wood-Vasey}, {Wu},
  {Xin}, {Yoachim}, \& {Zhan}}]{lsst}
{Ivezi{\'c}}, {\v{Z}}., {Kahn}, S.~M., {Tyson}, J.~A., {et~al.} 2019, \apj,
  873, 111, \dodoi{10.3847/1538-4357/ab042c}

\bibitem[{{Keeton} {et~al.}(2005){Keeton}, {Kuhlen}, \& {Haiman}}]{keeton05}
{Keeton}, C.~R., {Kuhlen}, M., \& {Haiman}, Z. 2005, \apj, 621, 559,
  \dodoi{10.1086/427722}

\bibitem[{{Khramtsov} {et~al.}(2019){Khramtsov}, {Sergeyev}, {Spiniello},
  {Tortora}, {Napolitano}, {Agnello}, {Getman}, {de Jong}, {Kuijken},
  {Radovich}, {Shan}, \& {Shulga}}]{khramtsov19}
{Khramtsov}, V., {Sergeyev}, A., {Spiniello}, C., {et~al.} 2019, \aap, 632,
  A56, \dodoi{10.1051/0004-6361/201936006}

\bibitem[{{Kochanek}(2004)}]{kochanek04}
{Kochanek}, C.~S. 2004, \apj, 605, 58, \dodoi{10.1086/382180}

\bibitem[{{Lang} {et~al.}(2016){Lang}, {Hogg}, \& {Mykytyn}}]{tractor}
{Lang}, D., {Hogg}, D.~W., \& {Mykytyn}, D. 2016, {The Tractor: Probabilistic
  astronomical source detection and measurement}, Astrophysics Source Code
  Library, record ascl:1604.008.
\newblock \doeprint{1604.008}

\bibitem[{{Lanusse} {et~al.}(2018){Lanusse}, {Ma}, {Li}, {Collett}, {Li},
  {Ravanbakhsh}, {Mandelbaum}, \& {P{\'o}czos}}]{cmudeeplens}
{Lanusse}, F., {Ma}, Q., {Li}, N., {et~al.} 2018, \mnras, 473, 3895,
  \dodoi{10.1093/mnras/stx1665}

\bibitem[{{Lawrence} {et~al.}(2007){Lawrence}, {Warren}, {Almaini}, {Edge},
  {Hambly}, {Jameson}, {Lucas}, {Casali}, {Adamson}, {Dye}, {Emerson},
  {Foucaud}, {Hewett}, {Hirst}, {Hodgkin}, {Irwin}, {Lodieu}, {McMahon},
  {Simpson}, {Smail}, {Mortlock}, \& {Folger}}]{ukidss}
{Lawrence}, A., {Warren}, S.~J., {Almaini}, O., {et~al.} 2007, \mnras, 379,
  1599, \dodoi{10.1111/j.1365-2966.2007.12040.x}

\bibitem[{{Lemon} {et~al.}(2020){Lemon}, {Auger}, {McMahon}, {Anguita},
  {Apostolovski}, {Chen}, {Fassnacht}, {Melo}, {Motta}, {Shajib}, {Treu},
  {Agnello}, {Buckley-Geer}, {Schechter}, {Birrer}, {Collett}, {Courbin},
  {Rusu}, {Abbott}, {Allam}, {Annis}, {Avila}, {Bertin}, {Brooks}, {Burke},
  {Carnero Rosell}, {Carrasco Kind}, {Carretero}, {Costanzi}, {da Costa}, {De
  Vicente}, {Desai}, {Eifler}, {Flaugher}, {Frieman}, {Garc{\'\i}a-Bellido},
  {Gaztanaga}, {Gerdes}, {Gruen}, {Gruendl}, {Gschwend}, {Gutierrez},
  {Honscheid}, {James}, {Kim}, {Krause}, {Kuehn}, {Kuropatkin}, {Lahav},
  {Lima}, {Lin}, {Maia}, {March}, {Marshall}, {Menanteau}, {Miquel}, {Palmese},
  {Paz-Chinch{\'o}n}, {Plazas}, {Roodman}, {Sanchez}, {Schubnell}, {Serrano},
  {Smith}, {Soares-Santos}, {Suchyta}, {Tarle}, \& {Walker}}]{lemon20}
{Lemon}, C., {Auger}, M.~W., {McMahon}, R., {et~al.} 2020, \mnras, 494, 3491,
  \dodoi{10.1093/mnras/staa652}

\bibitem[{{Lemon} {et~al.}(2022){Lemon}, {Anguita}, {Auger}, {Courbin},
  {Galan}, {McMahon}, {Neira}, {Oguri}, {Schechter}, {Shajib}, \&
  {Treu}}]{lemon22}
{Lemon}, C., {Anguita}, T., {Auger}, M., {et~al.} 2022, arXiv e-prints,
  arXiv:2206.07714.
\newblock \doarXiv{2206.07714}

\bibitem[{{Lemon} {et~al.}(2019){Lemon}, {Auger}, \& {McMahon}}]{lemon19}
{Lemon}, C.~A., {Auger}, M.~W., \& {McMahon}, R.~G. 2019, \mnras, 483, 4242,
  \dodoi{10.1093/mnras/sty3366}

\bibitem[{{Lemon} {et~al.}(2017){Lemon}, {Auger}, {McMahon}, \&
  {Koposov}}]{lemon17}
{Lemon}, C.~A., {Auger}, M.~W., {McMahon}, R.~G., \& {Koposov}, S.~E. 2017,
  \mnras, 472, 5023, \dodoi{10.1093/mnras/stx2094}

\bibitem[{{Lemon} {et~al.}(2018){Lemon}, {Auger}, {McMahon}, \&
  {Ostrovski}}]{lemon18}
{Lemon}, C.~A., {Auger}, M.~W., {McMahon}, R.~G., \& {Ostrovski}, F. 2018,
  \mnras, 479, 5060, \dodoi{10.1093/mnras/sty911}

\bibitem[{{Marocco} {et~al.}(2021){Marocco}, {Eisenhardt}, {Fowler},
  {Kirkpatrick}, {Meisner}, {Schlafly}, {Stanford}, {Garcia}, {Caselden},
  {Cushing}, {Cutri}, {Faherty}, {Gelino}, {Gonzalez}, {Jarrett}, {Koontz},
  {Mainzer}, {Marchese}, {Mobasher}, {Schlegel}, {Stern}, {Teplitz}, \&
  {Wright}}]{catwise}
{Marocco}, F., {Eisenhardt}, P. R.~M., {Fowler}, J.~W., {et~al.} 2021, \apjs,
  253, 8, \dodoi{10.3847/1538-4365/abd805}

\bibitem[{{Marshall} {et~al.}(2020){Marshall}, {Mechtley}, {Windhorst},
  {Cohen}, {Jansen}, {Jiang}, {Jones}, {Wyithe}, {Fan}, {Hathi}, {Jahnke},
  {Keel}, {Koekemoer}, {Marian}, {Ren}, {Robinson}, {R{\"o}ttgering}, {Ryan},
  {Scannapieco}, {Schneider}, {Schneider}, {Smith}, \& {Yan}}]{marshall20}
{Marshall}, M.~A., {Mechtley}, M., {Windhorst}, R.~A., {et~al.} 2020, \apj,
  900, 21, \dodoi{10.3847/1538-4357/abaa4c}

\bibitem[{{Mason} {et~al.}(2015){Mason}, {Treu}, {Schmidt}, {Collett},
  {Trenti}, {Marshall}, {Barone-Nugent}, {Bradley}, {Stiavelli}, \&
  {Wyithe}}]{mason15}
{Mason}, C.~A., {Treu}, T., {Schmidt}, K.~B., {et~al.} 2015, \apj, 805, 79,
  \dodoi{10.1088/0004-637X/805/1/79}

\bibitem[{{McGreer} {et~al.}(2016){McGreer}, {Eftekharzadeh}, {Myers}, \&
  {Fan}}]{mcgreer16}
{McGreer}, I.~D., {Eftekharzadeh}, S., {Myers}, A.~D., \& {Fan}, X. 2016, \aj,
  151, 61, \dodoi{10.3847/0004-6256/151/3/61}

\bibitem[{{McGreer} {et~al.}(2013){McGreer}, {Jiang}, {Fan}, {Richards},
  {Strauss}, {Ross}, {White}, {Shen}, {Schneider}, {Myers}, {Brandt}, {DeGraf},
  {Glikman}, {Ge}, \& {Streblyanska}}]{simqso}
{McGreer}, I.~D., {Jiang}, L., {Fan}, X., {et~al.} 2013, \apj, 768, 105,
  \dodoi{10.1088/0004-637X/768/2/105}

\bibitem[{{McMahon} {et~al.}(2013){McMahon}, {Banerji}, {Gonzalez}, {Koposov},
  {Bejar}, {Lodieu}, {Rebolo}, \& {VHS Collaboration}}]{vhs}
{McMahon}, R.~G., {Banerji}, M., {Gonzalez}, E., {et~al.} 2013, The Messenger,
  154, 35

\bibitem[{{More} {et~al.}(2016){More}, {Oguri}, {Kayo}, {Zinn}, {Strauss},
  {Santiago}, {Mosquera}, {Inada}, {Kochanek}, {Rusu}, {Brownstein}, {da
  Costa}, {Kneib}, {Maia}, {Quimby}, {Schneider}, {Streblyanska}, \&
  {York}}]{more16}
{More}, A., {Oguri}, M., {Kayo}, I., {et~al.} 2016, \mnras, 456, 1595,
  \dodoi{10.1093/mnras/stv2813}

\bibitem[{{Oguri} \& {Marshall}(2010)}]{om10}
{Oguri}, M., \& {Marshall}, P.~J. 2010, \mnras, 405, 2579,
  \dodoi{10.1111/j.1365-2966.2010.16639.x}

\bibitem[{{Oguri} {et~al.}(2006){Oguri}, {Inada}, {Pindor}, {Strauss},
  {Richards}, {Hennawi}, {Turner}, {Lupton}, {Schneider}, {Fukugita}, \&
  {Brinkmann}}]{oguri06}
{Oguri}, M., {Inada}, N., {Pindor}, B., {et~al.} 2006, \aj, 132, 999,
  \dodoi{10.1086/506019}

\bibitem[{{Paraficz} {et~al.}(2018){Paraficz}, {Rybak}, {McKean}, {Vegetti},
  {Sluse}, {Courbin}, {Stacey}, {Suyu}, {Dessauges-Zavadsky}, {Fassnacht}, \&
  {Koopmans}}]{paraficz18}
{Paraficz}, D., {Rybak}, M., {McKean}, J.~P., {et~al.} 2018, \aap, 613, A34,
  \dodoi{10.1051/0004-6361/201731250}

\bibitem[{{Peng} {et~al.}(2002){Peng}, {Ho}, {Impey}, \& {Rix}}]{galfit}
{Peng}, C.~Y., {Ho}, L.~C., {Impey}, C.~D., \& {Rix}, H.-W. 2002, \aj, 124,
  266, \dodoi{10.1086/340952}

\bibitem[{Prochaska {et~al.}(2020)Prochaska, Hennawi, Westfall, Cooke, Wang,
  Hsyu, Davies, Farina, \& Pelliccia}]{pypeit}
Prochaska, J.~X., Hennawi, J.~F., Westfall, K.~B., {et~al.} 2020, Journal of
  Open Source Software, 5, 2308, \dodoi{10.21105/joss.02308}

\bibitem[{{Reis} {et~al.}(2019){Reis}, {Baron}, \& {Shahaf}}]{prf}
{Reis}, I., {Baron}, D., \& {Shahaf}, S. 2019, \aj, 157, 16,
  \dodoi{10.3847/1538-3881/aaf101}

\bibitem[{{Runnoe} {et~al.}(2012){Runnoe}, {Brotherton}, \& {Shang}}]{runnoe12}
{Runnoe}, J.~C., {Brotherton}, M.~S., \& {Shang}, Z. 2012, \mnras, 422, 478,
  \dodoi{10.1111/j.1365-2966.2012.20620.x}

\bibitem[{{Scaramella} {et~al.}(2021){Scaramella}, {Amiaux}, {Mellier},
  {Burigana}, {Carvalho}, {Cuillandre}, {Da Silva}, {Derosa}, {Dinis},
  {Maiorano}, {Maris}, {Tereno}, {Laureijs}, {Boenke}, {Buenadicha}, {Dupac},
  {Gaspar Venancio}, {G{\'o}mez-{\'A}lvarez}, {Hoar}, {Alvarez}, {Racca},
  {Saavedra-Criado}, {Schwartz}, {Vavrek}, {Schirmer}, {Aussel}, {Azzollini},
  {Cardone}, {Cropper}, {Ealet}, {Garilli}, {Gillard}, {Granett}, {Guzzo},
  {Hoekstra}, {Jahnke}, {Kitching}, {Meneghetti}, {Miller}, {Nakajima},
  {Niemi}, {Pasian}, {Percival}, {Sauvage}, {Scodeggio}, {Wachter}, {Zacchei},
  {Aghanim}, {Amara}, {Auphan}, {Auricchio}, {Awan}, {Balestra}, {Bender},
  {Bodendorf}, {Bonino}, {Branchini}, {Brau-Nogue}, {Brescia}, {Candini},
  {Capobianco}, {Carbone}, {Carlberg}, {Carretero}, {Casas}, {Castander},
  {Castellano}, {Cavuoti}, {Cimatti}, {Cledassou}, {Congedo}, {Conselice},
  {Conversi}, {Copin}, {Corcione}, {Costille}, {Courbin}, {Degaudenzi},
  {Douspis}, {Dubath}, {Duncan}, {Dusini}, {Farrens}, {Ferriol}, {Fosalba},
  {Fourmanoit}, {Frailis}, {Franceschi}, {Franzetti}, {Fumana}, {Gillis},
  {Giocoli}, {Grazian}, {Grupp}, {Haugan}, {Holmes}, {Hormuth}, {Hudelot},
  {Kermiche}, {Kiessling}, {Kilbinger}, {Kohley}, {Kubik}, {K{\"u}mmel},
  {Kunz}, {Kurki-Suonio}, {Ligori}, {Lilje}, {Lloro}, {Mansutti}, {Marggraf},
  {Markovic}, {Marulli}, {Massey}, {Maurogordato}, {Melchior}, {Merlin},
  {Meylan}, {Mohr}, {Moresco}, {Morin}, {Moscardini}, {Munari}, {Nichol},
  {Padilla}, {Paltani}, {Peacock}, {Pedersen}, {Pettorino}, {Pires}, {Poncet},
  {Popa}, {Pozzetti}, {Raison}, {Rebolo}, {Rhodes}, {Rix}, {Roncarelli},
  {Rossetti}, {Saglia}, {Schneider}, {Schrabback}, {Secroun}, {Seidel},
  {Serrano}, {Sirignano}, {Sirri}, {Skottfelt}, {Stanco}, {Starck},
  {Tallada-Cresp{\'\i}}, {Tavagnacco}, {Taylor}, {Teplitz}, {Toledo-Moreo},
  {Torradeflot}, {Trifoglio}, {Valentijn}, {Valenziano}, {Verdoes Kleijn},
  {Wang}, {Welikala}, {Weller}, {Wetzstein}, {Zamorani}, {Zoubian}, {Andreon},
  {Baldi}, {Bardelli}, {Boucaud}, {Camera}, {Fabbian}, {Farinelli},
  {Graci{\'a}-Carpio}, {Maino}, {Medinaceli}, {Mei}, {Neissner}, {Polenta},
  {Renzi}, {Romelli}, {Rosset}, {Sureau}, {Tenti}, {Vassallo}, {Zucca},
  {Baccigalupi}, {Balaguera-Antol{\'\i}nez}, {Battaglia}, {Biviano}, {Borgani},
  {Bozzo}, {Cabanac}, {Cappi}, {Casas}, {Castignani}, {Colodro-Conde},
  {Coupon}, {Courtois}, {Cuby}, {de la Torre}, {Desai}, {Di Ferdinando},
  {Dole}, {Fabricius}, {Farina}, {Ferreira}, {Finelli}, {Flose-Reimberg},
  {Fotopoulou}, {Galeotta}, {Ganga}, {Gozaliasl}, {Hook}, {Keihanen},
  {Kirkpatrick}, {Liebing}, {Lindholm}, {Mainetti}, {Martinelli}, {Martinet},
  {Maturi}, {McCracken}, {Metcalf}, {Morgante}, {Nightingale}, {Nucita},
  {Patrizii}, {Potter}, {Riccio}, {S{\'a}nchez}, {Sapone}, {Schewtschenko},
  {Schultheis}, {Scottez}, {Teyssier}, {Tutusaus}, {Valiviita}, {Viel},
  {Vriend}, \& {Whittaker}}]{euclid}
{Scaramella}, R., {Amiaux}, J., {Mellier}, Y., {et~al.} 2021, arXiv e-prints,
  arXiv:2108.01201.
\newblock \doarXiv{2108.01201}

\bibitem[{{Schechter} {et~al.}(2017){Schechter}, {Morgan}, {Chehade},
  {Metcalfe}, {Shanks}, \& {McDonald}}]{schechter17}
{Schechter}, P.~L., {Morgan}, N.~D., {Chehade}, B., {et~al.} 2017, \aj, 153,
  219, \dodoi{10.3847/1538-3881/aa6899}

\bibitem[{{Schindler} {et~al.}(2018){Schindler}, {Fan}, {McGreer}, {Yang},
  {Wang}, {Green}, {Garavito-Camargo}, {Huang}, {O'Donnell}, {Patej}, {Pucha},
  {Rees}, \& {Spalding}}]{schindler18}
{Schindler}, J.-T., {Fan}, X., {McGreer}, I.~D., {et~al.} 2018, \apj, 863, 144,
  \dodoi{10.3847/1538-4357/aad2dd}

\bibitem[{{Sevilla-Noarbe} {et~al.}(2018){Sevilla-Noarbe}, {Hoyle},
  {March{\~a}}, {Soumagnac}, {Bechtol}, {Drlica-Wagner}, {Abdalla},
  {Aleksi{\'c}}, {Avestruz}, {Balbinot}, {Banerji}, {Bertin}, {Bonnett},
  {Brunner}, {Carrasco-Kind}, {Choi}, {Giannantonio}, {Kim}, {Lahav}, {Moraes},
  {Nord}, {Ross}, {Rykoff}, {Santiago}, {Sheldon}, {Wei}, {Wester}, {Yanny},
  {Abbott}, {Allam}, {Brooks}, {Carnero-Rosell}, {Carretero}, {Cunha}, {da
  Costa}, {Davis}, {de Vicente}, {Desai}, {Doel}, {Fernandez}, {Flaugher},
  {Frieman}, {Garcia-Bellido}, {Gaztanaga}, {Gruen}, {Gruendl}, {Gschwend},
  {Gutierrez}, {Hollowood}, {Honscheid}, {James}, {Jeltema}, {Kirk}, {Krause},
  {Kuehn}, {Li}, {Lima}, {Maia}, {March}, {McMahon}, {Menanteau}, {Miquel},
  {Ogando}, {Plazas}, {Sanchez}, {Scarpine}, {Schindler}, {Schubnell}, {Smith},
  {Smith}, {Soares-Santos}, {Sobreira}, {Suchyta}, {Swanson}, {Tarle},
  {Thomas}, {Tucker}, {Walker}, \& {DES Collaboration}}]{sn18}
{Sevilla-Noarbe}, I., {Hoyle}, B., {March{\~a}}, M.~J., {et~al.} 2018, \mnras,
  481, 5451, \dodoi{10.1093/mnras/sty2579}

\bibitem[{{Shajib} {et~al.}(2020){Shajib}, {Birrer}, {Treu}, {Agnello},
  {Buckley-Geer}, {Chan}, {Christensen}, {Lemon}, {Lin}, {Millon}, {Poh},
  {Rusu}, {Sluse}, {Spiniello}, {Chen}, {Collett}, {Courbin}, {Fassnacht},
  {Frieman}, {Galan}, {Gilman}, {More}, {Anguita}, {Auger}, {Bonvin},
  {McMahon}, {Meylan}, {Wong}, {Abbott}, {Annis}, {Avila}, {Bechtol}, {Brooks},
  {Brout}, {Burke}, {Carnero Rosell}, {Carrasco Kind}, {Carretero},
  {Castander}, {Costanzi}, {da Costa}, {De Vicente}, {Desai}, {Dietrich},
  {Doel}, {Drlica-Wagner}, {Evrard}, {Finley}, {Flaugher}, {Fosalba},
  {Garc{\'\i}a-Bellido}, {Gerdes}, {Gruen}, {Gruendl}, {Gschwend}, {Gutierrez},
  {Hollowood}, {Honscheid}, {Huterer}, {James}, {Jeltema}, {Krause},
  {Kuropatkin}, {Li}, {Lima}, {MacCrann}, {Maia}, {Marshall}, {Melchior},
  {Miquel}, {Ogando}, {Palmese}, {Paz-Chinch{\'o}n}, {Plazas}, {Romer},
  {Roodman}, {Sako}, {Sanchez}, {Santiago}, {Scarpine}, {Schubnell}, {Scolnic},
  {Serrano}, {Sevilla-Noarbe}, {Smith}, {Soares-Santos}, {Suchyta}, {Tarle},
  {Thomas}, {Walker}, \& {Zhang}}]{shajib20}
{Shajib}, A.~J., {Birrer}, S., {Treu}, T., {et~al.} 2020, \mnras, 494, 6072,
  \dodoi{10.1093/mnras/staa828}

\bibitem[{{Shen} {et~al.}(2021){Shen}, {Chen}, {Hwang}, {Liu}, {Zakamska},
  {Oguri}, {Li}, {Lazio}, \& {Breiding}}]{shen21}
{Shen}, Y., {Chen}, Y.-C., {Hwang}, H.-C., {et~al.} 2021, Nature Astronomy, 5,
  569, \dodoi{10.1038/s41550-021-01323-1}

\bibitem[{{Shen} {et~al.}(2022){Shen}, {Hwang}, {Oguri}, {Chen}, {Di Matteo},
  {Ni}, {Bird}, {Zakamska}, {Liu}, {Chen}, \& {Kratter}}]{shen22}
{Shen}, Y., {Hwang}, H.-C., {Oguri}, M., {et~al.} 2022, arXiv e-prints,
  arXiv:2208.04979.
\newblock \doarXiv{2208.04979}

\bibitem[{{Silverman} {et~al.}(2020){Silverman}, {Tang}, {Lee}, {Hartwig},
  {Goulding}, {Strauss}, {Schramm}, {Ding}, {Riffel}, {Fujimoto}, {Hikage},
  {Imanishi}, {Iwasawa}, {Jahnke}, {Kayo}, {Kashikawa}, {Kawaguchi}, {Kohno},
  {Luo}, {Matsuoka}, {Matsuda}, {Nagao}, {Oguri}, {Ono}, {Onoue}, {Ouchi},
  {Shimasaku}, {Suh}, {Suzuki}, {Taniguchi}, {Toba}, {Ueda}, \&
  {Yasuda}}]{silverman20}
{Silverman}, J.~D., {Tang}, S., {Lee}, K.-G., {et~al.} 2020, \apj, 899, 154,
  \dodoi{10.3847/1538-4357/aba4a3}

\bibitem[{{Sluse} {et~al.}(2012){Sluse}, {Hutsem{\'e}kers}, {Courbin},
  {Meylan}, \& {Wambsganss}}]{sluse12}
{Sluse}, D., {Hutsem{\'e}kers}, D., {Courbin}, F., {Meylan}, G., \&
  {Wambsganss}, J. 2012, \aap, 544, A62, \dodoi{10.1051/0004-6361/201219125}

\bibitem[{{Steinborn} {et~al.}(2016){Steinborn}, {Dolag}, {Comerford},
  {Hirschmann}, {Remus}, \& {Teklu}}]{steinborn16}
{Steinborn}, L.~K., {Dolag}, K., {Comerford}, J.~M., {et~al.} 2016, \mnras,
  458, 1013, \dodoi{10.1093/mnras/stw316}

\bibitem[{{Tang} {et~al.}(2021){Tang}, {Silverman}, {Ding}, {Li}, {Lee},
  {Strauss}, {Goulding}, {Schramm}, {Kawinwanichakij}, {Xavier Prochaska},
  {Hennawi}, {Imanishi}, {Iwasawa}, {Toba}, {Kayo}, {Oguri}, {Matsuoka},
  {Onoue}, {Jahnke}, {Ichikawa}, {Hartwig}, {Kashikawa}, {Kawaguchi}, {Kohno},
  {Matsuda}, {Nagao}, {Ono}, {Ouchi}, {Shimasaku}, {Suh}, {Suzuki},
  {Taniguchi}, {Ueda}, \& {Yasuda}}]{tang21}
{Tang}, S., {Silverman}, J.~D., {Ding}, X., {et~al.} 2021, \apj, 922, 83,
  \dodoi{10.3847/1538-4357/ac1ff0}

\bibitem[{{Treu} {et~al.}(2018){Treu}, {Agnello}, {Baumer}, {Birrer},
  {Buckley-Geer}, {Courbin}, {Kim}, {Lin}, {Marshall}, {Nord}, {Schechter},
  {Sivakumar}, {Abramson}, {Anguita}, {Apostolovski}, {Auger}, {Chan}, {Chen},
  {Collett}, {Fassnacht}, {Hsueh}, {Lemon}, {McMahon}, {Motta}, {Ostrovski},
  {Rojas}, {Rusu}, {Williams}, {Frieman}, {Meylan}, {Suyu}, {Abbott},
  {Abdalla}, {Allam}, {Annis}, {Avila}, {Banerji}, {Brooks}, {Carnero Rosell},
  {Carrasco Kind}, {Carretero}, {Castander}, {D'Andrea}, {da Costa}, {De
  Vicente}, {Doel}, {Eifler}, {Flaugher}, {Fosalba}, {Garc{\'\i}a-Bellido},
  {Goldstein}, {Gruen}, {Gruendl}, {Gutierrez}, {Hartley}, {Hollowood},
  {Honscheid}, {James}, {Kuehn}, {Kuropatkin}, {Lima}, {Maia}, {Martini},
  {Menanteau}, {Miquel}, {Plazas}, {Romer}, {Sanchez}, {Scarpine}, {Schindler},
  {Schubnell}, {Sevilla-Noarbe}, {Smith}, {Smith}, {Soares-Santos}, {Sobreira},
  {Suchyta}, {Swanson}, {Tarle}, {Thomas}, {Tucker}, \& {Walker}}]{treu18}
{Treu}, T., {Agnello}, A., {Baumer}, M.~A., {et~al.} 2018, \mnras, 481, 1041,
  \dodoi{10.1093/mnras/sty2329}

\bibitem[{{Tsuzuki} {et~al.}(2006){Tsuzuki}, {Kawara}, {Yoshii}, {Oyabu},
  {Tanab{\'e}}, \& {Matsuoka}}]{tsuzuki06}
{Tsuzuki}, Y., {Kawara}, K., {Yoshii}, Y., {et~al.} 2006, \apj, 650, 57,
  \dodoi{10.1086/506376}

\bibitem[{{van der Walt} {et~al.}(2011){van der Walt}, {Colbert}, \&
  {Varoquaux}}]{numpy1}
{van der Walt}, S., {Colbert}, S.~C., \& {Varoquaux}, G. 2011, Computing in
  Science and Engineering, 13, 22, \dodoi{10.1109/MCSE.2011.37}

\bibitem[{{Vestergaard} \& {Osmer}(2009)}]{vo09}
{Vestergaard}, M., \& {Osmer}, P.~S. 2009, \apj, 699, 800,
  \dodoi{10.1088/0004-637X/699/1/800}

\bibitem[{{Vestergaard} \& {Peterson}(2006)}]{vp06}
{Vestergaard}, M., \& {Peterson}, B.~M. 2006, \apj, 641, 689,
  \dodoi{10.1086/500572}

\bibitem[{{Volonteri} {et~al.}(2016){Volonteri}, {Dubois}, {Pichon}, \&
  {Devriendt}}]{volonteri16}
{Volonteri}, M., {Dubois}, Y., {Pichon}, C., \& {Devriendt}, J. 2016, \mnras,
  460, 2979, \dodoi{10.1093/mnras/stw1123}

\bibitem[{{Wang} {et~al.}(2016){Wang}, {Wu}, {Fan}, {Yang}, {Yi}, {Bian},
  {McGreer}, {Yang}, {Ai}, {Dong}, {Zuo}, {Jiang}, {Green}, {Wang}, {Cai},
  {Wang}, \& {Yue}}]{wang16}
{Wang}, F., {Wu}, X.-B., {Fan}, X., {et~al.} 2016, \apj, 819, 24,
  \dodoi{10.3847/0004-637X/819/1/24}

\bibitem[{{Williams} {et~al.}(2018){Williams}, {Curtis-Lake}, {Hainline},
  {Chevallard}, {Robertson}, {Charlot}, {Endsley}, {Stark}, {Willmer},
  {Alberts}, {Amorin}, {Arribas}, {Baum}, {Bunker}, {Carniani}, {Crandall},
  {Egami}, {Eisenstein}, {Ferruit}, {Husemann}, {Maseda}, {Maiolino}, {Rawle},
  {Rieke}, {Smit}, {Tacchella}, \& {Willott}}]{jaguar}
{Williams}, C.~C., {Curtis-Lake}, E., {Hainline}, K.~N., {et~al.} 2018, \apjs,
  236, 33, \dodoi{10.3847/1538-4365/aabcbb}

\bibitem[{{Wong} {et~al.}(2020){Wong}, {Suyu}, {Chen}, {Rusu}, {Millon},
  {Sluse}, {Bonvin}, {Fassnacht}, {Taubenberger}, {Auger}, {Birrer}, {Chan},
  {Courbin}, {Hilbert}, {Tihhonova}, {Treu}, {Agnello}, {Ding}, {Jee},
  {Komatsu}, {Shajib}, {Sonnenfeld}, {Blandford}, {Koopmans}, {Marshall}, \&
  {Meylan}}]{wong19}
{Wong}, K.~C., {Suyu}, S.~H., {Chen}, G. C.~F., {et~al.} 2020, \mnras, 498,
  1420, \dodoi{10.1093/mnras/stz3094}

\bibitem[{{Worseck} {et~al.}(2014){Worseck}, {Prochaska}, {O'Meara}, {Becker},
  {Ellison}, {Lopez}, {Meiksin}, {M{\'e}nard}, {Murphy}, \&
  {Fumagalli}}]{worseck14}
{Worseck}, G., {Prochaska}, J.~X., {O'Meara}, J.~M., {et~al.} 2014, \mnras,
  445, 1745, \dodoi{10.1093/mnras/stu1827}

\bibitem[{{Wu} \& {Shen}(2022)}]{dr16q}
{Wu}, Q., \& {Shen}, Y. 2022, arXiv e-prints, arXiv:2209.03987.
\newblock \doarXiv{2209.03987}

\bibitem[{{Yang} {et~al.}(2019{\natexlab{a}}){Yang}, {Venemans}, {Wang}, {Fan},
  {Novak}, {Decarli}, {Walter}, {Yue}, {Momjian}, {Keeton}, {Wang},
  {Zabludoff}, {Wu}, \& {Bian}}]{yang19lens}
{Yang}, J., {Venemans}, B., {Wang}, F., {et~al.} 2019{\natexlab{a}}, \apj, 880,
  153, \dodoi{10.3847/1538-4357/ab2a02}

\bibitem[{{Yang} {et~al.}(2019{\natexlab{b}}){Yang}, {Wang}, {Fan}, {Wu},
  {Bian}, {Ba{\~n}ados}, {Yue}, {Schindler}, {Yang}, {Jiang}, {McGreer},
  {Green}, \& {Dye}}]{yang19}
{Yang}, J., {Wang}, F., {Fan}, X., {et~al.} 2019{\natexlab{b}}, \apj, 871, 199,
  \dodoi{10.3847/1538-4357/aaf858}

\bibitem[{{Yang} {et~al.}(2022){Yang}, {Fan}, {Wang}, {Lanzuisi}, {Nanni},
  {Cappi}, {Chartas}, {Dadina}, {Decarli}, {Jin}, {Keeton}, {Venemans},
  {Walter}, {Wang}, {Wu}, {Yue}, \& {Zabludoff}}]{yang22}
{Yang}, J., {Fan}, X., {Wang}, F., {et~al.} 2022, \apjl, 924, L25,
  \dodoi{10.3847/2041-8213/ac45f2}

\bibitem[{{York} {et~al.}(2000){York}, {Adelman}, {Anderson}, {Anderson},
  {Annis}, {Bahcall}, {Bakken}, {Barkhouser}, {Bastian}, {Berman}, {Boroski},
  {Bracker}, {Briegel}, {Briggs}, {Brinkmann}, {Brunner}, {Burles}, {Carey},
  {Carr}, {Castander}, {Chen}, {Colestock}, {Connolly}, {Crocker}, {Csabai},
  {Czarapata}, {Davis}, {Doi}, {Dombeck}, {Eisenstein}, {Ellman}, {Elms},
  {Evans}, {Fan}, {Federwitz}, {Fiscelli}, {Friedman}, {Frieman}, {Fukugita},
  {Gillespie}, {Gunn}, {Gurbani}, {de Haas}, {Haldeman}, {Harris}, {Hayes},
  {Heckman}, {Hennessy}, {Hindsley}, {Holm}, {Holmgren}, {Huang}, {Hull},
  {Husby}, {Ichikawa}, {Ichikawa}, {Ivezi{\'c}}, {Kent}, {Kim}, {Kinney},
  {Klaene}, {Kleinman}, {Kleinman}, {Knapp}, {Korienek}, {Kron}, {Kunszt},
  {Lamb}, {Lee}, {Leger}, {Limmongkol}, {Lindenmeyer}, {Long}, {Loomis},
  {Loveday}, {Lucinio}, {Lupton}, {MacKinnon}, {Mannery}, {Mantsch}, {Margon},
  {McGehee}, {McKay}, {Meiksin}, {Merelli}, {Monet}, {Munn}, {Narayanan},
  {Nash}, {Neilsen}, {Neswold}, {Newberg}, {Nichol}, {Nicinski}, {Nonino},
  {Okada}, {Okamura}, {Ostriker}, {Owen}, {Pauls}, {Peoples}, {Peterson},
  {Petravick}, {Pier}, {Pope}, {Pordes}, {Prosapio}, {Rechenmacher}, {Quinn},
  {Richards}, {Richmond}, {Rivetta}, {Rockosi}, {Ruthmansdorfer}, {Sandford},
  {Schlegel}, {Schneider}, {Sekiguchi}, {Sergey}, {Shimasaku}, {Siegmund},
  {Smee}, {Smith}, {Snedden}, {Stone}, {Stoughton}, {Strauss}, {Stubbs},
  {SubbaRao}, {Szalay}, {Szapudi}, {Szokoly}, {Thakar}, {Tremonti}, {Tucker},
  {Uomoto}, {Vanden Berk}, {Vogeley}, {Waddell}, {Wang}, {Watanabe},
  {Weinberg}, {Yanny}, {Yasuda}, \& {SDSS Collaboration}}]{sdss}
{York}, D.~G., {Adelman}, J., {Anderson}, John~E., J., {et~al.} 2000, \aj, 120,
  1579, \dodoi{10.1086/301513}

\bibitem[{{Yue} {et~al.}(2021{\natexlab{a}}){Yue}, {Fan}, {Yang}, \&
  {Wang}}]{yue21b}
{Yue}, M., {Fan}, X., {Yang}, J., \& {Wang}, F. 2021{\natexlab{a}}, \apjl, 921,
  L27, \dodoi{10.3847/2041-8213/ac31a9}

\bibitem[{{Yue} {et~al.}(2022{\natexlab{a}}){Yue}, {Fan}, {Yang}, \&
  {Wang}}]{yue22b}
---. 2022{\natexlab{a}}, \aj, 163, 139, \dodoi{10.3847/1538-3881/ac4cb0}

\bibitem[{{Yue} {et~al.}(2022{\natexlab{b}}){Yue}, {Fan}, {Yang}, \&
  {Wang}}]{yue22a}
---. 2022{\natexlab{b}}, \apj, 925, 169, \dodoi{10.3847/1538-4357/ac409b}

\bibitem[{{Yue} {et~al.}(2021{\natexlab{b}}){Yue}, {Yang}, {Fan}, {Wang},
  {Spilker}, {Georgiev}, {Keeton}, {Litke}, {Marrone}, {Walter}, {Wang}, {Wu},
  {Venemans}, \& {Zabludoff}}]{yue21a}
{Yue}, M., {Yang}, J., {Fan}, X., {et~al.} 2021{\natexlab{b}}, \apj, 917, 99,
  \dodoi{10.3847/1538-4357/ac0af4}

\end{thebibliography}
\bibliographystyle{aasjournal}

\appendix

\section{Information of Candidates with Follow-Up Spectroscopy}

Here we summarize all objects with follow-up spectroscopy.
The spectral types of the objects are determined by visual inspection.
Specifically,
if an object exhibits multiple broad emission lines, we confirm 
the object as a quasar and determine its redshift.
An object is reported as a galaxy if it appears to be extended in images
and has a spectrum consistent with a galaxy.
We determine the redshift of the galaxy only if there are prominent emission lines.
Similarly, an object is reported as a star if it appears to be a point source in images
and has a spectrum consistent with a star.
We do not determine the specific spectral types of stars.
If we cannot determine the type of the spectrum 
(usually due to insufficient S/N),
we report the object as inconclusive.
The spectra are available upon request to the corresponding author.



\startlongtable
\begin{deluxetable*}{cccccl}
\label{tbl:contam}
\tablecaption{Candidates with Follow-Up Spectroscopy}
\tablewidth{0pt}
\tablehead{\colhead{Name} & \colhead{RA} & \colhead{Dec} & \colhead{Priority} & \colhead{Spectral Type} &\colhead{Comment}}
\startdata
\hline
J0025--0145 & 00:25:26.83 & --01:45:32.51 & 1 & Quasar & Intermediately lensed, $z=5.07$ \\
J0032+0120 & 00:32:18.14 & +01:20:36.30 & 1 & Galaxy &  \\
J0054+0133 & 00:54:00.42 & +01:33:52.46 & 1 & Star & May be two stars blended \\
J0101+0034 & 01:01:31.34 & +00:34:12.07 & 1 & Star & May be a star and a galaxy blended \\
J0109--0524 & 01:09:13.82 & --05:24:01.52 & 1 & Inconclusive & No trace \\
J0118--1915 & 01:18:44.60 & --19:15:02.57 & 1 & Star & May be a star and a galaxy blended \\
J0124+0553 & 01:24:09.69 & +05:53:43.06 & 1 &  Star + Galaxy &  \\
J0125+0142 & 01:25:21.41 & +01:42:17.21 & 1 & Star & May be two stars and a galaxy blended \\
J0127--0701 & 01:27:48.00 & --07:01:07.84 & 1 & Star & \\
J0133--5804 & 01:33:07.19 & --58:04:26.91 & 1 & Quasar + Inconclusive & Quasar at $z=3.08$\\
J0134--2308 & 01:34:57.48 & --23:08:54.57 & 1 &  Star &  \\
J0151--2242 & 01:51:41.26 & --22:42:09.43 & 1 & Star + Galaxy &  \\
J0154--0250 & 01:54:56.09 & --02:50:54.87 & 1 & EmGal & $z=0.39$ \\
J0158+1541 & 01:58:03.99 & +15:41:50.08 & 1 & Star & \\
J0205+0357 & 02:05:01.41 & +03:57:07.02 & 1 & Star + Galaxy &  \\
J0212--1548 & 02:12:34.63 & --15:48:21.45 & 1 & Quasar & $z=4.75$, no sign of lensing \\
J0223--4709 & 02:23:06.75 & --47:09:02.72 & 1 & Quasar & $z=5.0$, no sign of lensing \\
J0234--0652 & 02:34:15.83 & --06:52:08.58 & 1 & Star + Galaxy &  \\
J0240--0019 & 02:40:25.00 & --00:19:36.52 & 1 & Star & Two stars blended but unresolved in the spectrum\\
J0256--1817 & 02:56:29.09 & --18:17:27.37 & 1 & AGN + Star & AGN at $z=1.22$ \\
J0307--0108 & 03:07:15.57 & --10:18:28.47 & 1 & Galaxy &  \\
J0311--2343 & 03:11:54.69 & --23:43:24.18 & 1 & Galaxy + Star &  \\
J0315--1710 & 03:15:33.62 & --17:10:05.34 & 1 & Quasar & $z=2.79$, no sign of lensing \\
J0348--0438 & 03:48:37.24 & --04:38:14.26 & 1 & Galaxy & \\
J0352--2145 & 03:52:10.62 & --21:45:45.03 & 1 & Quasar & $z=5.1$, no sign of lensing \\
J0402--4220 & 04:02:22.13 & --42:20:53.03 & 1 & Quasar & Lensed, $z=2.88$ \\
J0408--6442 & 04:08:49.69 & --64:42:32.41 & 1 &  Star & \\
J0449--3143 & 04:49:43.33 & --31:43:04.13 & 1 & Quasar + Inconclusive & Quasar at $z=2.07$, blended with a star-like point source \\
J0453--6059 & 04:53:33.73 & --60:59:54.79 & 1 & Star &  \\
J0504--3311 & 05:04:40.51 & --33:11:05.75 & 1 & Inconclusive & May be a FeLoBAL \\
J0527--2958 & 05:27:50.65 & --29:58:52.23 & 1 & Star &  \\
J0532--3005 & 05:32:20.43 & --30:05:22.33 & 1 & AGN & $z=0.36$ \\
J0803+3908 & 08:03:57.75 & +39:08:23.06 & 1 & Quasar  & Lensed $(z=2.98)$, also reported in \citet{lemon22} \\
J0841+2804 & 08:41:19.99 & +28:04:01.82 & 1 & Star &  \\
J0846+2752 & 08:46:00.51 & +27:52:43.06 & 1 & Star &  \\
J0848+2630 & 08:48:34.94 & +26:30:42.77 & 1 & Star &  \\
J0918--0220 & 09:18:43.37 & --02:20:07.41 & 1 & Quasar & Lensed $(z=0.8)$, also reported in \citet{lemon22} \\
J0927+0955 & 09:27:42.73 & +09:55:24.07 & 1 & Star &  \\
J0933--0752 & 09:33:50.65 & --07:52:26.71 & 1 & Galaxy &  \\
J0936--0837 & 09:36:43.21 & --08:37:50.65 & 1 & Star + Galaxy &  \\
J0954--0022 & 09:54:06.97 & --00:22:25.39 & 1 & Quasar & Lensed, $z=2.27$ \\
J1003+0715 & 10:03:04.20 & +07:15:03.95 & 1 & Star &  \\
J1034--0821 & 10:34:14.90 & --08:21:44.91 & 1 & Galaxy &  \\
J1051+0334 & 10:51:32.40 & +03:34:17.66 & 1 & Quasar & Lensed, $z=2.40$ \\
J1102+0109 & 11:02:41.96 & +01:09:40.79 & 1 & Star + Star &  \\
J1225+4831 & 12:25:18.67 & +48:31:16.14 & 1 & Quasar & NIQ, $z=3.08$ \\
J1240+2249 & 12:40:20.56 & +22:49:01.81 & 1 & Inconclusive & Unrecognized spectral features\\
J1250+2827 & 12:50:43.45 & +28:27:14.82 & 1 & Inconclusive & Unrecognized spectral features\\
J1302--0849 & 13:02:48.24 & --08:49:42.81 & 1 & EmGal & $z=0.75$ \\
J1304+1156 & 13:04:56.94 & +11:56:24.59 & 1 & Inconclusive & Very faint trace \\
J1305+2017 & 13:05:18.95 & +20:17:28.92 & 1 & Star & May be two stars blended \\
J1402+2052 & 14:02:52.54 & +20:52:01.24 & 1 & EmGal & $z=0.36$ \\
J1408+0422 & 14:08:33.76 & +04:22:28.04 & 1 & Quasar & Lensed $(z=3.005)$, also reported in \citet{lemon22} \\
J1415--0852 & 14:15:15.44 & --08:52:53.23 & 1 & Quasar + Star & Quasar at $z=4.5$ \\
J1432--0717 & 14:32:22.30 & --07:17:49.87 & 1 & EmGal + Galaxy & EmGal at $z=0.83$ \\
J1442+1540 & 14:42:08.59 & +15:40:42.63 & 1 & EmGal + Galaxy & EmGal at $z=0.45$ \\
J1507--0248 & 15:07:25.36 & --02:48:54.96 & 1 & Star + Star & Two M stars \\
J1619+0033 & 16:19:39.62 & +00:33:54.35 & 1 & Star & \\
J1633+1802 & 16:33:11.08 & +18:02:38.51 & 1 & Star & \\
J1634+1405 & 16:34:26.71 & +14:05:10.44 & 1 & Star & \\
J1658+2507 & 16:58:46.87 & +25:07:21.76 & 1 & Star & \\
J2037--4537 & 20:37:21.27 & --45:37:48.46 & 1 & Quasar & NIQ $(z=5.66)$, reported in \citet{yue21b} \\
J2045--0033 & 20:45:59.26 & --00:33:24.58 & 1 & Star &  \\
J2046--0822 & 20:46:35.96 & --08:22:30.11 & 1 & Star &  \\
J2051--1228 & 20:51:39.46 & --12:28:41.38 & 1 & Star &  \\
J2329--0522 & 23:29:15.10 & --05:22:35.92 & 1 & Quasar &  Pair $(z=4.85)$  \\\hline
J0044--2353 & 00:44:48.04 & --23:53:01.40 & 2 & Quasar &  $z=4.8$, no sign of lensing\\
J0504--2426 & 05:04:30.92 & --24:26:33.10 & 2 & Star &  \\
J0838+2644 & 08:38:27.29 & +26:44:08.09 & 2 & Inconclusive & Very faint trace \\
J0848+1119 & 08:48:33.79 & +11:19:29.38 & 2 & EmGal + EmGal & Both at $z=0.29$ \\
J0921--0027 & 09:21:17.02 & --00:27:43.69 & 2 & Star & May be two stars \\
J0958--0337 & 09:58:03.63 & --03:37:56.24 & 2 & EmGal & $z=0.36$ \\
J1002--0630 & 10:02:44.87 & --06:30:40.08 & 2 & Star &  \\
J1005--0627 & 10:05:56.69 & --06:27:42.33 & 2 & Galaxy & \\
J1009--0314 & 10:09:11.23 & --03:14:44.08 & 2 & Star & \\
J1534--0147 & 15:34:17.85 & --01:47:15.12 & 2 & EmGal + EmGal & Both at $z=0.33$ \\\hline
\enddata
\end{deluxetable*}



\end{document}